\documentclass[12pt]{article}

\usepackage{graphicx}
\graphicspath{{figure_png/}} 

\usepackage[textwidth=8em,textsize=small]{todonotes}
\usepackage{natbib}

\title{Bayesian Spanning Tree: 
                 Estimating the Backbone of the Dependence Graph}
        
\author{
    Leo L. Duan\thanks{Department of Statistics, University of Florida, Gainesville, FL, email: li.duan@ufl.edu}
    \; and \;
    David B. Dunson\thanks{Department of Statistical Science, Duke University, Durham, NC, email: dunson@duke.edu}
    }
    \date{}

\usepackage{bookmark}

\usepackage{subcaption}
\usepackage{float}

\usepackage{hyperref}
\hypersetup{
    pdfpagemode=FullScreen,
    }

\usepackage{amsmath}
\usepackage{amssymb}

\usepackage[ruled,vlined]{algorithm2e}

\usepackage{mhequ}
\usepackage{enumitem}\usepackage[T1]{fontenc}
\usepackage{geometry}
\usepackage{csquotes}
\usepackage{import}
\usepackage{ifoddpage}
\usepackage{relsize}

 \newtheorem{theorem}{Theorem}
 \newtheorem{lemma}{Lemma}
 \newtheorem{remark}{Remark}
 \newtheorem{proof}{Proof}

\renewcommand{\t}{\text}
\newcommand{\T}{^{\rm T}}

\newcommand{\pr}{\text{pr}}

\newcommand{\be}{\begin{equation*}
    \begin{aligned}}

    \newcommand{\ee}{ \end{aligned}
\end{equation*}
}

\newcommand{\bel}{\begin{equation}
    \begin{aligned}}

    \newcommand{\eel}{ \end{aligned}
\end{equation}
}

\begin{document}
                \maketitle

\begin{abstract}
In multivariate data analysis, it is 
often important to estimate a graph characterizing dependence among \(p\) variables. A popular strategy uses 
the non-zero entries in a \(p\times p\) covariance or precision matrix, typically requiring restrictive modeling assumptions for accurate graph recovery.
To improve model robustness, we instead focus on estimating the {\em backbone} of the dependence graph.  We use a spanning tree likelihood, based on a minimalist graphical model that is purposely overly-simplified.  Taking a Bayesian approach, we place a prior on the space of trees and quantify uncertainty in the graphical model.  In both theory and experiments, we show that this model does not require the population graph to be a spanning tree or the covariance to satisfy assumptions beyond positive-definiteness.  The model accurately recovers the backbone of the population graph at a rate competitive with existing approaches but with better robustness.
We show combinatorial properties of the spanning tree, which may be of independent interest, and develop an efficient Gibbs sampler for Bayesian inference.
Analyzing electroencephalography data using a Hidden Markov Model with each latent state modeled by a spanning tree, we show that results are much more interpretable compared with popular alternatives.
             \end{abstract}

\section{Introduction}

In multivariate data analysis, it is commonly of interest to make inferences on the dependence structure in a collection of $p$ random variables.  The covariance matrix provides a typical summary of pairwise dependence between variables, while the inverse of the covariance or precision matrix is used to characterize conditional dependence relationships.  To simplify inferences, a focus is commonly in inferring a dependence graph: 
\(G=(V, E_{G})\), with $V=\{1,\ldots,p\}$ nodes representing the $p$ variables and $E_G=\{e_s\}$ the set of edges. If $(j,k)$ is an edge in $E_G$, there is a dependence relationship between the two variables  \(y_{j}\) and \(y_{k}\).

There is a huge literature on graphical models, encompassing different types of dependence, mostly defined through the covariance among the $y_j$'s, $\Sigma_0$, or its inverse (precision), $\Omega_0$.
Popular examples include: assuming the variables follow a multivariate Gaussian distribution, (i) 
 $\Sigma_{0:j,k}=0$ implies $y_j$ and $y_k$ are statistically independent \citep{dempster1972covariance, cox1996multivariate} and (ii) 
  \(\Omega_{0:j,k}=0\) implies 
   $y_k$ are conditionally independent given all other variables; such graphs have become very popular due to the graphical lasso \citep{friedman2008sparse}. (iii) using the lower-triangular decomposition of $\Omega_0$ after some permutation \(C C^{\rm T}=P \Omega_0 P^{\rm T}\) (\(P\) is a permutation matrix), those non-zero $C_{j,k}$'s give a directed acyclic graph (DAG) as a sequential data generating scheme  \citep{rutimann2009high,cao2019posterior}. There is a rich related literature including more complex elaborations, such as graphs that change over time  \citep{basu2015regularized}, hyper-graphs \citep{roverato2002hyper}, and copula graphical models
 \citep{liu2012high}.

A major practical issue in inferring dependence graphs based on observations of multivariate vectors $y^{(i)}=(y^{(i)}_{1},\ldots,y^{(i)}_{p})^{\rm T}$, for $i=1,\ldots,n$, is that the number of possible graphs is immense for large $p$.  For example, for covariance graphs (i), there are $2^{p(p-1)/2}$ possible graphs, which clearly increases extremely rapidly with $p$. This creates two practical issues.  Firstly, even for moderate $p$, we cannot visit all possible graphs so it becomes challenging to identify the ``best'' graph that is most likely given the observed data.
  Secondly, even if we could identify one best graph, there is likely a large number of alternative graphs that are equally plausible given the 
   observed data.  Hence, whenever we estimate a dependence graph in more than a few variables, we inherently expect a large amount of uncertainty.  There are many available algorithms that deal with the first problem, ranging from the graphical lasso to thresholding the empirical covariance.  However, the resulting point estimates should be interpreted carefully given the second problem.

Most graphical selection procedures leverage on estimates of the covariance or precision, \(\hat\Sigma\) or \(\hat\Omega\).  To obtain fewer errors in graph estimation, one typically needs to first achieve high accuracy in  \(\hat\Sigma\) or \(\hat\Omega\). In large $p$ settings, this is  challenging.
 \cite{cai2016estimating} showed that the empirical covariance \(\hat\Sigma=S_n\) converges to the population \(\Sigma_0\) in \(\|\hat\Sigma - \Sigma_0 \|^2_{op}=O({p/n})\), with \(\|.\|_{op}\) the operator norm. Hence, the sample size \(n\) may need to be substantially larger than \(p\). To obtain an accurate estimate under the $n<p$ scenario, the true \(\Sigma_0\) has to satisfy restrictions. For example, \cite{bickel2008covariance} assumed 
 \(\Sigma_0\) is sparse and proposed a simple thresholding estimator.  \cite{bickel2008regularized} instead assume the off-diagonal elements in \(\Sigma_0\) between \(y_j\) and \(y_k\) decay with \(|j-k|\), and propose a banding/tapering estimator. \cite{yuan2007model} and \cite{friedman2008sparse} 
 instead supposed that the 
 true \(\Omega_0\) is very sparse, motivating the graphical lasso (glasso). For a survey of large $p$ covariance/precision estimators, see \cite{cai2016estimating1}. Roughly speaking, 
one can recover \(\Sigma_0\) (or $\Omega_0$) at an error rate diminishing at \(O(\log p/n)\) if the corresponding assumption holds for the true \(\Sigma_0\).
These assumptions are difficult to verify in practice and violations (e.g, the true graph is dense) may lead to poor performance.

This motivates us to consider a less ambitious task ---  if we cannot accurately recover the whole edge set $E_G$, can we estimate a smaller subset, as an important summary statistic of $G$? Intuitively, a useful edge subset corresponds to the ``backbone'' graph, in which we use as few edges as possible, while connecting as many nodes as permitted. 
This leads us to consider a classic graph/combinatorial statistic called the ``spanning tree'' \citep{kruskal1956shortest}, as the smallest graph that still connects all the nodes. This tree is commonly used to solve the ``traveling salesperson problem'', by finding a simple travel plan that approximately minimizes the total distance traveled between $p$ cities. Similarly, we can imagine a  ``minimal'' generative process: starting from one variable, we sequentially generate a 
 new variable, that only depends on one of the existing variables. This leads to a spanning tree  likelihood.

Following a Bayesian approach, 
 we equip the spanning tree with a prior distribution, allowing us to obtain a posterior distribution over all possible spanning trees, hence quantifying model uncertainty. Importantly, in both theory and numerical experiments, we demonstrate that this model does not require the population $G$ to be a spanning tree, nor the population covariance \(\Sigma_0\) to satisfy any assumptions besides positive-definiteness; yet, it can accurately recover the backbone of the population graph $G$, as the minimum spanning tree transform based on \(\Sigma_0\). Our theory is in the same spirit as the celebrated result of \cite{white1982maximum}, who studied the asymptotic behavior of a restricted model estimator when the data are generated from a different full model; as well as the more recent spiked covariance model \citep{donoho2018optimal} for the optimal estimation of the leading eigenvalues of $\Sigma_0$ using a restricted parameterization. In our case, the posterior distribution on the spanning tree concentrates rapidly (at a negative exponential rate in $n$) around the spanning tree summary of the true graph.  In contrast, if we try to obtain the full graph, we necessarily concentrate critically slowly unless we make overly strong assumptions.  

The minimum spanning tree has been used previously in a variety of statistical contexts. Examples include hypothesis testing \citep{friedman1979multivariate}, 
classification   
 \citep{juszczak2009minimum} and  network analysis \citep{tewarie2015minimum}. However, to our best knowledge, our proposed Bayesian approach is new,  and we present several interesting combinatorial properties regarding the tree, such as the rank constraint, graph-cut partition and the marginal connecting probability for each pair of nodes; they may be of independent interest to the broad statistical community.

The implementation of this model is maintained and accessible at \href{https://github.com/leoduan/Bayesian_spanning_tree}{\url{https://github.com/leoduan/Bayesian_spanning_tree}}.

\section{Bayesian Spanning Tree}

To provide some intuition about the spanning tree as the backbone of a graph, we first briefly review the solutions to the traveling salesperson problem, demonstrate the simplicity of the spanning tree and motivating its use in a generative model.

\subsection{Traveling on a Graph via the Spanning Tree}
Suppose we have a graph \(G=(V, E_{G})\) with nodes \(V=\{1,\ldots,p\}\) and edges \(E_{G} = \{e_1,\ldots, e_M\}\). Each edge is undirected \(e_s=(j,k)\equiv (k,j)\), and associated with a weight \(w_{j,k}=w_{k,j}\ge 0\).

Consider the traveling salesperson problem: suppose \(G\) is a connected graph --- for any two nodes \(j\) and \(k\), there is a \(\text{path}(j,k)=\{(j,l_1),(l_1,l_2),\ldots,(l_{m-1},l_m),(l_m,k)\}\) (a subset of \(E_G\)) that allows us to travel from \(j\) to \(k\). With each node representing a city and $w_{j,k}$ the distance between two cities, a salesperson wants to go to every city, while trying to minimize the total distance traveled. This is a combinatorial optimization problem:
 \[\min_{\mathcal I}Q(\mathcal I)=\min_{\mathcal I} \sum_{(j,k)\in \mathcal I}w_{j,k},\]
where  \(\mathcal I\) is the itinerary, as an ordered sequence of edges.

 Finding the optimal itinerary is a challenging problem. Nevertheless, we can consider a simpler problem that gives a close-to-optimal solution: since those $p$ cities can be connected via $p-1$ edges (roads), what if we first find the best $p-1$ edges with the shortest total distance, and then develop an itinerary on them?

\begin{figure}[H]\centering
\begin{subfigure}[t]{.48\textwidth}
    \includegraphics[width=.9\linewidth,height=6cm]{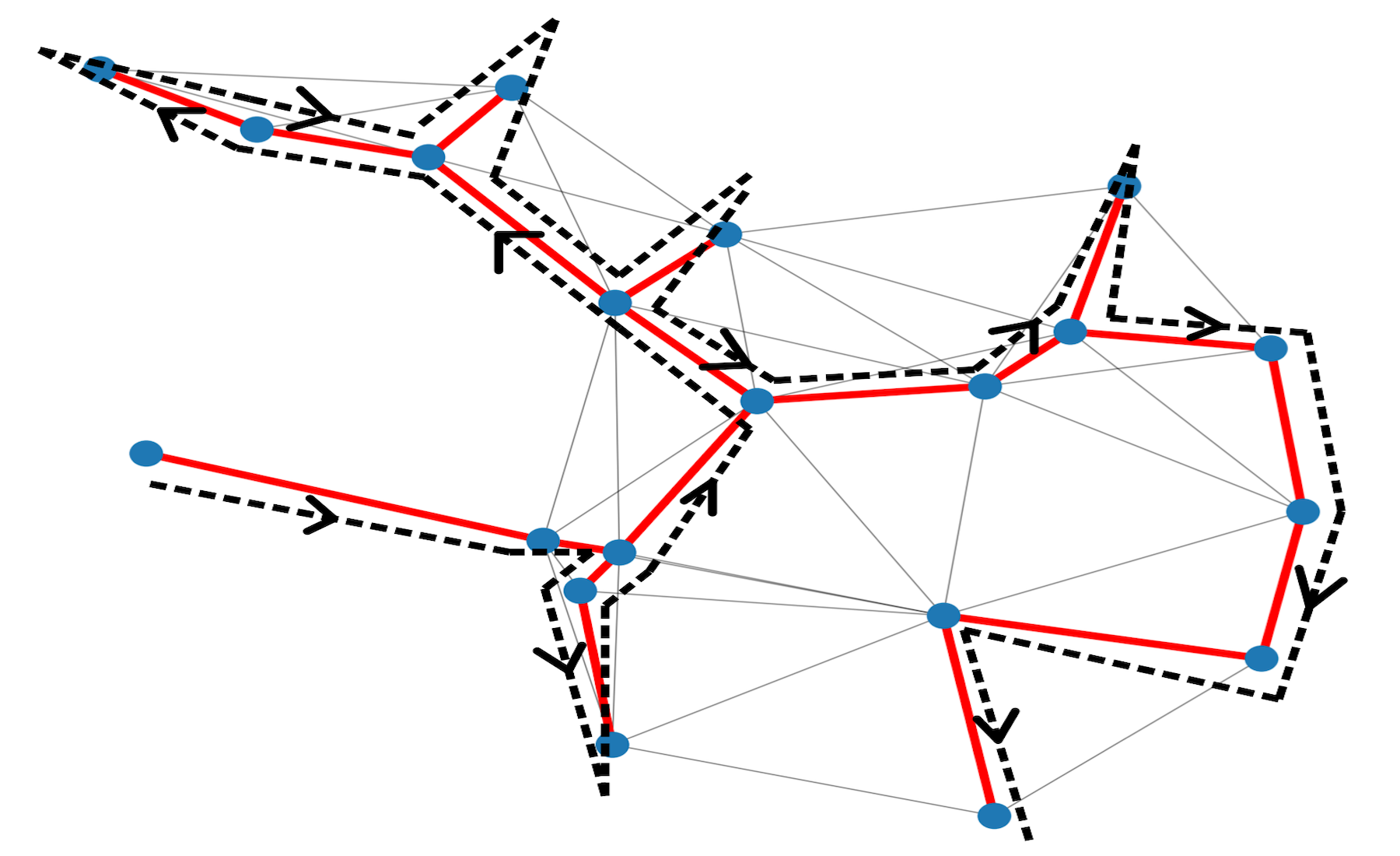}
    \caption{Traveling salesperson problem: reduce the total distance traveled to all cities [the nodes (blue dots), connected by edges (grey and red lines)]. The minimum spanning tree (red) gives a close-to-optimal solution.}
\end{subfigure}
\;
\begin{subfigure}[t]{.48\textwidth}
    \includegraphics[width=1\linewidth,height=5.7cm]{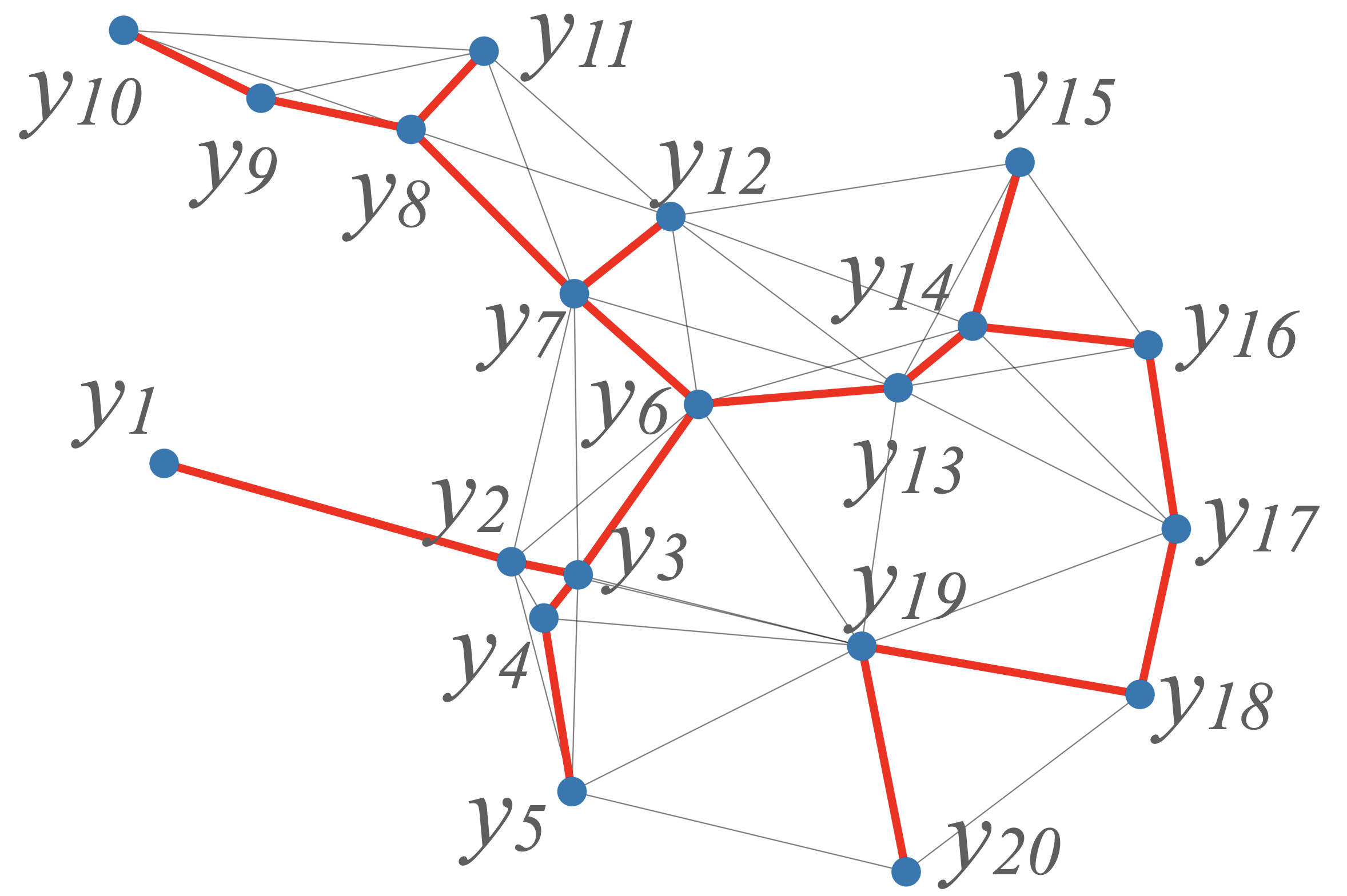}
    \caption{Generative graph conditioned on a spanning tree: a variable is generated when reaching a new node, as dependent on one of the existing variables.
    The tree is a random subgraph of the underlying graph.
    }
\end{subfigure}
 \caption{Illustration of the minimum spanning tree and its graphical model. \label{fig:mst}}
 \end{figure}

It is not hard to see that we only need to travel at most twice over each of those $p-1$ edges
 (shown in Figure~\ref{fig:mst}); hence we have a relaxed problem:
\be
\min_{\mathcal I}Q(\mathcal I) \le 2 \; {\min}_{T\in \mathcal T} \sum_{(j,k)\in
T} w_{j,k}
\ee
where $T$ is known as the  spanning tree:
\be
T= (V,E_T): E_T\subseteq E_G, |E_T|=p-1, T \t{ is connected},
\ee
that is, $T$ is the subgraph of \(G\) having the smallest number of edges, while still connecting all the nodes; and $\mathcal T$ is the collection of all the spanning trees of $G$.


Unlike the original problem, the relaxed one (also known as the ``minimum spanning tree'' problem) can be solved easily --- this is an M-convex problem [the discrete equivalent of convex \citep{murota1998discrete}]; hence simple greedy algorithms \citep{kruskal1956shortest,prim1957shortest} converge to the global optimum.


To link spanning trees to a graphical model, imagine that we  generate a new variable \(y_j\) each time we reach a new node, where each \(y_j\) depends on one of the existing \(y_{k}\)'s. Then the graph would become a spanning tree with likelihood
\bel\label{eq:graph_model_t}
\mathcal L(y; T) = \Pi(y_{1}) \prod_{j=2}^p \Pi \left[ y_{j}\mid y_{k}: k\in \{ 1,\ldots, j-1\}, (j,k)\in E_T \right].
\eel
This generative graph has two advantages: (i) it gives us a  simplified graph of $(p-1)$ edges that explain
the  key  ``dependencies'' (to be formalized later) among all the variables; (ii) 
the optimal point estimate $T$ is tractable.  


 \subsection{Bayesian Spanning Tree Model}
Slightly simplifying  \eqref{eq:graph_model_t}, we have the spanning tree graphical model:
\bel\label{eq:spanning_tree_likelihood}
\mathcal L(y; T,\theta) =  \Pi(  y_{1}) \prod_{e_s\in T}
 \Pi[y_{k} \mid  y_{j} , e_s = (j, k), \theta_s],
\eel
where \(y_{1}\) is the first variable generated that we refer to as the ``root'', we use \(e_s\in T\) as shorthand for \(e_s\in E_T\), and $\theta=\{\theta_s\}_{s=1}^{p-1}$ are the other parameters associated with the edges. We will refer to \eqref{eq:spanning_tree_likelihood} as the spanning tree likelihood, and will complete its specification in the next subsection.

We can improve the model flexibility by taking a Bayesian approach and assigning a prior on $T$. This has the appealing advantages of (i) enabling regularization on the tree through the prior specification, (ii) allowing us to obtain a set of trees in the high posterior probability region, as opposed to a single estimated tree, and (iii) as will be shown in Section 3.2, we can analytically marginalize out $T$ and obtain a marginal probability estimate of whether $(j,k)$ are connected.

We specify the tree prior in the following form:
\bel\label{eq:uniform_distribution}
\Pi_{0}(T) =  g(T) 1(T\in \mathcal T),
\eel
where $\mathcal T$ is the set of all spanning trees with $p$ nodes, and $g(T) \ge 0$ is a probability mass function that sums to one over $\mathcal T$; the set $\mathcal T$ is a combinatorial space, with a cardinalty $|\mathcal T|=p^{(p-2)}$ \citep{buekenhout1998number}. The posterior of $T$ is a discrete distribution:
\bel\label{eq:T_posterior}
\Pi(T\mid y,  \theta) = \frac{\mathcal L(y;T,\theta)\Pi_{0}(T)
}{\sum_{T'\in\mathcal T} \mathcal L(y;T',\theta)\Pi_{0}(T')}.
\eel
The marginal density of the root $\Pi(y_1)$ is canceled in $\Pi(T\mid y,  \theta) $, and hence we do not need to specify $\Pi(y_1)$ in the likelihood (the choice of $y_1$ as the root may impact $\Pi \left[ y_{j}\mid y_{k}:  (j,k),\theta_s \right]$, which will be addressed in the next subsection).

\subsection{Location Dependence and Likelihood Specification}
 For ease of exposition, we assume each $y^{(i)}_k \in \mathbb{R}$ is continuous, and denote the $n$ samples as  $\vec y_k = [y_{k}^{(1)},\ldots, y_{k}^{(n)}] \in \mathbb{R}^n$.
 Following a typical convention in graphical modeling \citep{wang2012bayesian,tan2015cluster}, we assume each $\vec y_k$ is centered and standardized.
To specify $ \Pi \big[ y_{k} \mid  y_{j} , e_s = (j,k), \theta_s \big]$, we consider the commonly used location-scale family density:
\bel\label{eq:div_likelihood}
& \Pi \big [ y_{k} \mid  y_{j} , e_s = (j, k), \sigma_s \big ] =  \frac{1}{\sigma_s^n}f\bigg(  \frac{\vec y_k-\vec y_j}{\sigma_s}\bigg),
\eel
where the conditional density of $y_k$ is centered on $y_j$, 
 $\sigma_s>0$ is a scale parameter associated with the edge $e_s$,
 and $f:\mathbb{R}^p\to [0,\infty)$ integrates to one over $\mathbb{R}^p$. 

In choosing a specific $f$, we focus on obtaining root exchangeability of this graphical model. In choosing a particular variable as the root node, we obtain a directed acyclic graph having a corresponding variable ordering.  However, given that the choice of root node is typically arbitrary, it is desirable to remove dependence on its choice from the resulting posterior distribution of $T$.  

Fortunately, this root exchangeability can be achieved as long as $f$ satisfies the symmetry about zero constraint: 
\bel
f(\vec x)=f(-\vec x)  \qquad \forall \vec x\in\mathbb R^n.
\eel
Changing the root choice corresponds to applying a particular permutation of the variable/node index $\{\pi(1),\ldots,\pi(p)\}$.  Considering the $e_s=(j,k)$ edge in the initial graph, after permutation we have $y_{\pi(j)}=y_j$, $y_{\pi(k)}=y_k$, and 
$\Pi \big [ y_{k} \mid  y_{j} , e_s = (j, k), \sigma_s \big ] $ replaced by $\Pi \big [ y_{\pi(j)} \mid  y_{\pi(k)} , e_s = (\pi(k), \pi(j)), \sigma_s \big ] $. Therefore, we have
\be
\Pi[ T_{(1,\ldots,n)}\mid (y_1,\ldots, y_p),  \theta ]  = \Pi[ T_{\pi(1),\ldots, \pi(p)} \mid y_{(\pi(1)},\ldots, y_{\pi(p))},  \theta],
\ee
where $T_{\pi(1),\ldots, \pi(p)}$ denotes the permuted graph obtained by replacing $k$ with $\pi(k)$ for each node, and $(j,k)$ with $(\pi(j),\pi(k))$ for each edge. Hence, the structure of the tree does not change, but only the node number labels changes.
\begin{remark}
The above exchangeability property is different from node exchangeability, where one would permute the variable/node index alone without changing the graph node index. For a more comprehensive discussion of graph exchangeability, see \cite{cai2016edge}.
\end{remark}

To satisfy such symmetry constraints, and simplify our theoretical developments, we focus on a simple Gaussian density, 
\bel\label{eq:gaussian_likelihood}
& \Pi \big [ y_{k} \mid  y_{j} , e_s = (j, k), \sigma_s \big ] =  \frac{1}{(2\pi\sigma_s^2)^{n/2} |R|^{1/2}} \exp \left[ - \frac{ (\vec y_k- \vec y_j)^{\rm T}R^{-1}(\vec y_k- \vec y_j)}{2\sigma^2_s} \right ],
\eel
where $R$ is an $n\times n$ positive definite matrix allowing correlation between the samples
 \(\vec y_{k}= \{ y^{(1)}_{k}, y^{(2)}_{k},\ldots, y^{(n)}_{k}\}\).
If the multivariate data samples are uncorrelated, we simply let $R=I_n$. In more complex settings, such as when the samples have temporal dependence, $R$ is a nuisance parameter, as our focus is on dependence across the outcomes and not the samples.  The approach for
handling $R$ necessarily depends on the sampling design; for time series one may use an AR-1 structure, for spatially indexed data one may use a spatially decaying correlation function, etc. For simplicity, we focus on using $R=I_n$.


We refer to \eqref{eq:gaussian_likelihood} as the Gaussian spanning tree likelihood.
\begin{remark}
Although \eqref{eq:gaussian_likelihood} may look similar to a regular multivariate Gaussian density, the key parameters in this likelihood are the choices of $(j,k)$'s in $T$, which determine which $(y_j -y_k)$'s  enter into the likelihood.
\end{remark}

With $\sigma_s$ varying across edges, it is important to choose an appropriate prior to regularize the spanning tree model.  With this goal in mind, we build on the popular global-local shrinkage prior framework 
 \citep{polson2010shrink} in the next subsection.

\subsection{Global-Local Shrinkage Prior for Tree Selection}
\begin{figure}[H]
 \centering
    \includegraphics[width=.5\linewidth]{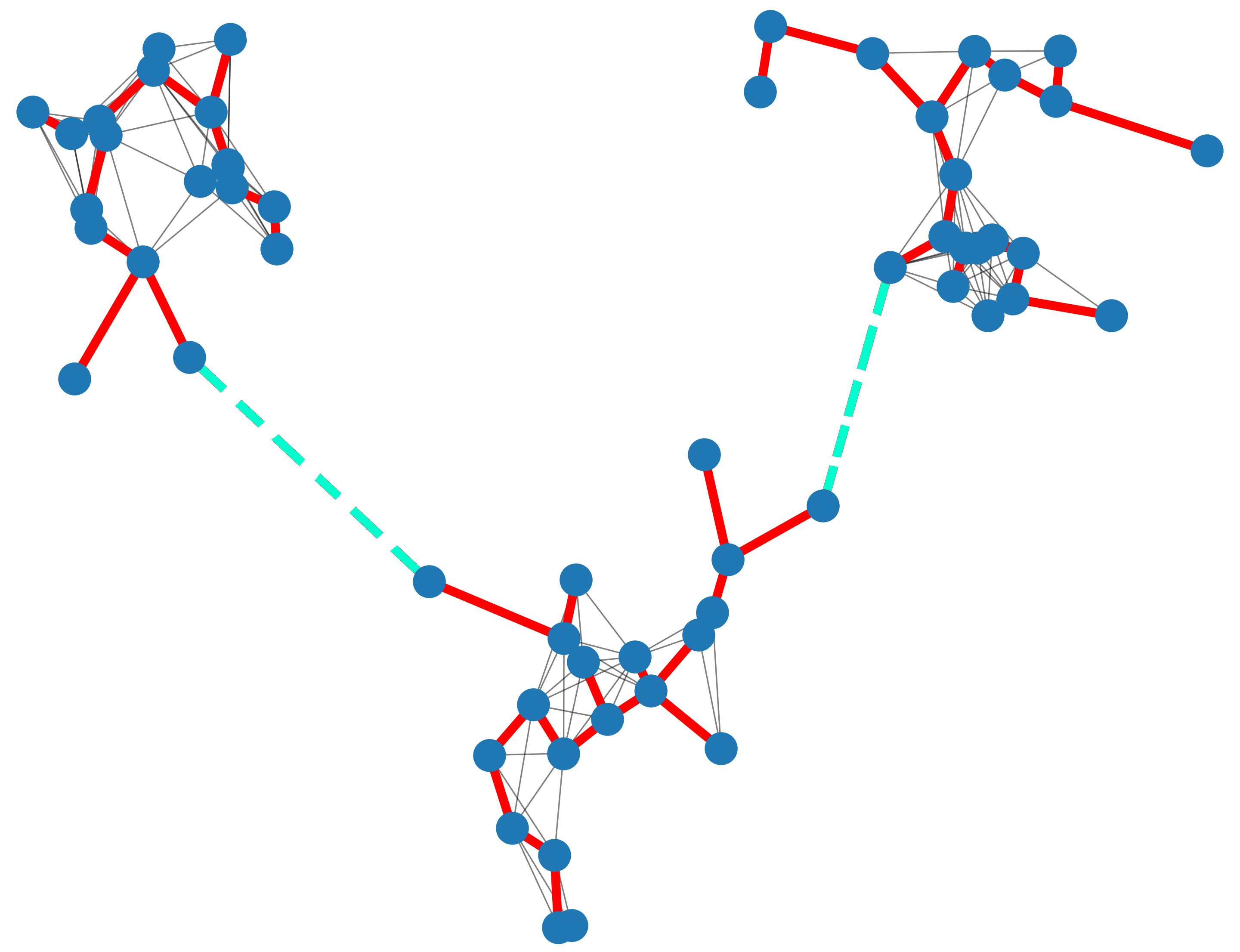}
 \caption{Illustration of the estimated spanning trees in the high posterior probability region.  The minimum spanning tree is shown in red and cyan, and the alternative trees are shown in grey and cyan. Two long edges (cyan) are necessary for forming a valid spanning tree.
   \label{fig:long_edge_illu}}
 \end{figure}

Letting $\Pi_0(T)$ denote the prior probability of tree $T$, the posterior probability of choosing a particular spanning tree under the Gaussian spanning tree likelihood is 
 \bel\label{eq:T_posterior_gaussian}
\t{Pr}(T=T_\star \mid y,  \theta) = \frac{
 \exp \left[ -(1/2)\sum_{(j,k)\in T_\star } \left( {  \|\vec y_k- \vec y_j\|^2}/{\sigma^2_s} \right ) \right ]
\Pi_{0}(T_\star)
}{\sum_{T'\in\mathcal T} 
\exp \left[ -(1/2) \sum_{(j,k)\in T' } \left( {  \|\vec y_k- \vec y_j\|^2}/{\sigma^2_s}  \right )\right ]
\Pi_{0}(T')}.
\eel
Intuitively, if most of the $\sigma_s$'s are small, then the posterior distribution will be dominated by trees $T^*$ having most $ \|\vec y_k- \vec y_j\|$'s  small. Hence, a prior favoring small $\sigma$'s will favor a smaller high probability region of spanning trees, leading to greater interpretability.
However, as shown in Figure~\ref{fig:long_edge_illu},  in order to form a valid spanning tree,
we may have to choose a few long edges with large $ \|\vec y_k- \vec y_j\|$; hence, the prior for $\sigma_s$ should ideally be concentrated at small values with heavy tails.

To satisfy both properties, we use a global-local prior as:
\bel\label{eq:global_local}
\sigma_s = \lambda_s \tau, \qquad \lambda_s\stackrel{iid}\sim \Pi_\lambda, \qquad\tau \sim \Pi_\tau.
\eel
where $\tau>0$ is the global scale with  $\Pi_\tau$ concentrated near zero, so that $\tau\approx 0$ provides an overall strong shrinkage; whereas $ \lambda_s>0$ is the local scale from $\Pi_\lambda$, a heavy-tailed distribution so that a few $\lambda_s$ can have very large values.


Although there are a broad variety of global-local priors that would suffice for our purposes, the generalized double Pareto \citep{armagan2013generalized} is particularly convenient due to the closed-form marginal.  We focus on the multivariate extension of 
 \cite{xu2015bayesian}, with $\lambda_s^2\sim \int\t{Ga} [\lambda_s^2;(n+1)/2,\kappa_s^2/2] \t{Ga}(\kappa_s;\alpha,1)\textup{d} \kappa_s $. Marginalizing over $\lambda_s$, we have (omitting a constant not involving $\alpha, \tau$, with complete details provided in the supplementary materials):
\bel\label{eq:gdp_marginal}
\Pi \big [ y_{k} \mid  y_{j} , e_s = (j, k), \tau \big ] \propto \frac{\Gamma(\alpha+n)}{\Gamma(\alpha)}\frac{1}{ \tau^n}\left(1+\frac{\|\vec y_j-\vec y_k\|_2}{\tau}\right)^{-(\alpha+n)}.
\eel
To favor a small global scale $\tau$ while being adaptive to the data, we use an informative exponential prior $\tau\sim \t{Exp}(1/\mu_\tau)$ with the prior mean set to $\mu_\tau =\min_{(j,k: j\neq k)}\|\vec y_j-\vec y_k\|_2/n$, as an empirical estimate of the smallest scale among vectors $(\vec y_j-\vec y_k)$'s. We choose $\alpha=5$ as a balanced choice between shrinkage and tail-robustness; details can found in \cite{armagan2013generalized}.
 
\subsection{Prior for the Tree }

We can obtain further regularization via the tree prior $\Pi_0(T)=g(T)1(T\in \mathcal T)$. We discuss a few choices here, ranging from informative to non-informative.

Perhaps the simplest informative choice is an edge-based prior, including information that certain edges are more likely to be in the graph through  
a $p\times p$ matrix containing $\eta_{j,k}\ge 0$, and letting 
\bel\label{eq:edge_prior}
\Pi_0(T \mid \eta) = z^{-1}(\eta)\prod_{(j,k)\in T} \eta_{j,k}, 
\eel
where $z(\eta)= \sum_{T\in \mathcal T}\prod_{(j,k)\in T} \eta_{j,k}$ is the normalizing constant. Those $(j,k)$ with larger $\eta_{j,k}$ will be more likely to be in $T$ {\em a priori}.  For example, in brain connection networks, one may favor connections between regions that are closer together spatially by letting 
 $\eta_{j,k}=\exp(-\|x_j-x_k\|_2)$ with $x_j$ the associated spatial coordinate. As another example,
 if one wants to block connections between $j$ and $k$, one can simply  set
 $\eta_{j,k}=0$.

Often we do not know which edges are more likely, but may have prior preferences for certain graph statistics. Here we consider the degree, as the total number of edges for each node $D_j =\sum_{k=1}^p 1[(j,k)\in T]$. To obtain a degree-based prior, we propose to set 
$\eta_{j,k}=v_j v_k$, leading to
\bel\label{eq:diri_prior}
\Pi_0(T \mid v_1,\ldots,v_p) = z^{-1}(\eta)  \prod_{j=1}^p v_j^{D_j },
\eel
with $(v_1,\ldots,v_p)$ encoding prior knowledge of which nodes have more edges, and the normalizing constant having closed form $z(\eta)=(\sum_{j=1}^p v_j)^{p-2}\prod_{j=1}^p v_j$; a proof is provided in the supplementary materials.

\begin{figure}[H]
\begin{subfigure}[t]{.45\textwidth}
\centering
    \includegraphics[width=.6\linewidth]{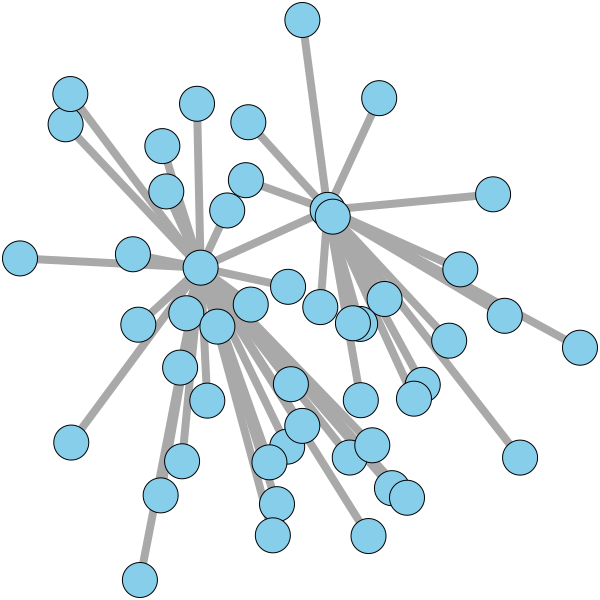}
    \caption{ A hub graph generated from Dirichlet degree-based tree prior, with high degrees in only a few nodes.}
\end{subfigure}\;
\begin{subfigure}[t]{.45\textwidth}\centering
    \includegraphics[width=.6\linewidth]{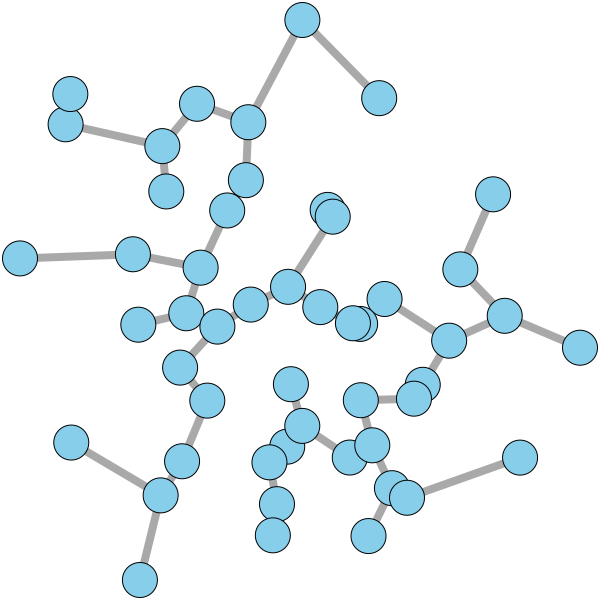}
    \caption{ A graph generated from the uniform tree prior, with similar degrees across nodes.}
\end{subfigure}
 \caption{Illustration of graphs generated from the Dirichlet degree-based tree prior and the non-informative uniform prior.  \label{fig:tree_prior_illu}}
 \end{figure}

We now explore the case in which the $v_j$s are assigned a 
Dirichlet hyper-prior
\bel
 (v_1,\ldots,v_p)\sim \t{Dir}(\alpha,\ldots,\alpha),
\eel
where $\alpha$ is the concentration parameter. Since $\sum_{j=1}^p v_j=1$, we have $\Pi_0(T,v_1,\ldots,v_p \mid \alpha)\propto v_j^{(D_j+\alpha-2)}$. Conjugacy allows us to integrate out $v=(v_1,\ldots,v_p)$ and obtain the marginal prior on the degrees
\be
\Pi_0(T \mid \alpha) 
\propto {\prod_{j=1}^p \Gamma(D_j+\alpha-1)},
\ee
where $\sum_{j=1}^p D_j = 2(p-2)$. Due to the rapid increase of the gamma function $\Gamma(x)\approx  \sqrt{2\pi x} (x/e)^x$, for small to moderate $\alpha$ (such as when $\alpha \ll p$), this prior is skewed towards having a few dominating large $D_j$'s --- as a  result, the graph will contain a few ``hubs'', each connecting to a large number of nodes [Figure~\ref{fig:tree_prior_illu}(a)].

Alternatively, as a non-informative prior, one could use the uniform distribution over the space of $\mathcal T$, $\Pi_0(T)= 1/p^{(p-2)}1(T\in \mathcal T)$ [Figure~\ref{fig:tree_prior_illu}(b)]. We use this as the default throughout the article.

Regardless of the choice, all the above priors enjoy a simple form in the posterior distribution,  as a  product separable over the edges $\Pi(T \mid y) \propto \prod_{(j,k)\in T}  \big\{ \eta_{j,k} \Pi[y_{k} \mid  y_{j} , (j, k), \theta_s] \big\}$. This separable form allows us to easily update each edge in posterior computation, while leading to useful closed-form quantities in the marginal posterior, as presented below.

\section{Properties}

\subsection{Partition Function and Marginal Connecting Probability}
The spanning tree is supported in a large combinatorial space. Despite the high complexity, some quantities related to the marginal posterior distribution are available analytically in closed-form. For generality, we focus on the posterior distribution of $T$ in the following form:
\be
\t{pr}(T \mid y)
=
 \frac{ \prod_{(j,k)\in T}
  \exp(q_{j,k})}{z_q},
\ee
where  $\exp(q_{j,k}) = \eta_{j,k} \Pi[y_{k} \mid  y_{j} , (j, k), \theta_s] $ as shown above. The denominator $z_q$ is commonly known as the partition function:
\bel\label{eq:partition_function}
z_q = \sum_{T \in \mathcal T} \prod_{(j,k)\in T}
 \exp(q_{j,k}).
\eel
Letting $A_q=\{\exp(q_{j',k'})\}_{(j',k')}$, with $A_{q:j,j}=0$ in the diagonal, we show in the following Theorem that $z_q$ can be computed easily. 
All proofs are deferred to the supplementary materials.
 \begin{theorem}[Kirchhoff's matrix tree theorem for partition function]\label{thm:kirchoff}
 Let $A_q$ be defined as above, $D_{q}=\t{diag} \{\sum_{k\neq j}\exp(q_{j,k})\}_{j}$,
 $L_q = D_q-A_q$ be the Laplacian matrix, and $J$ be a matrix of ones. Then
 $$z_q = \t{det}(L_q + J/p^2).$$
 \end{theorem}

In summarizing the posterior distribution of $T$, it is useful to examine the chance that a particular $(j,k)$ is picked by the spanning tree.
In particular, we focus on the marginal posterior probability that edge $(j,k)$ is in $T$, which corresponds to the sum of the posterior probabilities of all trees having edge $(j,k)$. Remarkably these marginal posterior edge probabilities are available in closed form.  Differentiating the log-partition function, we have 
\be
\t{pr}[T \ni (j,k)\mid y]
& =\sum_{T\in \mathcal T}\mathbf{1}[(j,k)\in T ] \t{pr}(T\mid y)\\
& =
 \frac{  \sum_{\t{all } T \ni (j,k)} \exp(q_{j,k})  \prod_{(h,l)\in T:(h,l)\neq j,k}
  \exp(q_{h,l})}{z_q} \\
  & = \frac{ \partial \log z_q}{\partial q_{j,k}} \\
  & =   (\Omega^L_{j,j}+\Omega^L_{k,k} - 2 \Omega^L_{j,k}) \exp(q_{j,k}),
\ee
where $\Omega^L = (L_q + J/p^2)^{-1}$. We refer
$\t{pr}[T \ni (j,k)\mid y]$ as the ``marginal connecting probability'' for $(j,k)$.

\subsection{Connectivity Guarantee via Matrix Rank Constraint\label{sec:incidence}}
The parameter $T$ needs to satisfy the connected graph constraint, which may appear challenging to enforce computationally -- if we want to propose an update to $T$ during a sampling algorithm, how can we ensure that the proposal is still a spanning tree? A na\"ive way would be checking consecutively over the edges in \(E_T\) to ensure the connectivity --- this  procedure is commonly known as ``graph traversal''. 

We propose an alternative to bypass the need for graph traversal. Consider
 the incidence matrix \(B\), a \(p\times (p-1)\) matrix that records the {\em node-to-edge}
relationship. For \(s=1,\ldots, p-1\), if \(e_s=(j,k)\), we set \(B_{j,s}=1\), \(B_{k,s}=-1\) and all other \(B_{l,s}=0\) for \(l\neq j\) or \(k\).
The matrix $B$ is useful, because a graph of \(p\) nodes is connected if and only if the rank of its incidence matrix \(\t{rank}(B)=p-1\)  [Theorem 2.3 of \cite{bapat2010graphs}]; that is, $B$ is of full column  rank. 
Therefore, we can convert the combinatorially constrained space into a simple rank-constrained problem:
\be
\mathcal T = \{T: B\t{ is the incidence matrix of }T, \; \t{rank}(B)=p-1\}.
\ee

We show the full-rankness of $B$ implies an appealing combinatorial property, which allows us to quickly find the ``graph cut partition'' related to each edge. To formalize, in a spanning tree, removing an edge $e_s=(j,k)$ will create two disconnected components; we want to find the graph cut partition as
 the two disjoint sets of vertices:
\be
\t{Cut}[(j,k)] &=(V_1, V_2) \\
 \t{ such that }  & j \in V_1, k \in V_2, \\
 & V_1\cup V_2= V, \;\; V_1\cap V_2= \varnothing\\
  & G(V_1) \t{ connected}, G(V_2) \t{ connected},
\ee
where $G(V_k)$ is the sub-graph of $G$ containing only the nodes in $V_k$.
 Finding $\t{Cut}[(j,k)]$ is non-trivial -- in a brute-force approach, one starts from node \(j\), traversing the edges \(E_T\setminus \{(j,k)\}\) and adding all the visited nodes to \(V_1\); after visiting all the nodes accessible in the path from $j$, one assigns the remaining nodes to \(V_2\).

The rank constraint can significantly reduce the burden ---  due to the full column rank, the projection of any column \(\vec B_s\) (corresponding to an edge \(e_s\)) into the nullspace of the others would be a non-zero vector. Interestingly, the output of this projection  contains only two unique values, allowing us to directly find \(V_1\) and \(V_2\). 
\begin{theorem}[Traversal-free solution to find the graph cut partition] \label{thm:disconnect}
Denote the \(s\)th column of \(B\) by \(\vec B_s\), and other columns by \(B_{[-s]}\).
  Then the \(p\)-element
vector
 \[\vec \beta_s = \{ I- B_{[-s]}(B_{[-s]}^{\rm T} B_{[-s]})^{-1} B_{[-s]}^{\rm T}\}
\vec B_s,\]
 contains only two unique values \(\beta^*_{1,s}, \beta^*_{2,s}\in \mathbb{R}\) with \(\beta^*_{1,s}\neq \beta^*_{2,s}\).
The Cut$(e_s)=(V_1,V_2)$ can be found using \(V_1 =\{
j: \beta_{j,s}=\beta^*_{1,s}\}\) and \(V_2 =\{ j: \beta_{j,s}=\beta^*_{2,s}\}\).
\end{theorem}
  
 \begin{remark}
The matrix inversion \((B_{[-s]}^{\rm T} B_{[-s]})^{-1}\) can be computationally costly for large \(p\) with a complexity at $O(p^3)$. To address this issue, we develop an efficient algorithm that can extract \((B_{[-s]}^{\rm T} B_{[-s]})^{-1}\) from $(B^{\rm T}B)^{-1}$, and update the value of $(B^{\rm T}B)^{-1}$ when a column of $B$ changes. This allows us to only evaluate  $(B^{\rm T}B)^{-1}$ for one time during the algorithm initialization, and reduce the cost of computing $\vec \beta_s$  to $O(p)$ during the posterior sampling.
The details are provided in the supplementary materials.
\end{remark}

\section{Posterior Computation}

We develop an efficient Gibbs sampler for sampling from the posterior distribution over spanning trees.
To facilitate fast exploration of the high posterior probability region, we develop: (i) a graph update step that can rapidly change the shape of the spanning tree; (ii) an initialization that gives the  approximate posterior mode of the spanning tree.

%

\subsection{Gibbs sampling with the Cut-and-Reconnect Step}

\subsubsection{Update $T$}
At the current state of $T$ with \(E_T= \{e_1,\ldots, e_{p-1}\}\), to make an update to the spanning tree, we propose a ``cut-and-reconnect'' step for each edge \(e_s=(j,k)\): we first remove this edge and find its graph cut partition \(\t{Cut}(e_s)=(V_1,V_2)\), and then sample a new edge $(j',k')$ across $V_1$ and $V_2$, so that we obtain a new spanning tree $T^*$.

The transition that replaces $(j,k)$ with $(j',k')$ is reversible and has a simple multinomial probability:
$$ \pr [ (j',k') \in T^* \mid E_T \setminus (j,k)]
= \frac{\exp(q_{j',k'})}{\sum_{j\in V_1,k\in V_2} \exp(q_{j,k})}.
$$
Repeating this for $s=1,\ldots,p-1$ rapidly changes the shape of the tree.

\begin{figure}[H]
\begin{subfigure}[t]{.45\textwidth}
    \includegraphics[width=.9\linewidth]{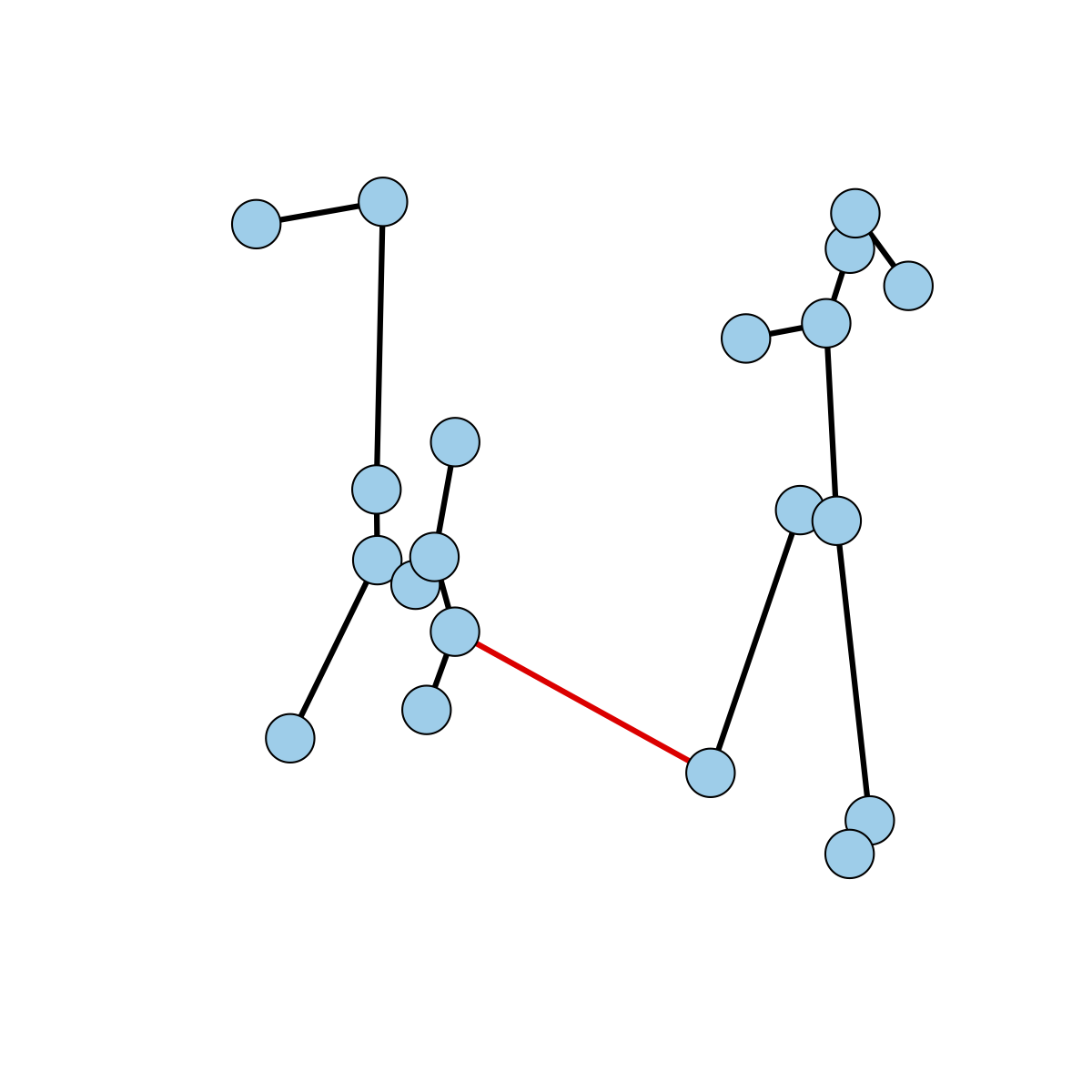}
    \caption{Cutting an edge (red) gives two disconnected subgraphs (with node sets \(V_1\) and \(V_2\)), calculated using the Theorem~\ref{thm:disconnect}.}
\end{subfigure}
\quad
\begin{subfigure}[t]{.45\textwidth}
    \includegraphics[width=.9\linewidth]{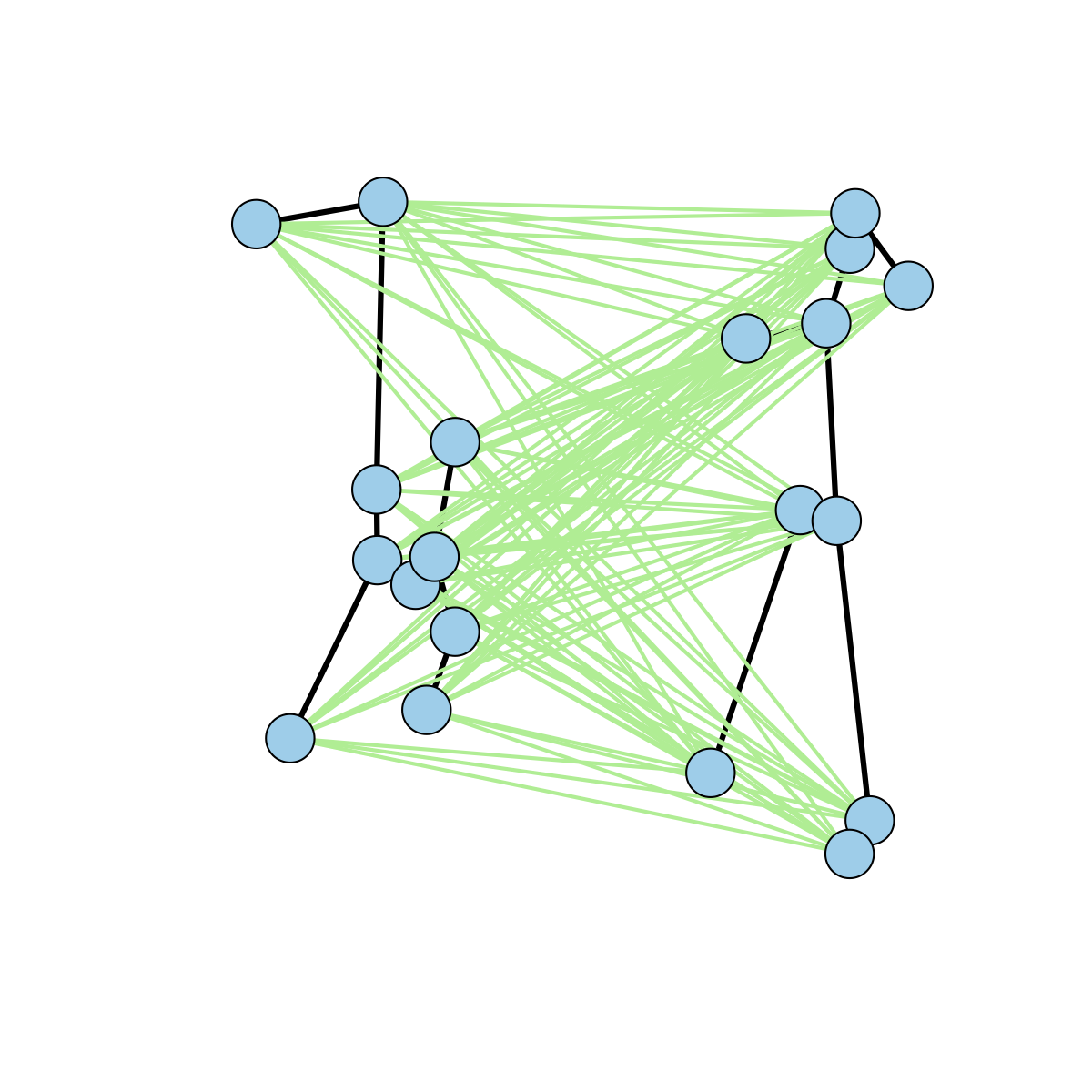}
    \caption{Drawing a new edge from the multinomial distribution with \(|V_1|\times |V_2|\) choices (green) and form a new tree.}
\end{subfigure}
 \caption{At each step, the cut-and-reconnect step explores a change of edge over a large number of candidates, leading to rapid state exploration in the spanning tree space.
 \label{fig:algo}}
 \end{figure}

 To understand why this edge-based update can efficiently explore the space of $\mathcal T$, we compare it with an alternative node-based update ---   removing a node $j$ from the graph and reattaching it to one of the other nodes. For the node-based update, there are only at most \((p-1)\) candidates in the multinomial draw; whereas for the edge-based update, there are \(|V_1|(p-|V_1|)\)  candidates, which has an order up to \(O(p^2)\). To illustrate the large number of candidates in the edge-based update, we plot a diagram in Figure~\ref{fig:algo}.

\subsubsection{Update the other parameters}
Using the marginal density in \eqref{eq:gdp_marginal}, the parameters can be updated via 
\begin{itemize}
            \item Sample $\tau = |\tilde \tau|$ using random-walk Metropolis, by proposing $\tilde \tau^* \sim \t{Uniform}(\tilde\tau- \delta, \tilde\tau + \delta)$ with $\delta>0$ a tuning parameter, and accepting with probability:
        \be 
\min \bigg\{1, 
        \frac
        {\prod_{(j,k)\in T} \left[\frac{1}{|\tilde \tau^*|^n}\left(1+\frac{\|\vec y_j-\vec y_k\|_2}{|\tilde \tau^*|}\right)^{-(\alpha+n)}  \right ]\exp(-|\tilde \tau^*|/\mu_\tau )}
                {\prod_{(j,k)\in T} \left[\frac{1}{|\tilde \tau|^n}\left(1+\frac{\|\vec y_j-\vec y_k\|_2}{|\tilde \tau|}\right)^{-(\alpha+n)}  \right ]\exp(-|\tilde \tau|/\mu_\tau )}
                \bigg\}.
        \ee
     \item If
the degree-based prior in \eqref{eq:diri_prior} is used,
     update $(v_1,\ldots, v_p)\sim \t{Dir}(D_1+\alpha-1  ,\ldots,D_p+\alpha-1)$.
\end{itemize}
For the first step, we use an adaptation period  at the beginning of the MCMC algorithm to tune $\delta>0$, so that the acceptance rate of this step is around $0.3$.

In the above algorithm, updating one edge of the tree has a computational complexity of $O(p)$; hence sampling all edges has a complexity of $O(p^2)$. To further accelerate the algorithm, one could use the random scan Gibbs sampler \citep{levine2006optimizing}, which in each iteration randomly chooses and updates a subset of the edges, hence reducing the complexity to $O(p)$.

In terms of the computational time, for a graph containing $p=200$ nodes, sampling a $1,000$ steps takes about 2 minutes using Python on a quad-core laptop. The Markov chain mixes rapidly, and we show diagnostics in the supplementary materials.

\begin{remark}
As an alternative to sampling, if the marginal connecting probability is the main interest, one can obtain a fast  estimate of $\t{pr}[T\ni (j,k)\mid y, \hat\tau]$ using a reasonable point estimate of $\tau$. For example, one could first find the conditional posterior mode of the spanning tree $\hat T$  (presented in the next subsection), then use the maximizer in the density \eqref{eq:gdp_marginal} given the tree: $\hat\tau = \alpha \sum_{(j,k)\in \hat T}\|y_j-y_k\|_2/[n(p-1)]$. Computing the marginal connecting probability (as well as the other quantities such as the partition function) is almost instantaneous and takes at most a few seconds for $p=10^4$.
\end{remark}

\subsection{Conditional Posterior Mode Estimation}
Our next goal is to find a good initialization for the spanning tree; for this purpose, we will use the posterior mode  of the spanning tree given
some initial estimate of \((\tau,\eta)\). In this article, we initialize $\tau$ at $\mu_\tau$, and all $\eta_{j,k}=1/p^2$.
Denote the adjacency matrix of the tree by \(A=\{a_{j,k}\}\), \(a_{j,k}=1\) if \((j,k)\in T\), \(a_{j,k}=0\) otherwise; and the space of all adjacency matrices for spanning trees by $\mathcal A_T$.
Then the posterior mode of the adjacency matrix is
\bel\label{eq:mode}
\underset{A \in \mathcal A_T}{\arg\max}\sum_{j>k} a_{j,k}  q_{j,k}, \qquad \t{with } q_{j,k} =  -(\alpha+n) \log\left(1+\frac{\|\vec y_j-\vec y_k\|_2}{ \tau}\right) + \log \eta_{j,k}.
\eel

Therefore, the mode is equivalent to the minimum spanning tree in a complete and weighted graph, with the edge weight \( (- q_{j,k})\).
There are several algorithms that can quickly find the globally optimal solution \citep{prim1957shortest,dijkstra1959note,kruskal1956shortest, karger1995randomized}.
We choose to present Prim's algorithm due to its simplicity.

\begin{algorithm}[H]
\SetAlgoLined
 Initialize \(U=\{1\}\), $E_T = \varnothing$ and \(\bar U= \{1,\ldots, p\}\setminus U\)\;
 \While{ \(|U|<p\)}{
 Find \((j,k) = \underset{ (l,l')\in (U, \bar U) }{\arg\max}  q_{l,l'}\), and $(j,k)$ into $E_T$.\\
 Move \(l'\) from \(\bar U\) to \(U\).
  }
 \caption{Finding the conditional posterior mode \(\hat A\)}
\end{algorithm}
To explain this algorithm, we initialize the tree as a single node \(\{1\}\), then add one edge at a time; each time, 
 among the edges that connect the current tree and the remaining nodes, we pick the one with the largest $q_{l,l'}$. This is a greedy algorithm that finds the locally optimal solution at each step; nevertheless, since the minimum spanning tree problem is M-convex [a discrete extension of continuous convexity], the greedy algorithm is guaranteed to converge to the global optimum [see Chapter 6.7 of \cite{murota1998discrete}].

\section{Theoretic Study}
In this section, we provide a more theoretical exposition on our spanning tree model.

\subsection{Connection to Gaussian Graphical Models}

We now focus on the Gaussian spanning tree model, and compare it with Gaussian graphical models. Following  \cite{Ravikumar2011}, we assume $y^{(i)}=(y^{(i)}_1,\ldots, y^{(i)}_p)$ is a zero-mean random vector with covariance $\Sigma_0$. We denote the empirical covariance by \(S_n =\sum_{i=1}^n y^{(i)} y^{(i){\rm T}}/n\).

   The incidence matrix $B$  can be  used as a contrast matrix for computing $\vec y_j-\vec y_k$. Therefore, we have a matrix representation for the posterior:
\bel\label{eq:gaussian_lik_mat}
 \mathcal L(y; T, \theta) \Pi_0(T) & \propto |\Psi|^{-n/2}\exp\left \{ -\frac{1}{2}\sum_{i=1}^n\t{tr}{ (y^{(i){\rm T}}B\Psi^{-1} B^{\rm T} y^{(i)}  } )+  \t{tr}(A \log\eta) \right\}\\
 & =  |\Psi|^{-n/2}\exp\left \{ -\frac{n}{2} \t{tr}{ (B \Psi^{-1} B^{\rm T} S_n } )+  \t{tr}(A \log\eta) \right\},
 \eel
 where \(\Psi = \t{diag} (\sigma^2_1,\ldots, \sigma^2_{p-1})\), $\log \eta$ is calculated element-wise on $\eta_{j,k}$, and $A$ is the adjacency matrix for $T$.
For a tractable theoretic analysis, we will treat $\eta$ as fixed with all $|\log \eta_{j,k}|$ finite.

We can immediately see some similarity of \eqref{eq:gaussian_lik_mat} to regularized Gaussian graphical models, such as the one associated with the graphical lasso \citep{friedman2008sparse}, with a regularized likelihood  proportional to  \(|\hat\Omega|^{-1/2} \exp\{ -n/2 \t{tr}(\hat\Omega{S_n})-\lambda|\hat\Omega|_{1,1}\)\}, and $\hat\Omega$  the precision matrix assumed to be sparse, as induced by the $(1,1)$-matrix norm.

Indeed, we will show that  \eqref{eq:gaussian_lik_mat} is equivalent to a {regularized} Gaussian graphical model, except the structure of the precision matrix is determined by a spanning tree. Consider a weighted spanning tree, with weighted adjacency \((A_{\phi})_{j,k}=\sigma^{-2}_s\)
 for the edge \(e_{s}=(j,k)\) and  \((A_{\phi})_{j,k}=0\) if $(j,k)$ is not an edge; and denote its degree matrix by $D_\phi$.
\begin{lemma}
    The Laplacian matrix  \(L_\phi = D_\phi- A_\phi\) has \(L_\phi= B\Psi^{-1}B^{\rm T}\).
\end{lemma}

 Using the spectral property of the Laplacian, the smallest eigenvalue  \(\lambda_{(1)}( L_\phi)=0\)  with its  eigenvector \(\vec 1/{\sqrt{p}}\); and the number of zero eigenvalues is equal to the number of isolated subgraph(s) --- since the spanning tree is connected, there is only one subgraph hence only one  eigenvalue equal to zero.
Therefore, it is not hard to see the matrix
\be
 \tilde \Omega = L_\phi+ \epsilon J
\ee
is strictly positive definite with \(\epsilon>0\), which can be viewed as a precision matrix. Further, we have the following identity.

\begin{lemma}
    With \(\tilde \Omega = L_\phi + \epsilon J\), we have \(|\hat\Omega| = p^2 \epsilon |\Psi^{-1}|\).
\end{lemma}
Therefore, \eqref{eq:gaussian_lik_mat} becomes (omitting constant):
\be
 \mathcal L(y; T, \theta) \Pi_0(T) & \propto  |\tilde\Omega|^{-n/2}\exp \left \{ -\frac{n}{2} \t{tr}({S_n}\tilde\Omega) \right\} \exp \{ \t{tr}(A_T \log\eta) \},
\ee
which is a restricted Gaussian graphical model with the precision matrix parameterized by the spanning tree.

\subsection{Convergence of the Tree}
With the connection to Gaussian graphical models established, a natural question is what is the advantage of the spanning tree-based model, compared to other less restricted models?

Other than immense computational advantages, the key advantage is that we do not require any assumption on the population $\Sigma_0$ to accurately recover the minimum spanning tree based on \(\Sigma_0\). In comparison, most of the existing approaches require specific strong assumptions on 
\(\Sigma_0\) such as sparsity or norm constraints, unless \(n\gg p\).

For the ease of analysis, we consider the non-informative prior $\Pi_0(T)\propto 1$. Integrating out $\sigma^2_s$ over the generalized double Pareto prior and  examining the Prim's algorithm, we see the following equivalence: 
\be
  \underset{ (j,k)\in (U, \bar U) }{\arg\min}  (\alpha+n) \log\left(1+\frac{\|\vec y_j-\vec y_k\|_2}{ \tau}\right)=
    \underset{ (j,k)\in (U, \bar U) }{\arg\min} \|\vec y_j-\vec y_k\|_2^2,
\ee
due to the monotonicity of the function $(\alpha+n)\log(1+x/\tau)$ in $x>0$, for any $\tau>0$. As we can verify that $\|\vec y_j-\vec y_k\|_2^2 /n = S_{n:j,j} + S_{n:k,k}- 2 S_{n:j,k}$, with \(S_{n:j,k}\) the \((j,k)\)th element of the empirical covariance \(S_n\), we have the posterior mode of the tree as
 \bel\label{eq:mode_tree}
 \hat T= \arg\min_{T\in \mathcal T}\sum_{(j,k)\in T}  W_{n:j,k}, \qquad  W_{n:j,k}=S_{n:j,j} + S_{n:k,k}- 2 S_{n:j,k}.
\eel
It is easy to see that
as $n\to\infty$,  \(S_n\to \Sigma_0\) in probability and the posterior mode will converge to
\bel\label{eq:oracle_spanning_tree}
T_0 = \arg\min_{T\in \mathcal T} \sum_{(j,k)\in T} W_{0:j,k}, \quad
W_{0:j,k}=\Sigma_{0:j,j} + \Sigma_{0:k,k}- 2\Sigma_{0:j,k}.
\eel
That is, asymptotically, our model recovers the minimum spanning tree of $W_0$, providing accurate partial information about the population covariance \(\Sigma_0\) .

The next crucial question is, how fast does \eqref{eq:mode_tree} converge to \eqref{eq:oracle_spanning_tree}?  At a finite $n$, to successfully recover \(T_0\) at $\hat T$, we only need the {\em ordering} in
\(\{W_{n:j,k}\}_{(j,k)}\) to partly match the one in \(\{ W_{0:j,k}\}_{(j,k)}\). Intuitively, this condition is  much easier to meet, compared to having 
\(\|\hat\Omega-\Sigma^{-1}_{0}\|\approx 0\) as in other approaches using a full covariance/precision matrix estimation.

We now formalize this intuition.
For  the required ordering condition, we first state an important property of the minimum spanning tree. For generality,
we consider the case when the minimum spanning tree may be not unique (that is,
there could be multiple equivalent solutions in \eqref{eq:oracle_spanning_tree}).

 \begin{theorem}[Path strict optimality]
For a complete graph with edge weights \(\{W_{j,k}\}_{j,k}\), denote \(\{ T^{(1)}_0,T^{(2)}_0, \ldots, T^{(M)}_0\}\) as the set of all the minimum spanning trees.  Any edge outside the trees \((h,l)\not \in \cup_{m=1}^M {T^{(m)}_0}\) is longer than every edge on the tree path    \((j,k)\in T_{m}:(j,k)\in\t{path}(h,l)\)
for \(m=1,\ldots,M\); that is, 
we have \(W_{h,l}>W_{j,k}\) strictly. 
 \end{theorem}

\begin{figure}[H]
\begin{subfigure}[t]{1\textwidth}
\centering
    \includegraphics[width=.5\linewidth]{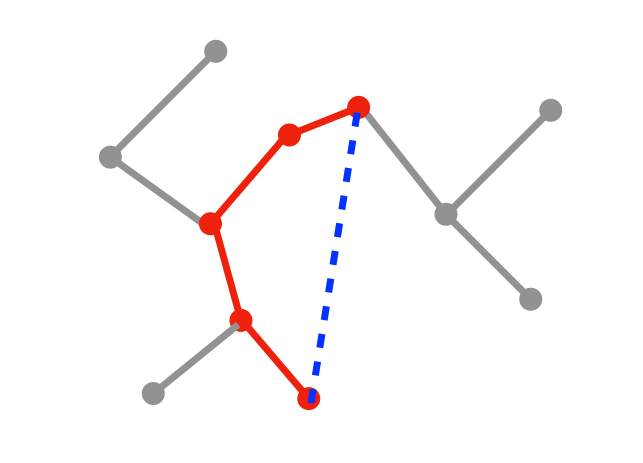}
    \caption{Path optimality of the minimum spanning tree (shown in solid lines): any  edge not on the minimum spanning tree (blue) is  strictly longer than every edge  in the tree path (red) connecting the two nodes.}
\end{subfigure}
\begin{subfigure}[t]{1\textwidth}
\centering
    \includegraphics[width=.6\linewidth]{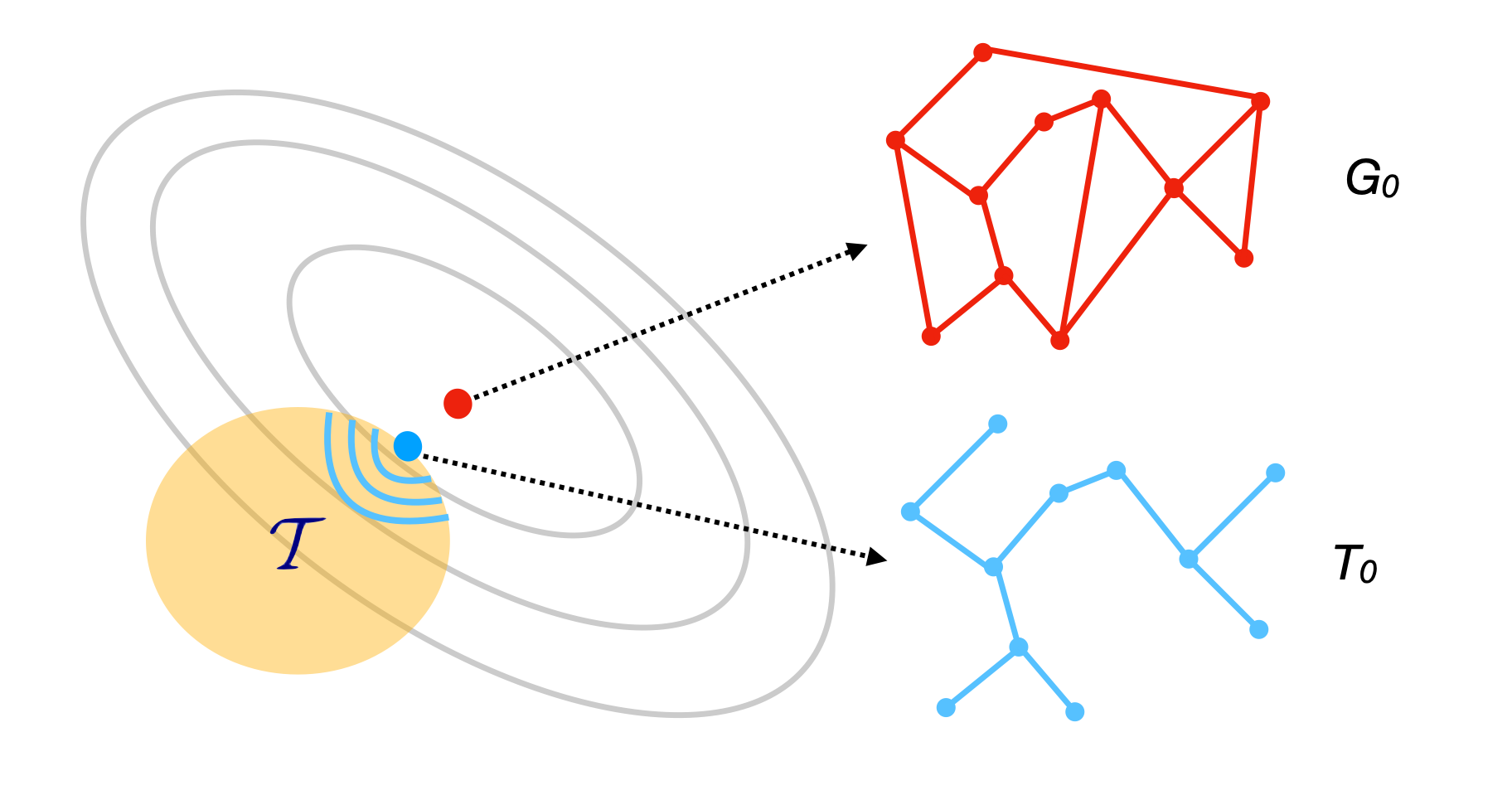}
    \caption{
   Intuition about the faster convergence of the restricted graph model: the oracle graph $G_0$ sits in the unrestricted graph space. Due to the high dimension $p \gg n$, the covariance/precision matrix-based estimator has a large uncertainty (grey contours) and critically slow convergence rate.  The spanning tree model is in a restricted parameter space $\mathcal T$ (orange), where the tree estimator has a much smaller uncertainty (blue contours) and faster convergence to $T_0$ --- the minimum spanning tree transform of $G_0$.}
\end{subfigure}
 \caption{Illustration of the theory results. \label{fig:theory}}
 \end{figure}

Using the above theorem, we can define a ``separability constant'' $\delta>0$ to quantify how separable the minimum spanning tree(s) is from the other spanning trees:
\be
 & \delta = \min_{(h,l,j,k)\in \mathcal I}(W_{h,l}-W_{j,k}),\\
 & \t{where }  \mathcal I = \bigg \{(h,l,j,k):  (h,l)\not \in \cup_{m=1}^M {T^{(m)}_0}, (j,k)\in T_{0}^{m}:(j,k)\in\t{path}(h,l)\\
 & \qquad \qquad \qquad \t{ for } m=1,\ldots,M \bigg\}.
\ee

We use Figure~\ref{fig:theory}(a) to explain the above separability constant --- let $(h^*,l^*,j^*,k^*)$ be the indices so that we reached $W_{h^*,l^*} - W_{j^*,k^*}  =\delta$, with $(j^*,k^*)$ on one of the minimum spanning trees $T^{(1)}_0$ and $(h^*,l^*)$ not on any of the $T^{(m)}_0$. If we remove the edge $(j^*,k^*)$, the $\t{path}(h^*,l^*)$ will be disrupted; hence the graph cut will lead to $h^* \in V_1$ and $l^* \in V_2$. As a result, adding $(h^*,l^*)$ will form a new spanning tree. This new tree is sub-optimal in terms of having the total weights $\delta$ larger than $T_0^{(1)}$.

We are now ready to state the convergence rate for the posterior mode.

\begin{theorem}
Assume \(y^{(i)}\stackrel{iid}\sim \mathcal F\), \(i=1,\ldots,n\), with \(\mathcal F\)  a \(p\)-variate distribution with mean \(\vec 0\) and covariance matrix \(\Sigma_0\), and each \(y^{(i)}_j\) sub-Gaussian with bound parameter \(\lambda\). Denote the set of all minimum spanning trees based on \(\Sigma_0\) as in \eqref{eq:oracle_spanning_tree} by \(\mathcal T_0=\{ T^{(1)}_0,T^{(2)}_0,
\ldots, T^{(M)}_0\}\). Denoting
a posterior mode of the spanning tree by \(\hat T\), 
    \[
    \t{pr} ( \hat T \not\in \mathcal T_0 )
    \le \frac{2}{3}
 \exp\bigg\{-\frac{n \delta^2 }{8(\beta_0^2+ \beta_{0}\delta )}
 + 3\log p\bigg\},
    \]
    where \(\beta_0=2(11\lambda^2)^3/(v^2)\) and \(v^2 = \min_{j,k} [ \mathbb{E}(\vec y_j-\vec y_k)^4-W_{0:j,k}^2]\).
\end{theorem}
We do not impose a Gaussian assumption on \(\mathcal F\), but just the tail concentration condition on \(y\).
The probability of falling outside of \(\mathcal T_0\) drops to zero rapidly, at an exponentially decaying rate in the separability constant \(\delta\) and sample size \(n\). Therefore, we only need  \(n\gg \log p\) without requiring any conditions on $\Sigma_0$ or its associated graph $G_0$. Figure~\ref{fig:theory}(b) illustrates the intuition.

\section{Numerical Experiments}

\subsection{Comparing Point Estimates with Existing Approaches}

We compare our Bayesian spanning tree model against a few popular graph estimation approaches: the thresholding estimator on the absolute empirical correlation; the graphical lasso with a chosen \(\alpha\), as the multiplier to the \(l_{1,1}\)-norm of the precision matrix; the graphical lasso with  \(\alpha\) chosen by cross-validation  (as implemented in the scikit-learn package). We record the graph edge estimate \((j,k)\) as where \(\hat\Sigma_{j,k}\neq 0\) in the thresholding estimator, and \(\hat\Omega_{j,k}\neq 0\) in the graphical lasso.

First, we  consider the common assumption that the graph is very sparse. We use the scikit-learn package to generate a precision matrix $\Omega_0$,
with sparsity level set at $3\%$ non-zero values, and non-zero correlations of magnitude between $(0.3,0.9)$. At $p=200$, this leads to approximately $600$ edges in each experiment. Then we simulate repeats of the data \((y^{(i)}_1,\ldots,y^{(i)}_p)\sim \t{N}(0, \Omega^{-1}_0)\) for \(i=1,\ldots,n\) and obtain graph estimates $\hat G$ from each method.
Each setting is repeated over different \(n\)'s, and for each \(n\) the
mean of \(10\) experiments is shown with the 95\% confidence interval for each reported quantity.
 
Denote the oracle graph by $G_0$ and its minimum spanning tree by $T_0$.
 Ideally, we want the graph estimate to fully cover the backbone subgraph $\hat G \supseteq T_0$, while having $\hat G \subseteq G_0$ so that we do not obtain too many falsely positive edge estimates. Therefore, a useful benchmark for the estimation error is $| T_0 \setminus \hat G| + | \hat G \setminus G_0|$; we show the details in the supplementary materials.

\begin{figure}[H]
\begin{subfigure}[t]{.3\textwidth}
    \includegraphics[width=.8\linewidth]{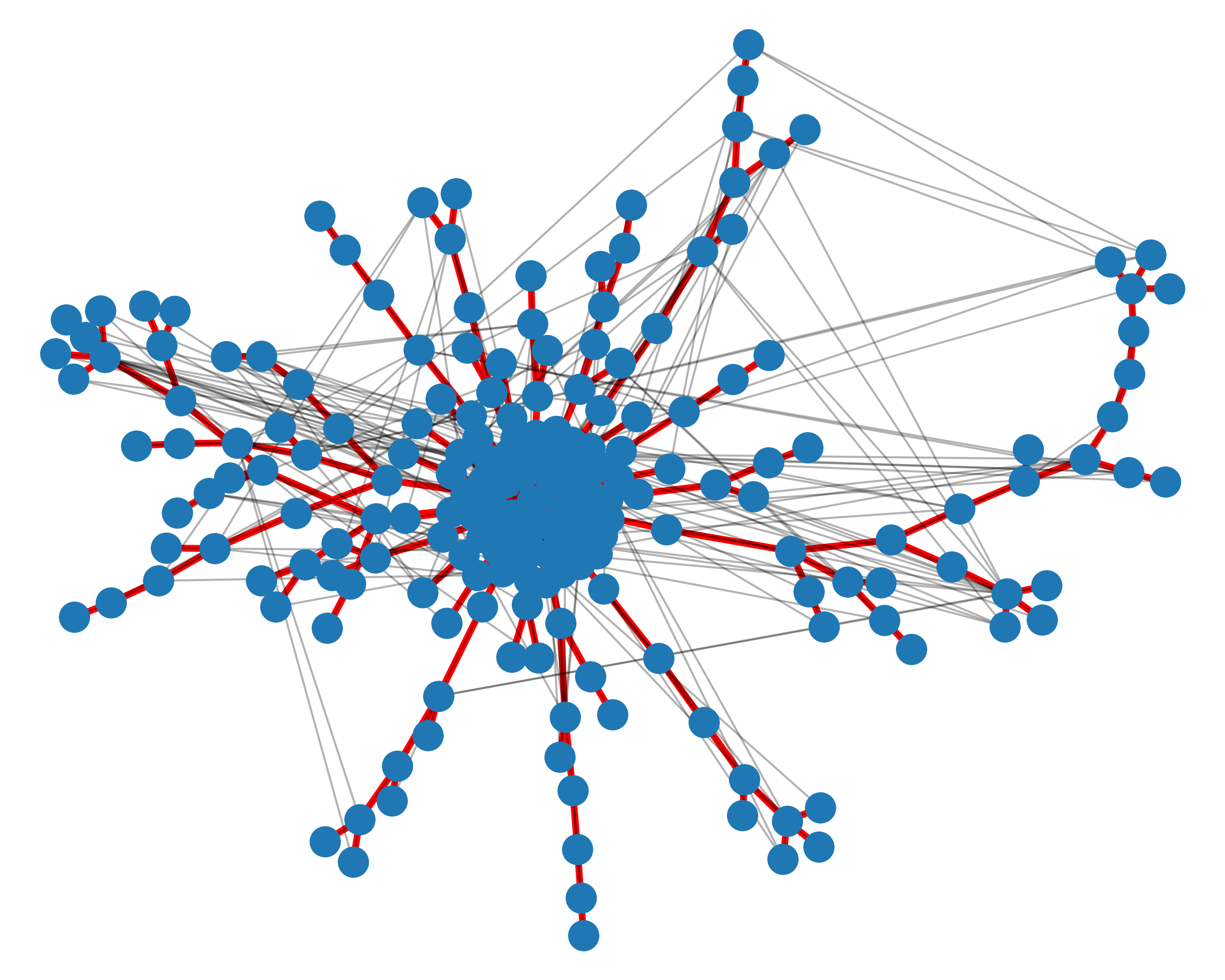}
    \caption{Oracle graph, with its minimum spanning tree shown in red.}
\end{subfigure}
\quad
\begin{subfigure}[t]{.3\textwidth}
    \includegraphics[width=.8\linewidth]{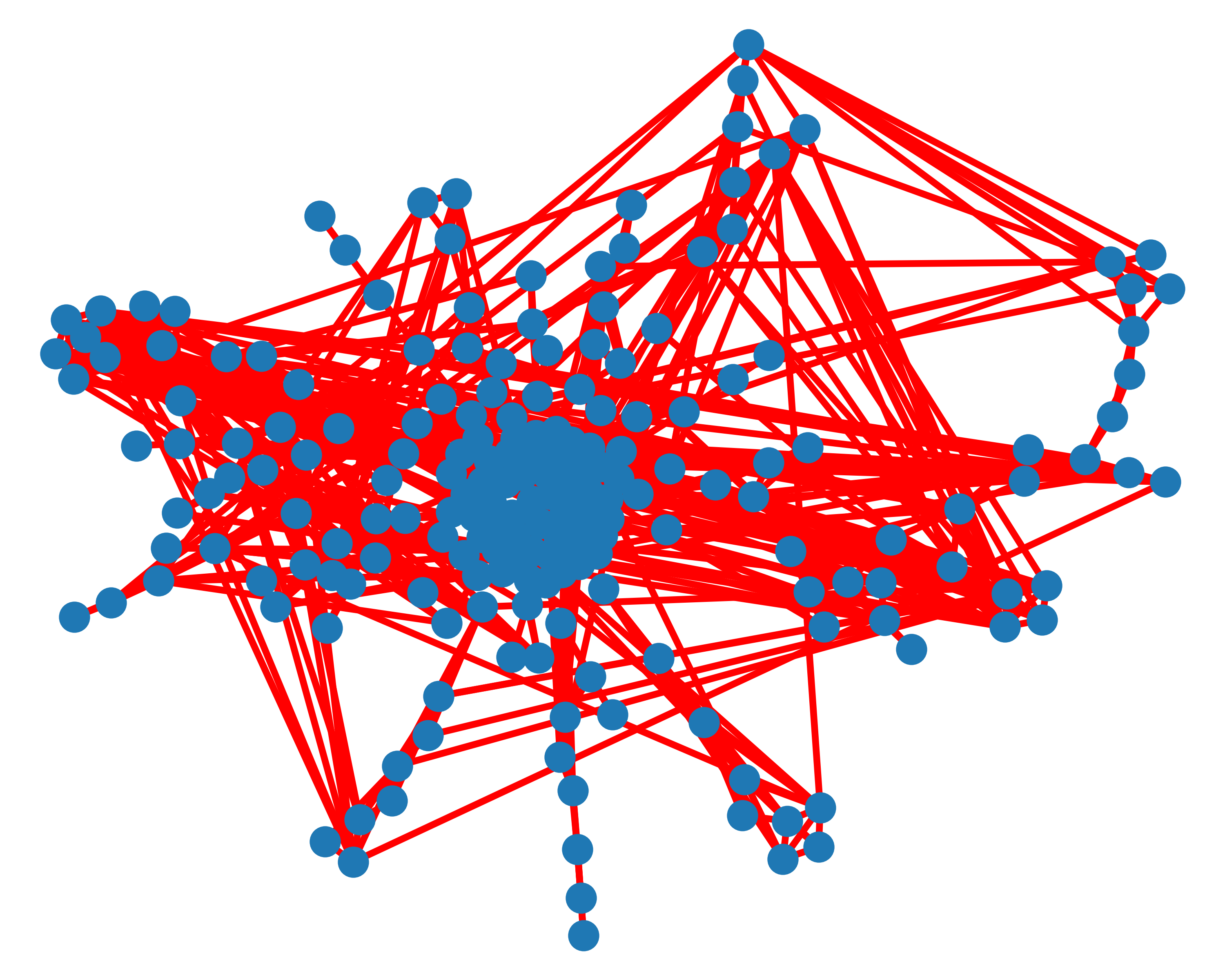}
    \caption{Graphical lasso with \(\alpha\) chosen by cross-validation.}
\end{subfigure}
\quad
\begin{subfigure}[t]{.3\textwidth}
    \includegraphics[width=.8\linewidth]{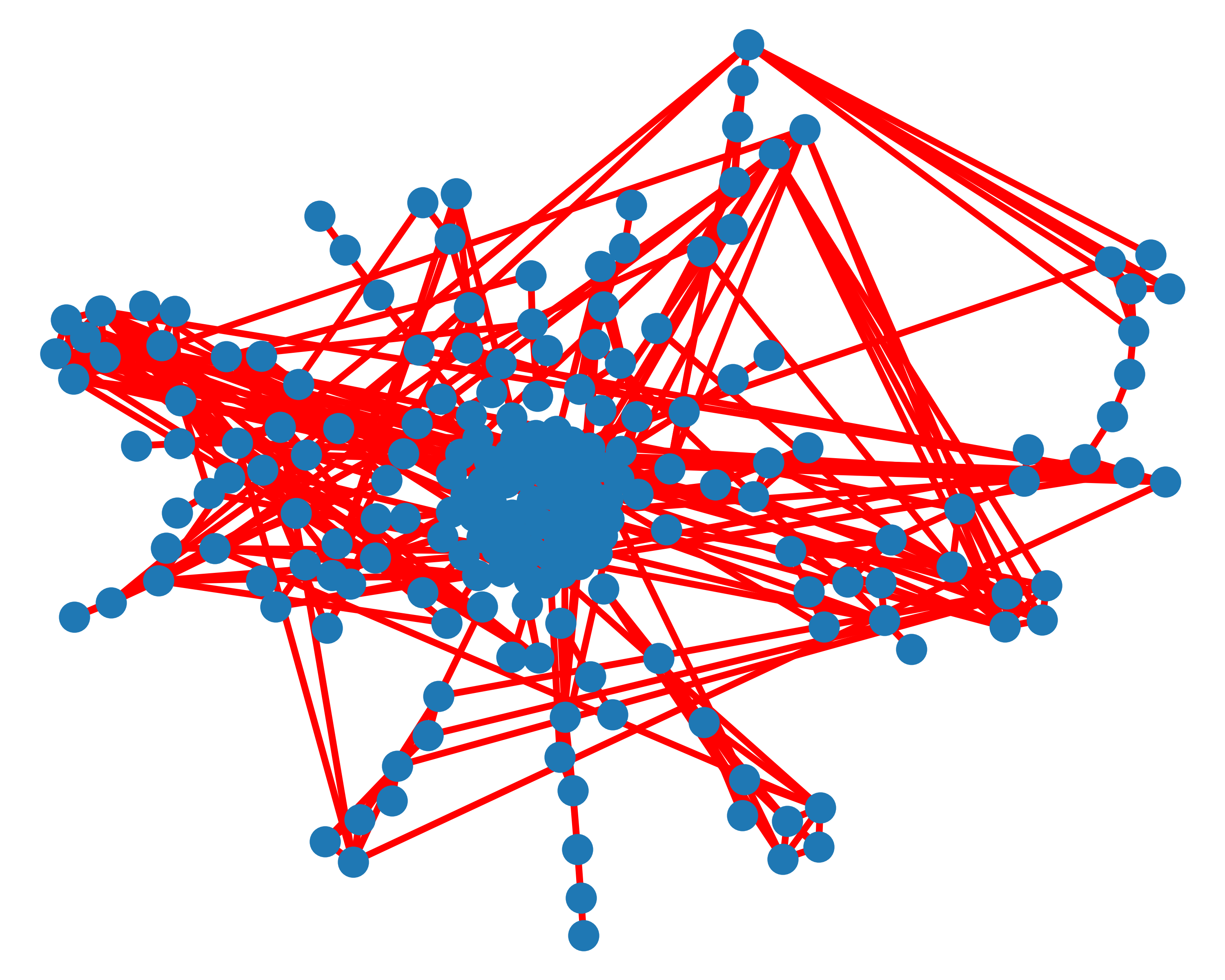}
    \caption{Graphical lasso using \(\alpha=0.8\).}
\end{subfigure}
\\
\begin{subfigure}[t]{.3\textwidth}
    \includegraphics[width=.8\linewidth]{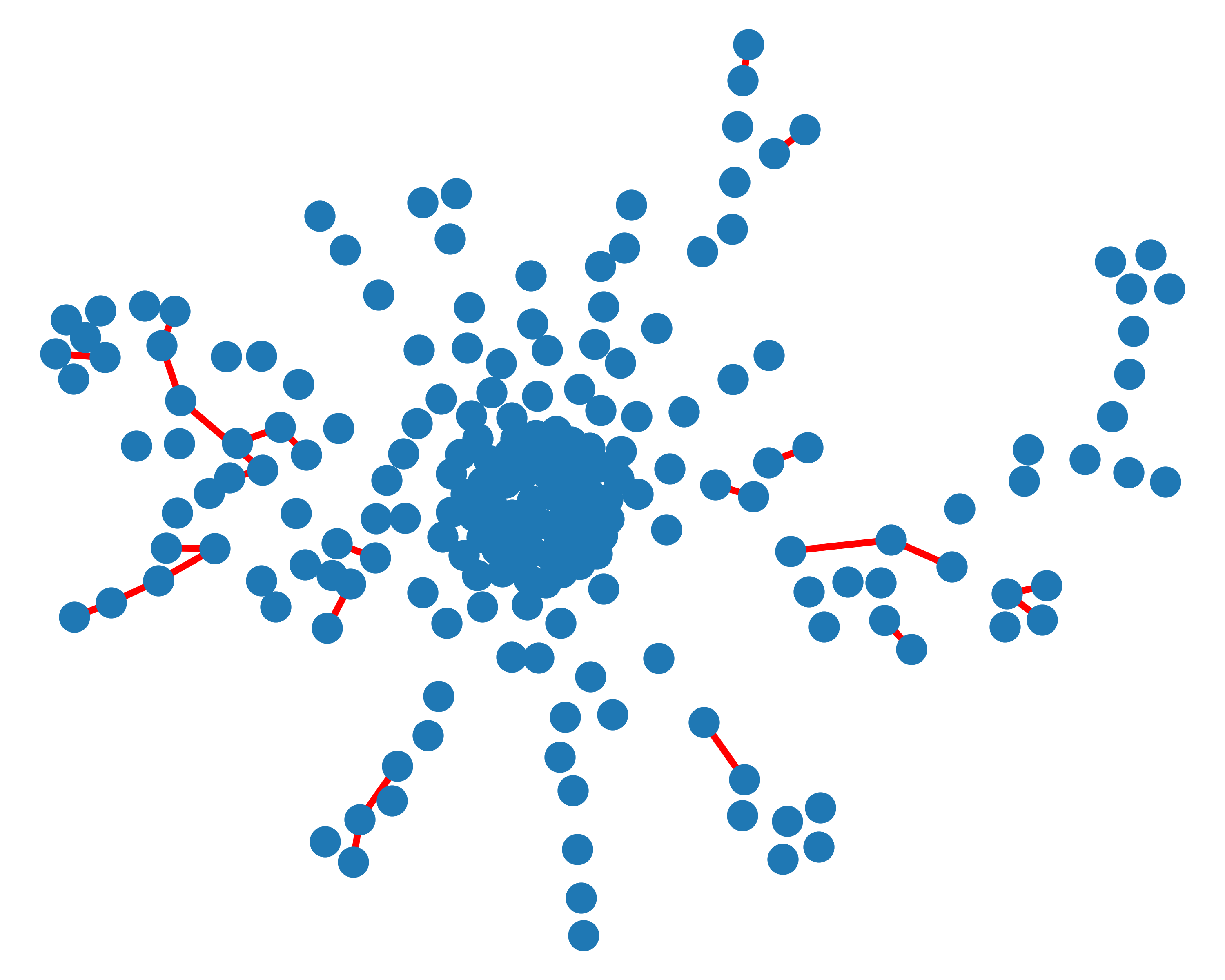}
    \caption{Thresholding absolute correlation at \(0.9\).}
\end{subfigure}
\quad
\begin{subfigure}[t]{.3\textwidth}
    \includegraphics[width=.8\linewidth]{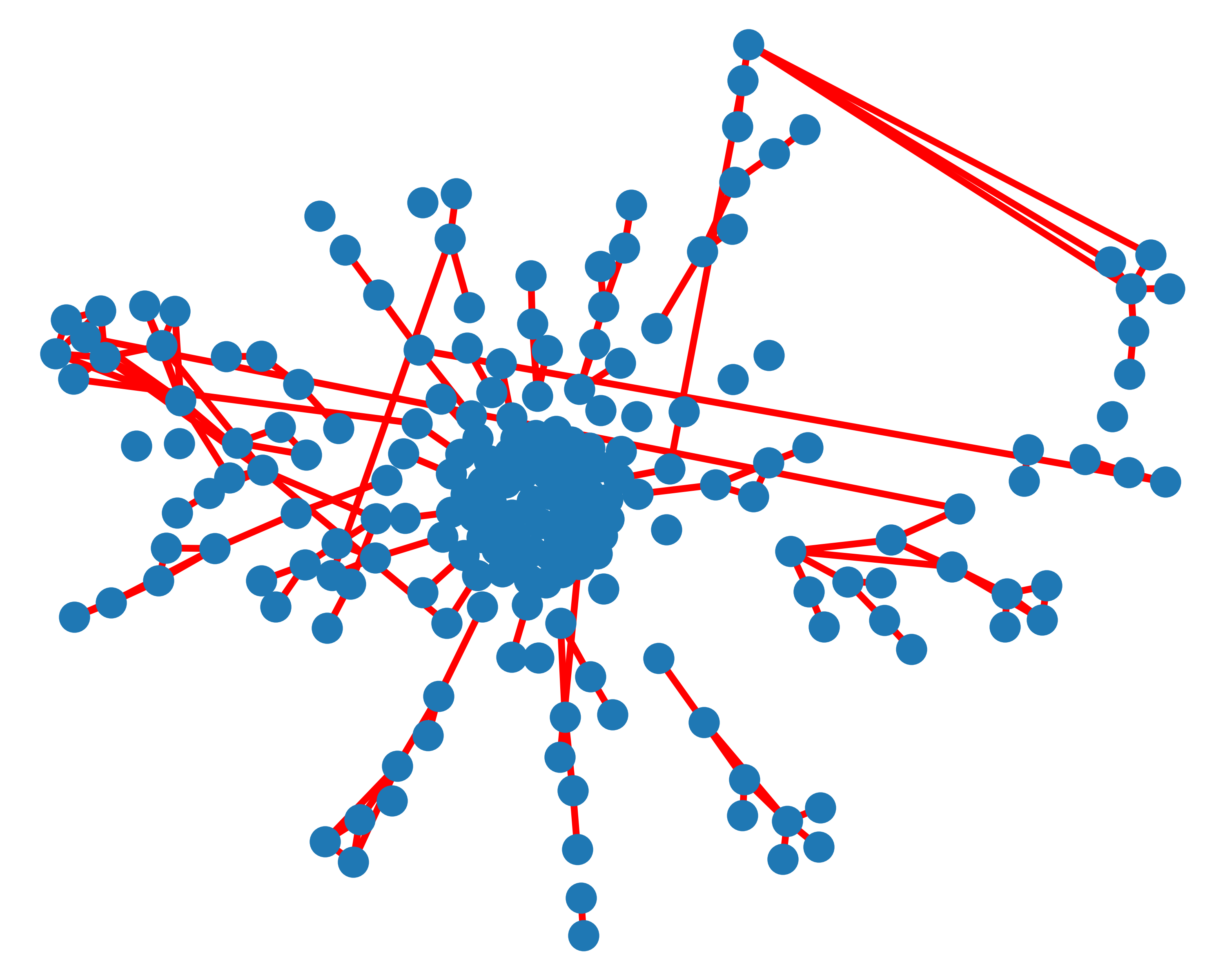}
    \caption{Thresholding absolute correlation at \(0.5\).}
\end{subfigure}
\quad
\begin{subfigure}[t]{.3\textwidth}
    \includegraphics[width=.8\linewidth]{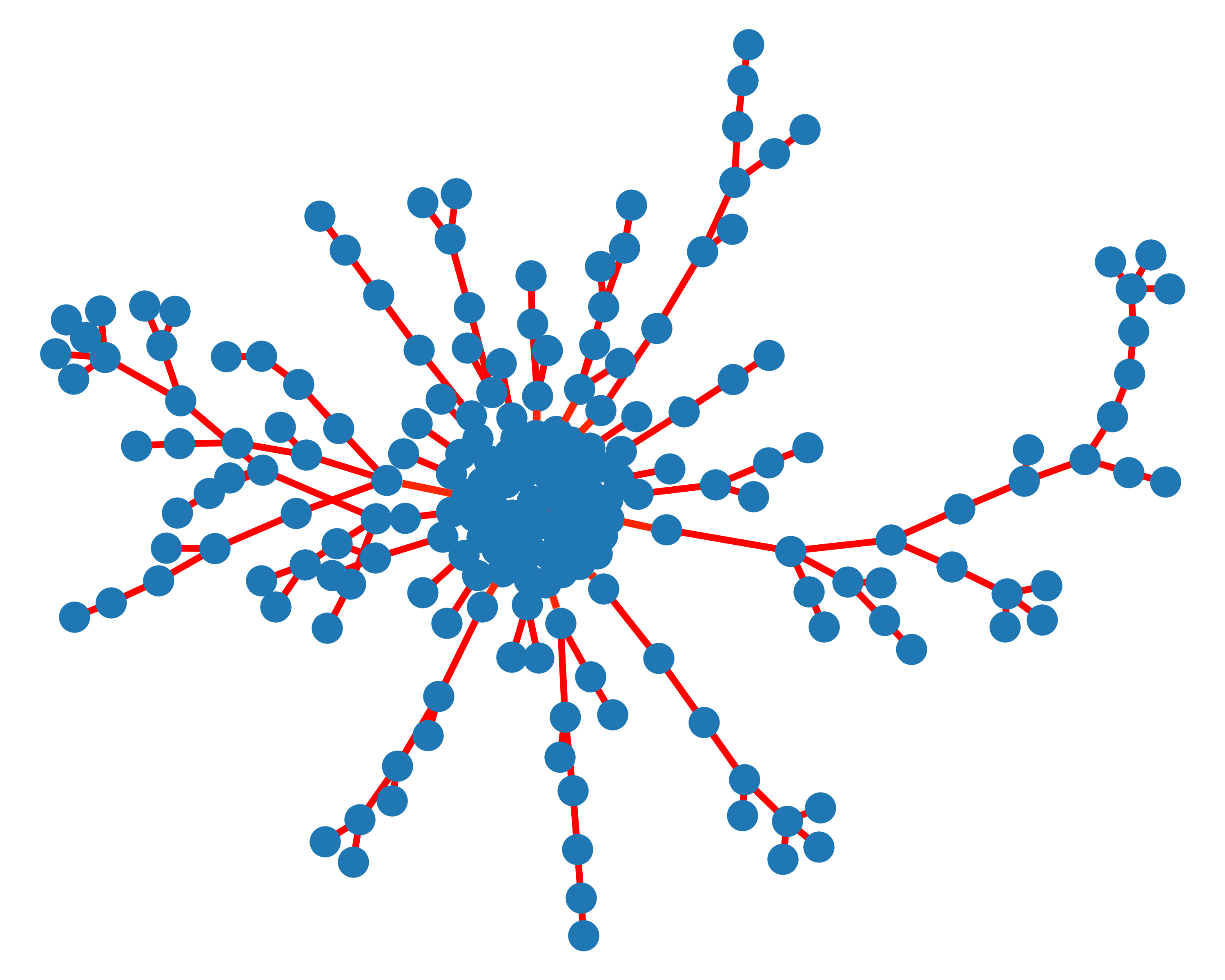}
    \caption{Bayesian spanning tree (posterior mode shown).}
\end{subfigure}\\
\begin{subfigure}[t]{1\textwidth}
    \includegraphics[width=1\linewidth, trim={7cm 0 6cm 0},clip]{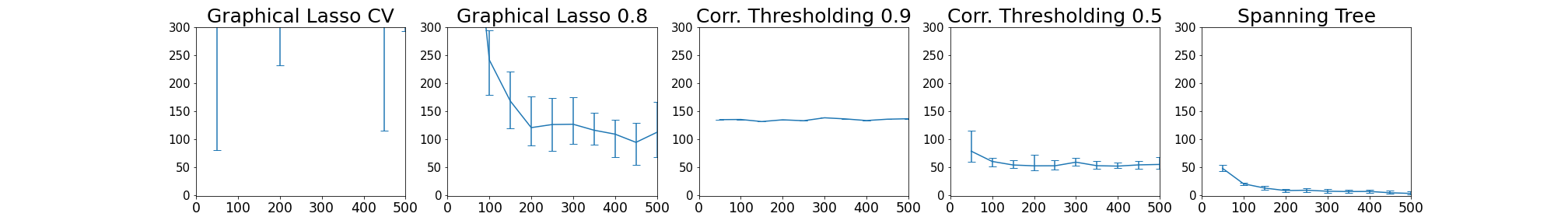}
    \caption{Graph estimation error, when the oracle has about $600$ edges over $200$ nodes.}
\end{subfigure}
\begin{subfigure}[t]{1\textwidth}
    \includegraphics[width=1\linewidth, trim={7cm 0 6cm 0},clip]{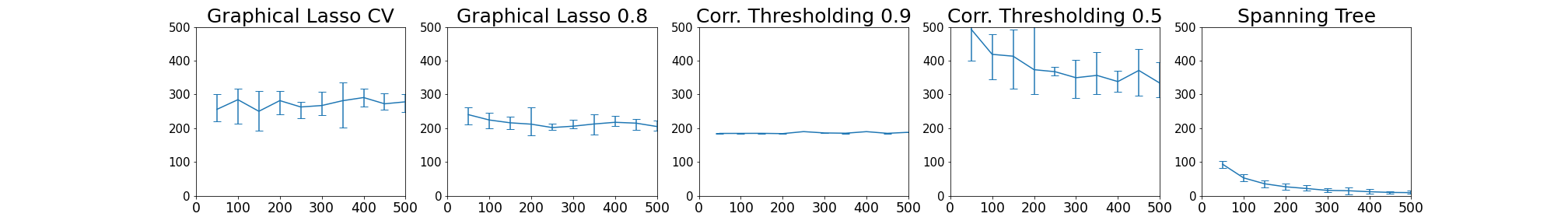}
    \caption{Graph estimation error, when the oracle has about $4,000$ edges over $200$ nodes.}
\end{subfigure}
 \caption{Simulated experiments on  graph estimation. Panels (b-f) are point estimates obtained at $n=100$ for the oracle graph with about $600$ edges.
 \label{fig:sim3}}
 \end{figure}

 As shown in Figure~\ref{fig:sim3}, the graphical lasso using cross-validation produces the largest number of edges; while its estimates cover $T_0$, they also contain too many falsely positive $|\hat G \setminus G_0|$. Empirically tuning \(\alpha=0.8\) in the graphical lasso somewhat reduces this problem. The correlation thresholding estimator  using \(|\rho|_{j,k} \ge 0.5\) (as a common ``default'' choice in practice) seems to produce the best results among the existing approaches. 
The Bayesian spanning tree shows a competitive performance to the best existing approach. At the same time, it shows a low error for estimating $T_0$: at $n\ge 100$, it almost perfectly recovers all the edges in $T_0$. 

Next, we slightly change the experiment setting by increasing the denseness of the oracle graph. We set the sparsity level to $0.2$, which leads to a graph with about $4,000$ edges over $200$ nodes. This time, all the existing approaches show much larger estimation error, likely due to the breakdown of the sparsity assumption. On the other hand, the Bayesian spanning tree still maintains a good performance, with the estimation error rapidly decreasing in $n$.   For conciseness, we provide all the details in the supplementary materials.

\subsection{Uncertainty Quantification of the Graph Estimates}

We now demonstrate the capability of the uncertainty quantification of our Bayesian spanning tree model. 

\begin{figure}[H]
\begin{subfigure}[t]{.3\textwidth}
    \includegraphics[width=1\linewidth]{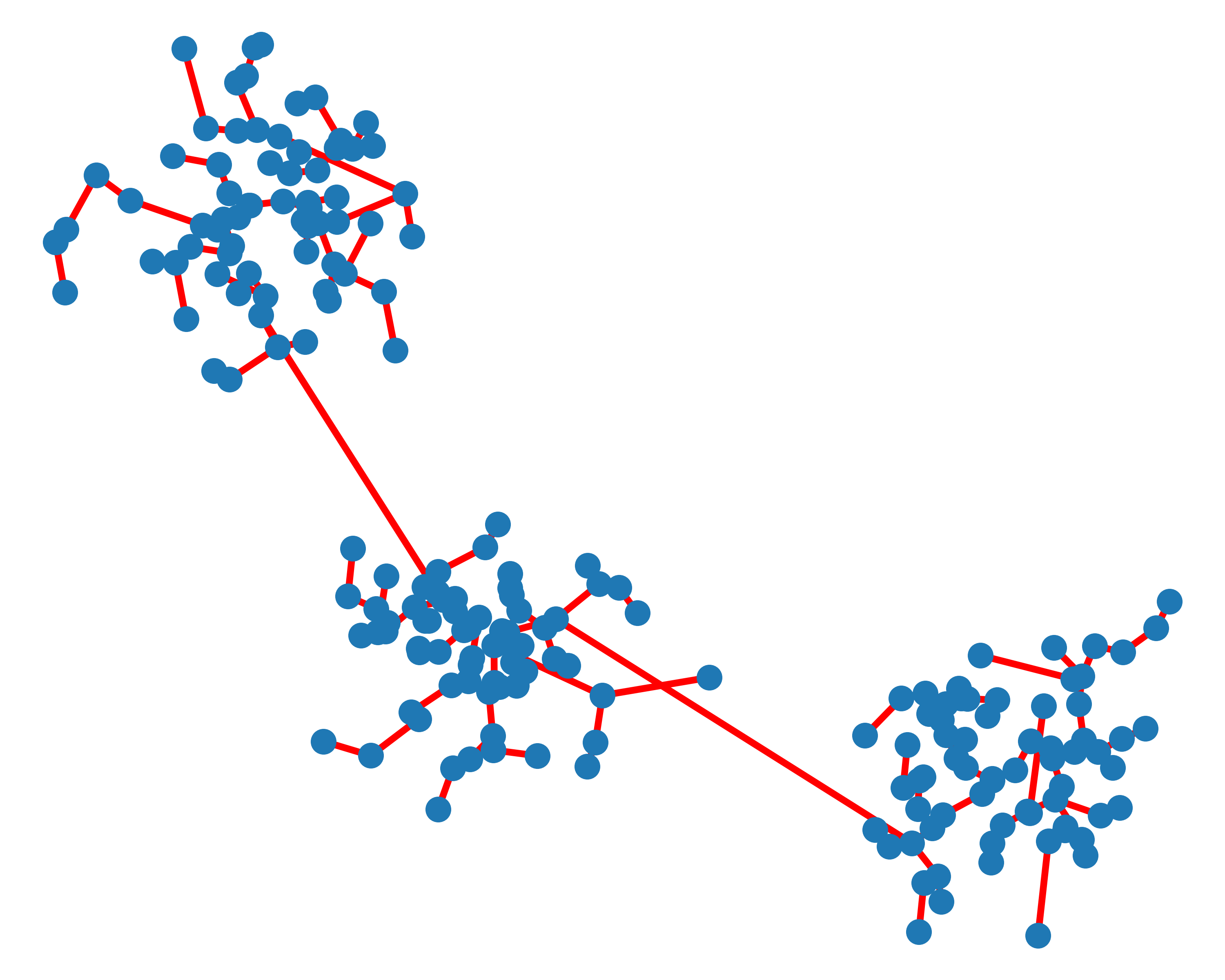}
    \caption{One spanning tree sampled from the posterior distribution.}
\end{subfigure}
\quad
\begin{subfigure}[t]{.3\textwidth}
    \includegraphics[width=1\linewidth]{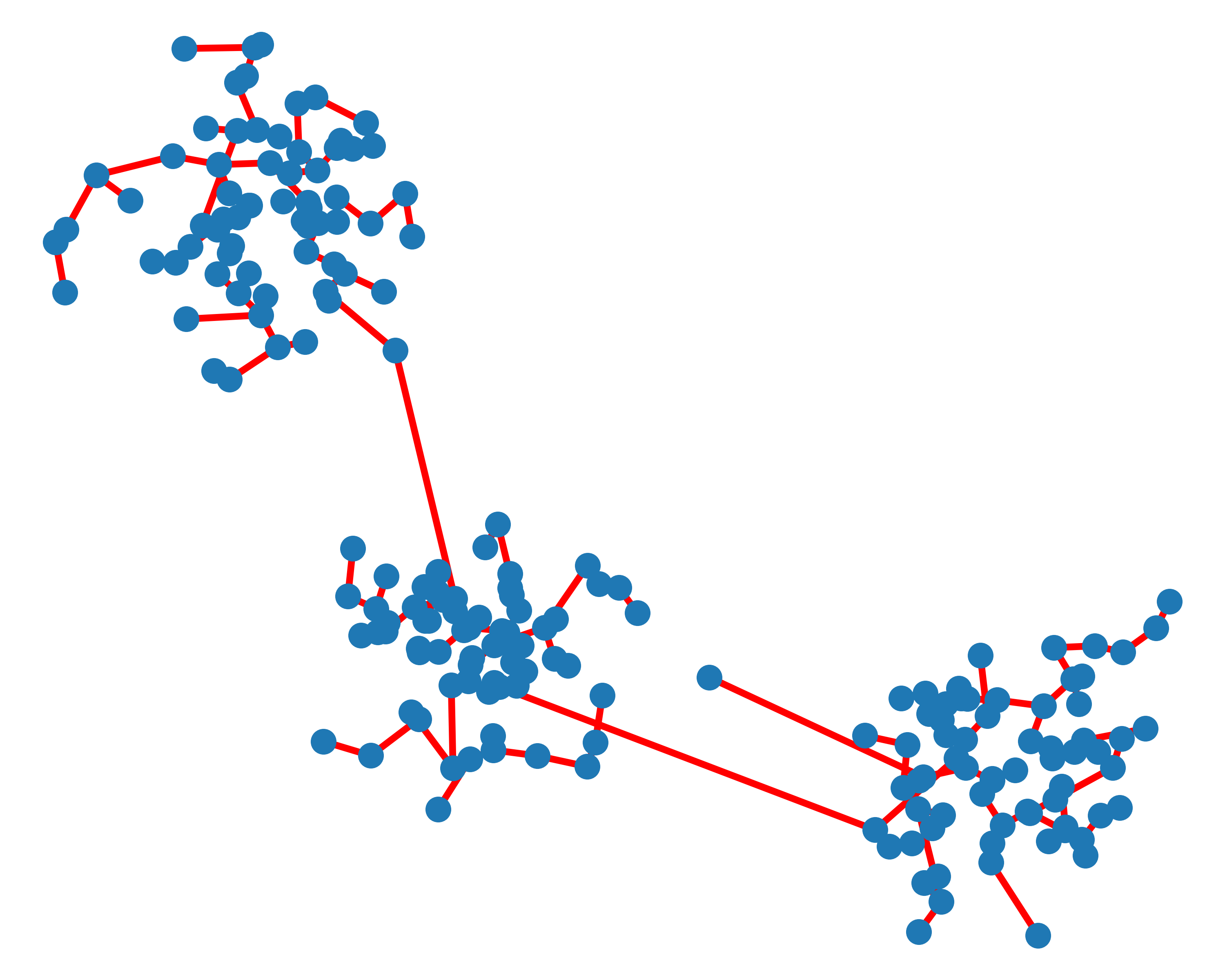}
    \caption{Another spanning tree sampled from the posterior distribution.}
\end{subfigure}
\quad
\begin{subfigure}[t]{.3\textwidth}
    \includegraphics[width=1\linewidth]{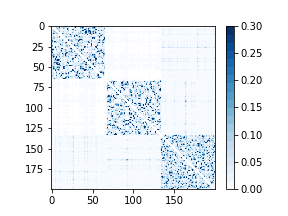}
    \caption{The matrix of marginal connecting probabilities $\t{pr}[T \ni (j,k)\mid y]$.}
\end{subfigure}
 \caption{High uncertainty of spanning trees when estimating a latent position graph with 3 communities, each formed by a bivariate Gaussian.
 \label{fig:blobs}}
 \end{figure}

First, we consider the common example of a latent position graph \citep{hoff2002latent} associated with three communities. In the latent space, each community is a group of points generated from a bivariate Gaussian. As shown in Figure~\ref{fig:blobs}, the likely spanning tree $T$ is the one containing three component trees, each spanning the points within a community, and two long edges binding the three trees together.

There is a large amount of uncertainty in this model, as can be seen via comparing Panels (a) and (b): (i) within a community, each point has a large number of other points in its neighborhood, hence there are multiple ways to form a tree with high posterior probability (that is, we have a low separability constant $\delta$); (ii) when connecting two communities together, these candidate long edges do not differ much in the density \eqref{eq:gdp_marginal} (due to the near polynomial density tail of generalized double Pareto), hence they are almost equally likely to enter $T$. As shown in Panel (c), most of the edges have a relatively low marginal connecting probability $\t{pr}[T \ni (j,k)\mid y]$, indicating the posterior probability of $T$ is scattered over a large number of different trees.

\begin{figure}[H]
\begin{subfigure}[t]{.3\textwidth}
    \includegraphics[width=1\linewidth]{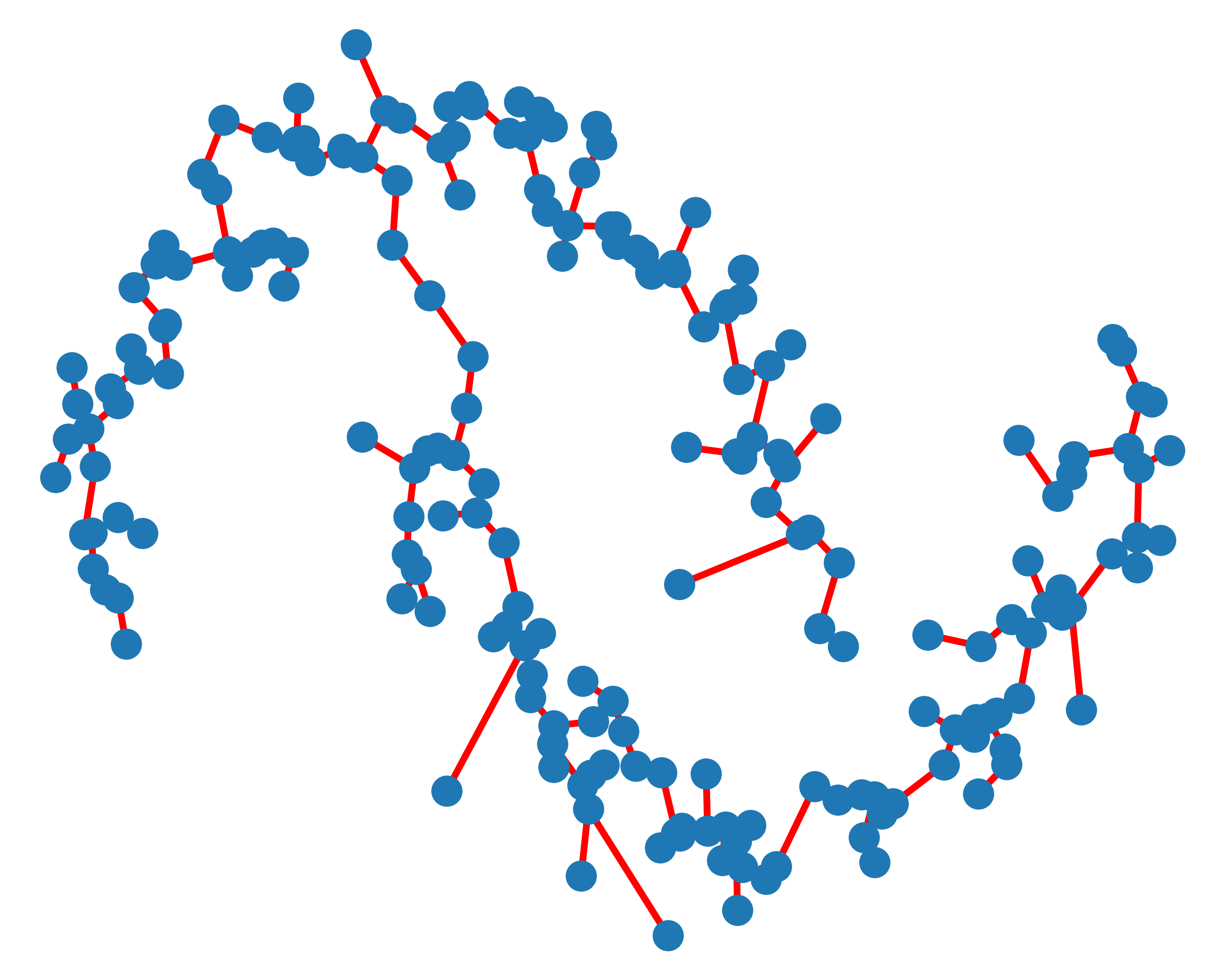}
    \caption{One spanning tree sampled from the posterior distribution.}
\end{subfigure}
\quad
\begin{subfigure}[t]{.3\textwidth}
    \includegraphics[width=1\linewidth]{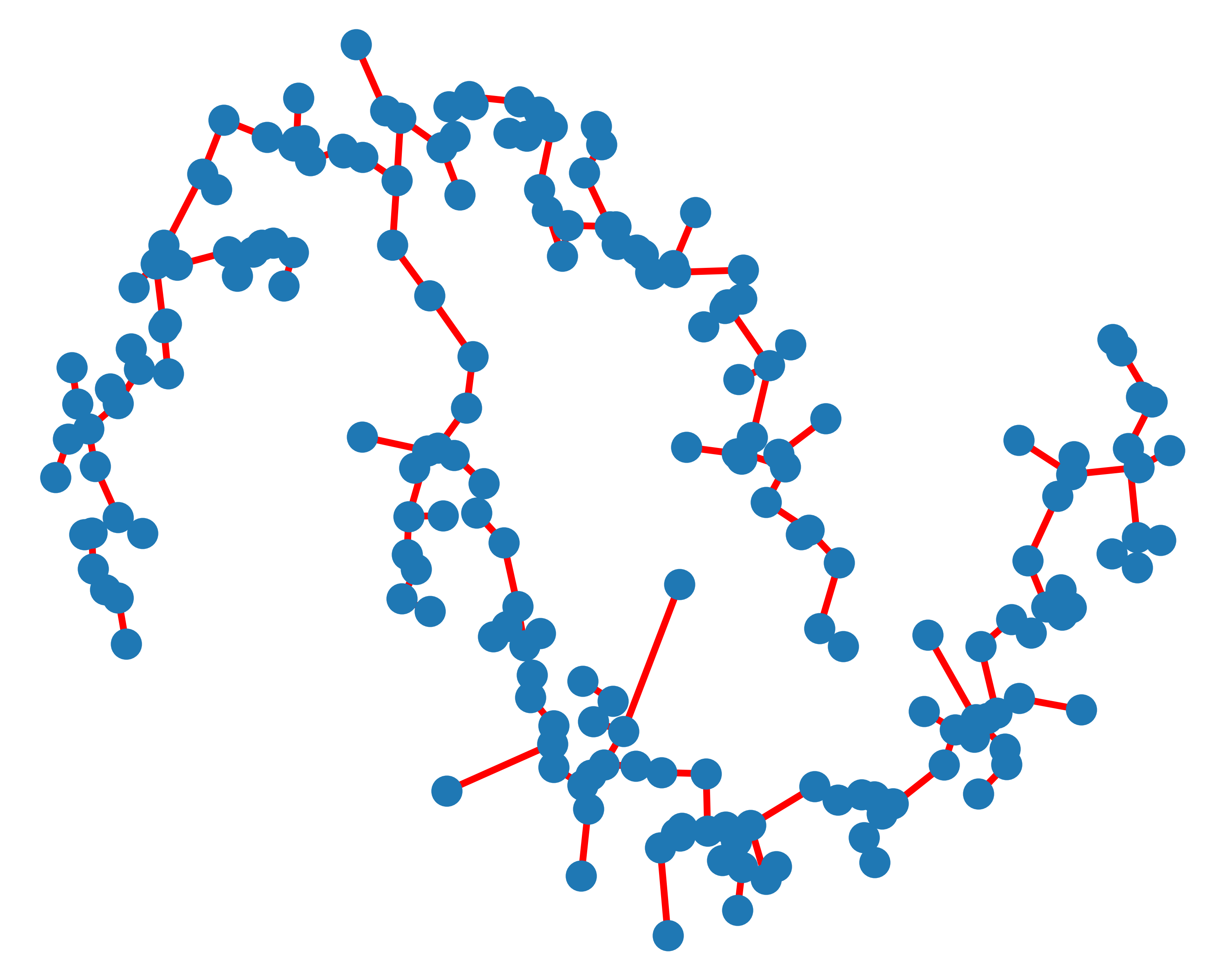}
    \caption{Another spanning tree sampled from the posterior distribution.}
\end{subfigure}
\quad
\begin{subfigure}[t]{.3\textwidth}
    \includegraphics[width=1\linewidth]{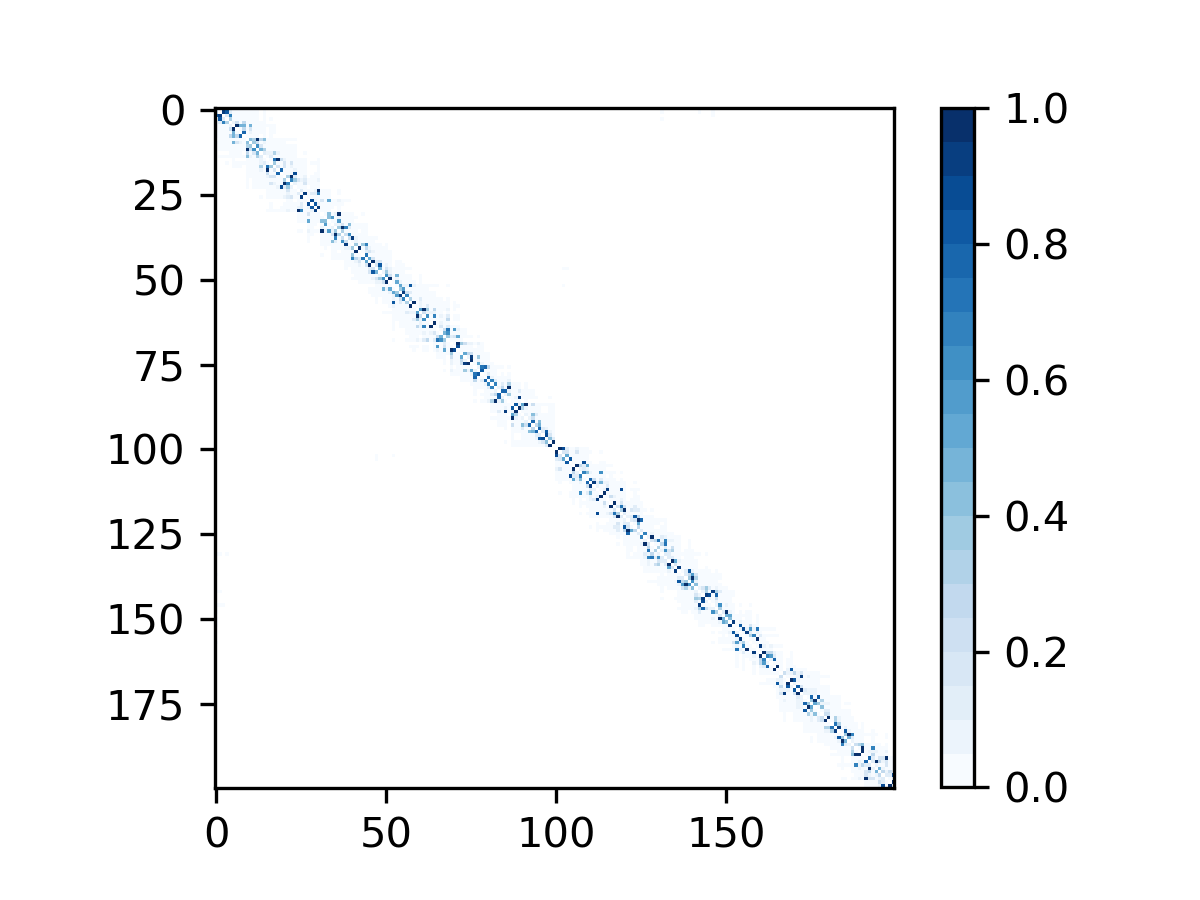}
    \caption{The matrix of marginal connecting probabilities $\t{pr}[T \ni (j,k)\mid y]$.}
\end{subfigure}
 \caption{Low uncertainty of spanning trees for when estimating a graph formed by the two-moon manifold.
 \label{fig:two_moons}}
 \end{figure}

 Next, we move to the case of a graph formed near two manifolds. We use the two-moon example provided in the scikit-learn package. As shown in Figure~\ref{fig:two_moons}, each point has only one or a few points in the neighborhood, therefore, the posterior distribution of $T$ is highly concentrated near the posterior mode. As a result, the two random posterior samples do not seem to differ much; and we have high values of $\sum_T\t{pr}[(j,k)\in T,T \mid y]$ near the diagonal of the matrix.
  
\section{Spanning Tree Modeling of Brain Networks}

We use our Bayesian spanning tree model to analyze data from a neuroscience study on  human working memory. The study involves 20 human subjects participating in the  Sternberg verbal memory test: first, each subject reads a list of $g$ numbers on the screen, trying to memorize them for $2$ seconds; then with the numbers removed from the screen, the subject answers if a particular number was in the list shown earlier. During this process, electroencephalogram (EEG) signals are obtained from electrode channels placed over 128 regions of interest (ROIs) of each subject's brain, over $5$ seconds covering $512$ times points. The goal of this study is to not assess whether the subject correctly answers the question, but to find out how the brain acts differently when the subject is performing a simpler task with $g=2$ numbers versus a more challenging task with $g=6$ numbers.

We denote each EEG time series by $y^{[s,g,t]}_j$, as the signal  collected from the $j$th ROI for subject $s$ under the task load $g$ at time $t$. To flexibly model these time series, we use the following Hidden Markov Model, based on $K$ latent graph states, each modeled by a spanning tree  $( T^{(1)},\ldots, T^{(K)})$:

\be
&  y^{[s,g,t]}_1,  y^{[s,g,t]}_2,\ldots,   y^{[s,g,t]}_{128} \sim \t{BSTM}(\tilde T_{s,g,t}; \tau),\\
& \t{pr}[\tilde T_{s,g,0} = T^{(k)} ] = q_{0,k}, \\
& \t{pr}[\tilde T_{s,g,t} = T^{(k)} \mid \tilde T_{s,g,t-1} = T^{(k')} ] = q^{[g]}_{k',k} \t{ for } t>1,\\
& (q_{0,1}, \ldots, q_{0,K}) \sim \t{Dir}(0.5,\ldots, 0.5), \\
& (q^{[g]}_{k',1}, \ldots, q^{[g]}_{k',K}) \sim \t{Dir}(0.5,\ldots, 0.5) \t{ for } k'=1,\ldots,K,\\
& \Pi_0(T^{(k)}) \propto 1,
 \ee
 where BSTM$(\tilde T;\tau)$ represents the Bayesian spanning tree model with the tree $\tilde T$ and the density \eqref{eq:gdp_marginal} with the scale parameter $\tau$; 
$(q_{0,1}, \ldots, q_{0,K})$ are the initial probability distribution for the $K$ states. To enable borrowing of information across subjects and tasks, we let the parameters $\tau$, $q_{0,k}$'s and the dictionary of trees $T^{(k)}$'s to be shared across subjects and tasks. On the other hand, to characterize the difference between two tasks,
we set each $q^{[g]}_{k',k}$, the transition probability from state $k'$ to state $k$, to be different according to the task load $g=2$ or $6$.

We use the Dirichlet distribution with concentration parameter $0.5$ to induce approximate sparsity in the values of the initial and transition probabilities, and we set $K=20$ for as an upper bound. In posterior samples, we found the maximum number of states used by the model is only $5$, indicating $K=20$ is sufficient as an upper bound.

 \begin{figure}[H]
 \centering
 \begin{subfigure}[t]{.5\textwidth}
    \includegraphics[width=1\linewidth]{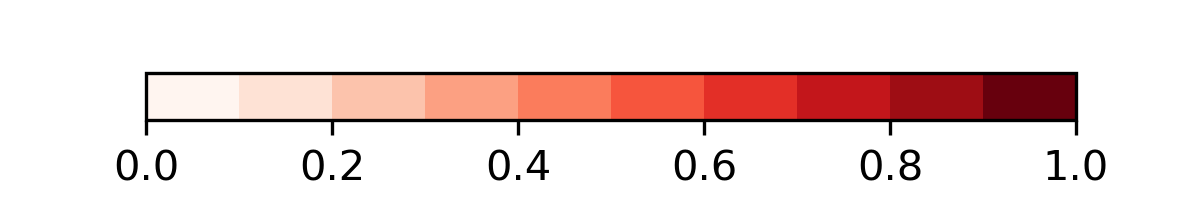}
\end{subfigure}\\
\begin{subfigure}[t]{.18\textwidth}
    \includegraphics[width=1\linewidth]{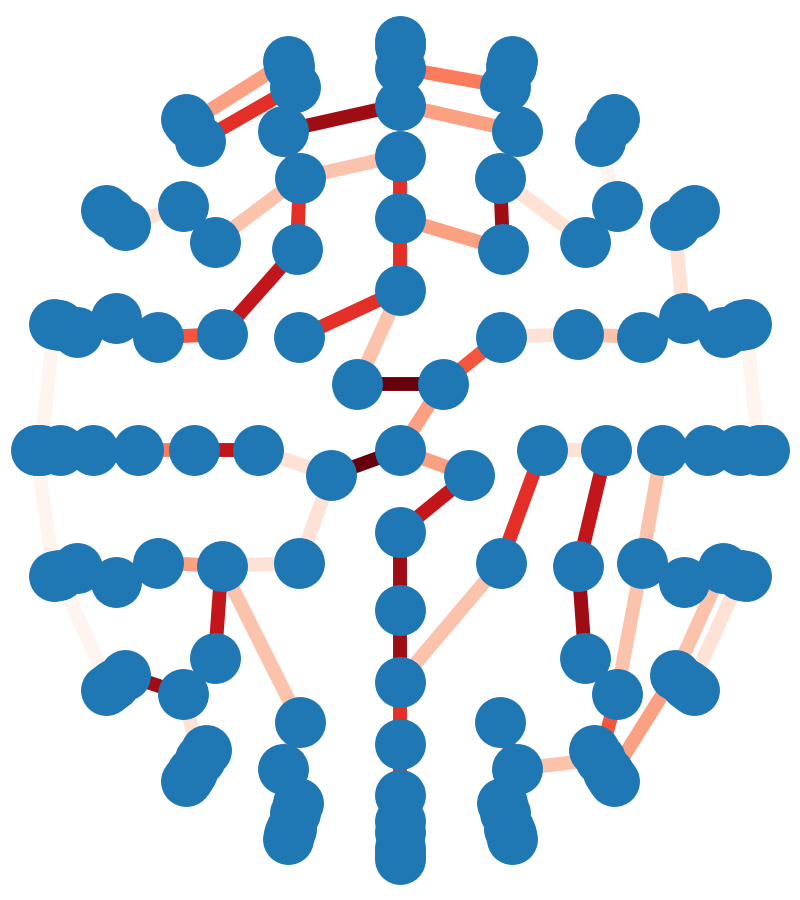}
    \caption{State 1}
\end{subfigure}
\;
\begin{subfigure}[t]{.18\textwidth}
    \includegraphics[width=1\linewidth]{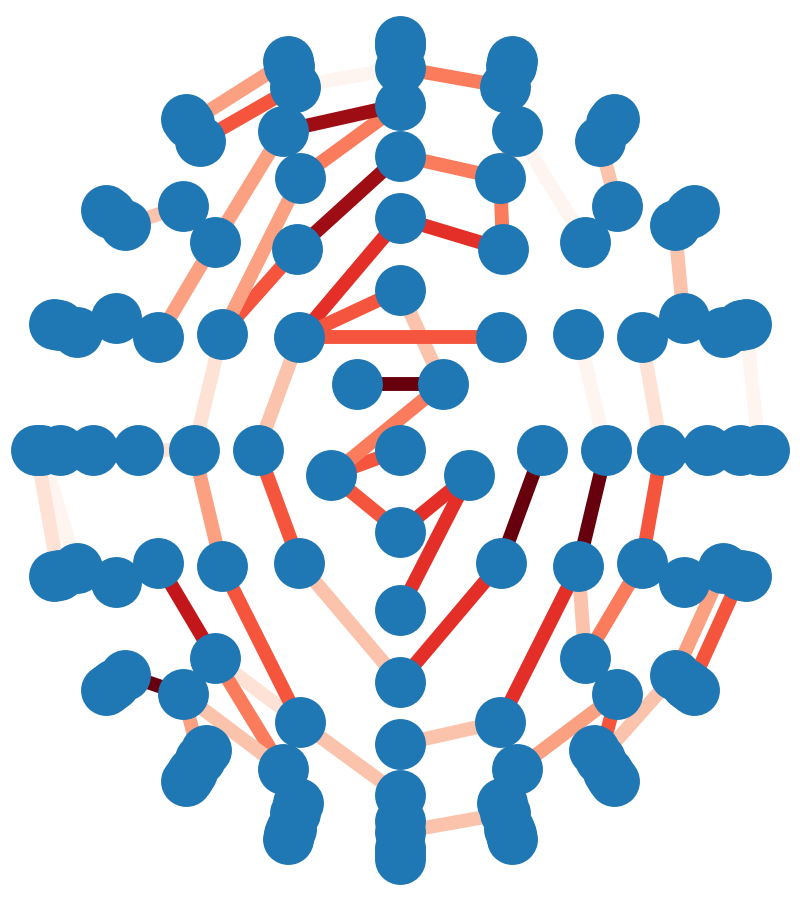}
    \caption{State 2}
\end{subfigure}
\;
\begin{subfigure}[t]{.18\textwidth}
    \includegraphics[width=1\linewidth]{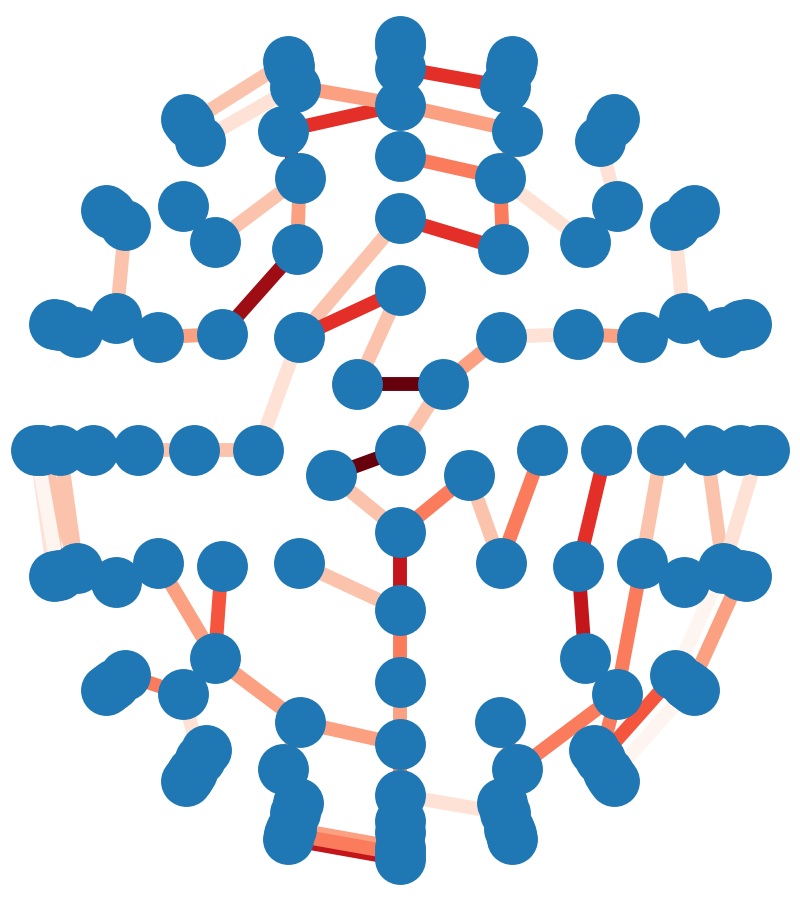}
    \caption{State 3}
\end{subfigure}
\;
\begin{subfigure}[t]{.18\textwidth}
    \includegraphics[width=1\linewidth]{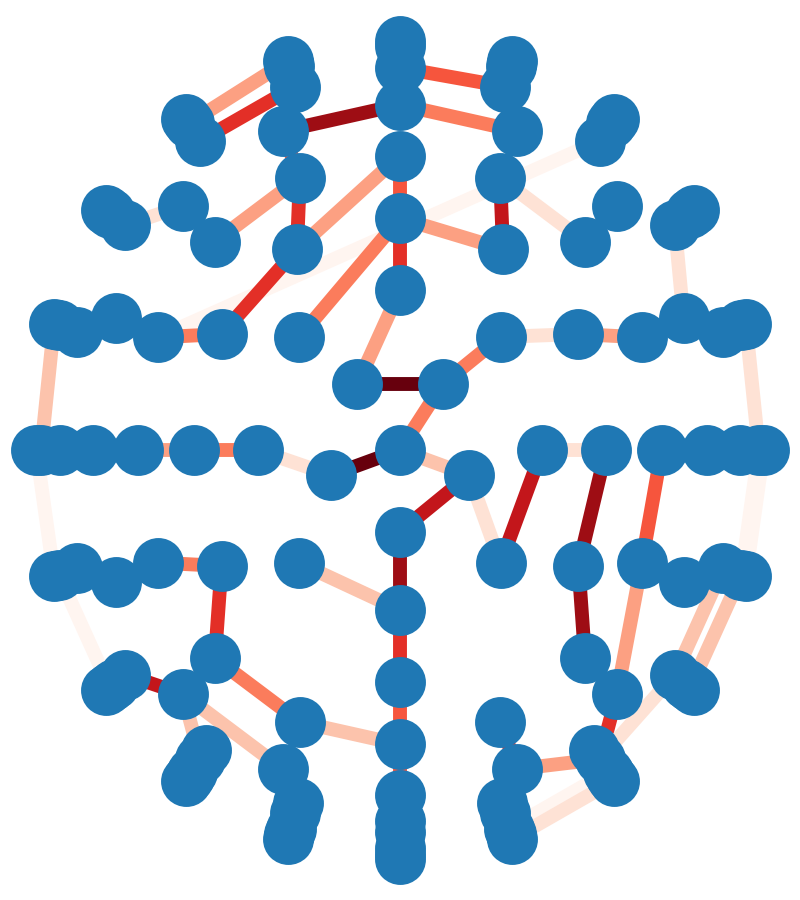}
    \caption{State 4}
\end{subfigure}
\;
\begin{subfigure}[t]{.18\textwidth}
    \includegraphics[width=1\linewidth]{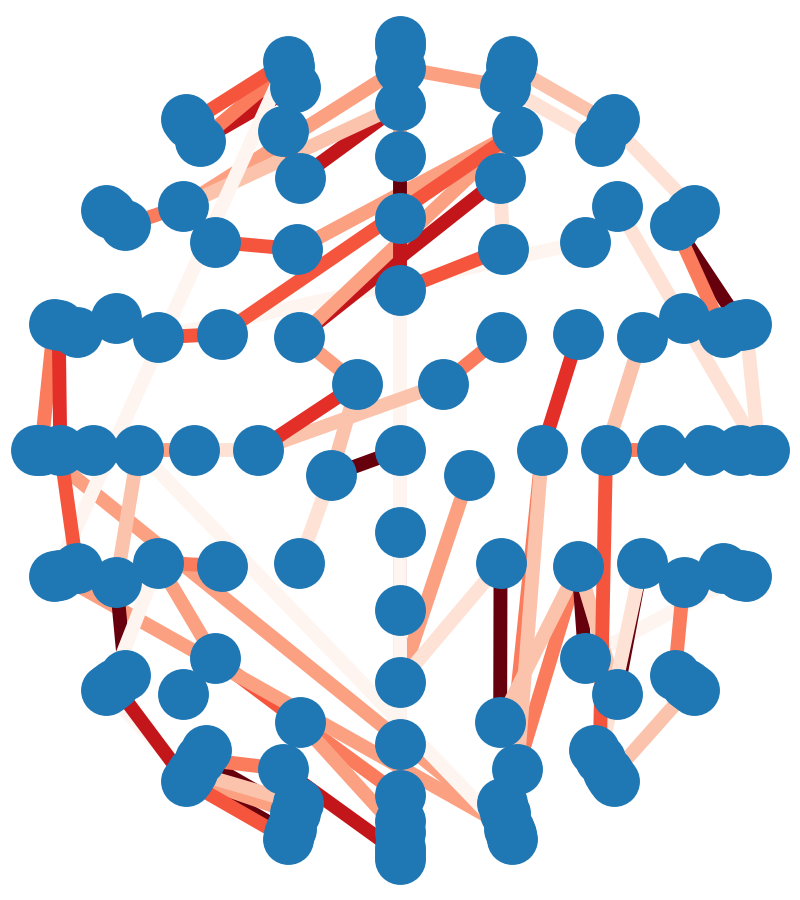}
      \caption{State 5}
\end{subfigure}\\
\begin{subfigure}[t]{.3\textwidth}
    \includegraphics[width=1\linewidth]{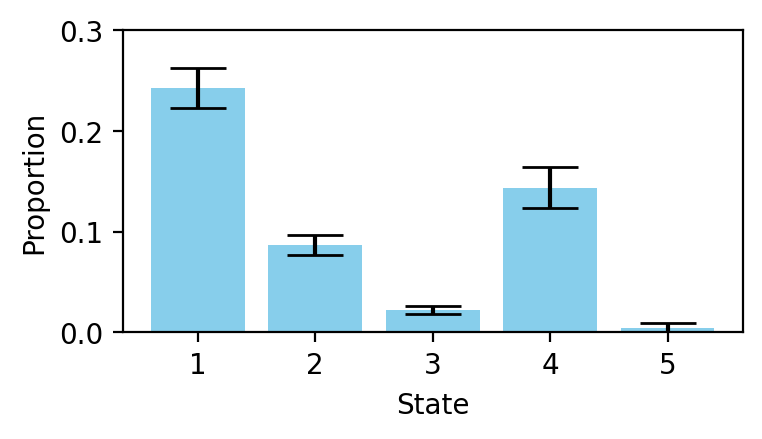}
      \caption{Distribution of state assignment for memory task load $g=2$.}
\end{subfigure}
\;
\begin{subfigure}[t]{.3\textwidth}
    \includegraphics[width=1\linewidth]{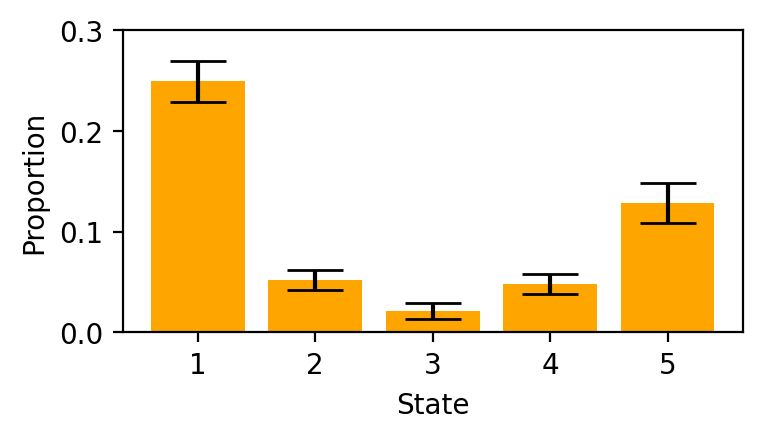}
          \caption{Distribution of state assignment for memory task load $g=6$.}
\end{subfigure}
\;
\begin{subfigure}[t]{.3\textwidth}
    \includegraphics[width=1\linewidth]{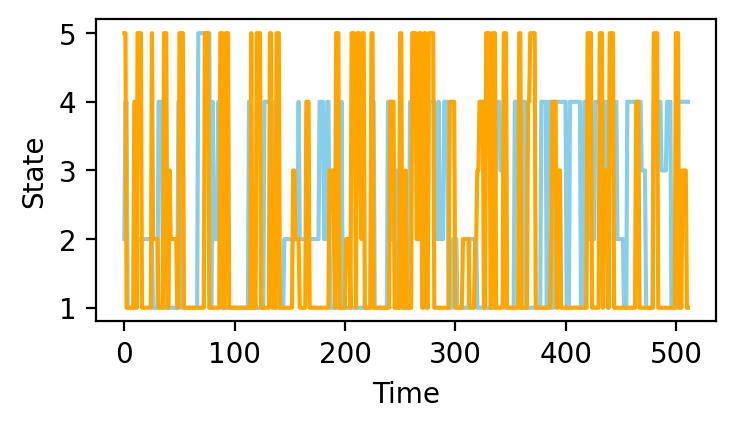}
          \caption{Comparison of the two time series in the state assignment for one subject.}
\end{subfigure}
 \caption{Results of analyzing the electroencephalogram (EEG) time series collected over 128 channels on the human brain, using the Hidden Markov Model with the spanning trees as the latent states. The posterior distribution contains five latent states (Panels a-e, viewed from the top of the head), where each panel shows the posterior mode spanning tree with the color representing the marginal connecting probability. Comparing the task of memorizing $g=2$ numbers (Panel f) to the one of $g=6$ numbers (Panel g), the latter involves more time spent in the State 5, during which the brain is more active and has more long-range connectivities over the brain (Panel e). Panel h shows a comparison in the times series of the latent state assignments, when a given subject is memorizing $g=2$ (blue) versus  $g=6$ (orange) numbers.
 \label{fig:eeg}}
 \end{figure}

We run our MCMC algorithm for $20,000$ iterations, and discard the initial $10,000$ as a burn-in.
Figure~\ref{fig:eeg} shows the results for the data analysis.  We plot the posterior mode spanning tree for each latent state, while showing the uncertainty on each edge using the marginal connecting probability.

The results are quite interpretable --- as shown in Panels a and d, the two dominating states correspond to having each ROI connect to another that is spatially close. There is separation between the front of the brain (upper part in each plot) and the rear (lower part).  Comparing the task of memorizing $g=2$ numbers (Panel f) against the one of $g=6$ numbers (Panel g), the latter involves more time spent in the State 5, during which the brain is more active and has more long-range connectivities over the brain (Panel e).

To validate our trained model, we use a previously reserved set of EEG data collected from another 20 subjects (hence 40 time series), and using our estimated model to classify whether a time series is likely to be collected under $g=2$ or $g=6$. Specifically, for each testing time series
$\tilde y^{[s,\tilde g,t]} = (\tilde y^{[s,\tilde g,t]}_1,  \tilde y^{[s,\tilde g,t]}_2,\ldots,   \tilde y^{[s,\tilde  g,t]}_{128})_t$, and for $g=2$ or $6$, we sample the latent state assignments given $\hat T^{(k)}$'s fixed at the posterior mode, $\hat  q_{0,k}$'s and $\hat q^{[g]}_{k',k}$'s fixed at the posterior mean from the trained model, over the course of $10,000$ MCMC iterations. We average the last $5,000$ iterations to compute the likelihood $ \Pi( \tilde y^{[s,\tilde g,t]} \mid  \hat T^{(k)}, \hat  q_{0,k}, \hat  q^{[g]}_{k',k})$ marginalizing out the state assignment. Comparing $g=2$ and $6$, we obtain the classification probability:
\be
\t{pr}(g=2 \mid .) = \frac{ \Pi( \tilde y^{[s,\tilde g,t]} \mid \hat  T^{(k)}, \hat  q_{0,k}, \hat  q^{[2]}_{k',k} \t{ all } k,k') }{ \Pi( \tilde y^{[s,\tilde g,t]} \mid \hat  T^{(k)},\hat  q_{0,k}, \hat  q^{[2]}_{k',k} \t{ all } k,k')
+  \Pi( \tilde y^{[s,\tilde g,t]} \mid  \hat  T^{(k)}, \hat  q_{0,k}, \hat q^{[6]}_{k',k} \t{ all } k,k')
}.
\ee
Using $g=2$ if $\t{pr}(g=2 \mid .)>0.5$ and $g=6$ otherwise, we obtain a low misclassification rate of $15\%$  when comparing with true $\tilde g$. In the  receiver operating characteristic curve, we calculate the area under the curve (AUC) and obtain $89\%$. This suggests that our model provides an adequate characterization the differences between the groups, given the low signal-to-noise ratio in EEG data.

To compare, we also run the same Hidden Markov Model except with each latent state modeled by a multivariate Gaussian distribution $\t{N}(\vec \mu_k,\Omega_k)$, with $\Omega_k$ estimated via the graphical lasso with $\alpha=0.5$. Setting $K$ at $20$, we obtain a competitive validation result with the misclassification rate at $15\%$ and AUC at $85\%$; however, a major drawback is that this Gaussian Hidden Markov Model involves $16$ latent states with nontrivial probabilities ($ \ge 5\%$), hence is much more difficult to interpret than the $5$ states from the spanning tree-based model. In addition, we test the Gaussian Hidden Markov Model with $K$ reduced to 10, and it leads to much worse validation performance, with the misclassification rate at $45\%$ and AUC at $53\%$, which is almost close to a random guess; similarly, we test $\alpha=0.1$ and $\alpha=1$ in the graphical lasso, and obtain similarly poor performance. Therefore, the Gaussian Hidden Markov Model is much less parsimonious than the spanning tree-based model.

\section{Discussion}
In this article, we propose to use the spanning tree as a restricted graph model to estimate the backbone of a latent graph. We study its mathematical properties and 
demonstrate good performances in both theory and applications in recovering important subsets of edges. There are several interesting extensions worth exploring. First, instead of using only one spanning tree, one could consider the union of multiple spanning trees as a more flexible graphical model; there are several open questions that need to be addressed, such as the identifiability issue due to the overlap of multiple trees as well as its finite sample recovery theory. Second, one could consider the spanning tree as the opposite extremity of the clique graph [a completely connected graph with $p(p-1)/2$ edges]; therefore, one could view the broad class of connected graphs as in some continuum between those two extremal graphs, potentially creating useful new graph models. Lastly, the spanning tree framework could be adopted in the community detection framework. As alluded to when we chose the edge density, one could further build a clustering algorithm by first finding a spanning tree, then cutting the longest few  edges to produce several isolated components. Such a clustering algorithm has been proposed in computer science \citep{jana2009efficient}, yet its mathematical and statistical properties remain largely unexplored.

\section*{Acknowledgement}
This work was partially supported by grants R01-MH118927-01 and R01ES027498 of the United States National Institutes of Health, and grant N00014-21-1-2510-01 of the United States Office of Naval Research.

\bibliographystyle{chicago}
\bibliography{reference}

\appendix

\section*{Supplementary Materials}

\subsection*{Proof of Theorem 1}
\begin{proof}
The proof slightly extends \cite{chaiken1978matrix}, which states
\be
z_q  = \frac{1}{p}\prod_{j=2}^p \lambda_{(j)}(L_q),
\ee
where $\lambda_{(2)}(L_q)\le \ldots \lambda_{(p)}(L_q)$  are the largest $p-1$ eigenvalues of $L_q$. Since the smallest eigenvalue of $L_q$ is $0$ with a corresponding eigenvector $1_p/\sqrt{p}$, adding $J/p/p$ to $L_q$ and taking the determinant yields the result.
\end{proof}

\subsection*{Proof of Theorem 2}

\begin{proof}

For ease of notation, let  $M=I- B_{[-s]}(B_{[-s]}^{\rm T} B_{[-s]})^{-1} B_{[-s]}^{\rm T}$. After removing edge $e_s$, we obtain two separated subgraphs, denoted by $G_1$ and $G_2$.
Considering two nodes $j_1$ and $j_2$ in the same connected subgraph (without loss of generality, in $G_1$), we use a $p$-element binary vector $\vec x(j_1,j_2)$ to represent auxiliary edge $(j_1,j_2)$, with the $j_1$th element equal to $1$ and $j_2$th element equal to  $-1$, and all other elements $0$.

We know there is a path in $G_1$ that connects the nodes $j_1$ and $j_2$. We can represent the auxiliary edge $\vec x(j_1,j_2)$ as a linear combination of the columns of $B_{[-s]}$ using the edges in the path. That is, there exists $a_h$'s taking value in $ \{-1,1\}$ such that:
$$\vec x(j_1,j_2) = \sum_{h: e_h\in \t{path}(j_1,j_2)} a_h \vec B_{h}.$$ 
This means that $\vec x(j_1,j_2)$ is in the column space of $B_{[-s]}$, therefore $M \vec x(j_1,j_2)= \vec 0$. 

Multiplying $M$ to $\vec x(j_1,j_2)$ corresponds to creating a contrast between the columns $j_1$ and $j_2$ of $M$. Hence if $j_1,j_2$ are in the same subgraph, $M_{l,j_1}=M_{l,j_2}$ for all $l=1,\ldots,p$; since $M$ is symmetric, $M_{j_1,l}=M_{j_2,l}$ for all $l=1,\ldots,p$.
Therefore, for two index sets $V_1 =\{j: j \in G_1\}$ and $V_2 =\{ k: k\in G_2\}$, the matrix $M$ can be divided into four blocks; within each the elements have the same value: $M_{j_1,j_2}=m_{1,1}$ for $j_1\in V_1,j_2\in V_1$,
$M_{j_2,j_1}=M_{j_1,j_2}=m_{1,2}$ for $j_1\in V_1,j_2\in V_2$,
$M_{j_1,j_2}=m_{2,2}$ for $j_1\in V_2,j_2\in V_2$.

Since $\vec B_s=\vec x(j,k)$ [or $-\vec x(j,k)$, we will use the former without loss of generality], we know:
$$M \vec B_s =  (M_{l,j}- M_{l,k})_{l=1,\ldots,p},$$
where $M_{l,j}- M_{l,k}= m_{1,1}-m_{1,2}$ if $l\in V_1$, and 
 $M_{l,j}- M_{l,k}=m_{1,2}-m_{2,2}$ if $l\in V_2$. That is, the vector $M \vec B_s$ is also partitioned into two parts according to $l\in V_1$ or $l\in V_2$. 

It remains to show those two values are distinct $m_{1,1}-m_{1,2} \neq m_{1,2}-m_{2,2}$. We use proof by contradiction. Supposing equality holds, we have $M \vec B_s = \vec 1_p c$, for some scalar $c\in\mathbb R$. Since $B=[B_{[-s]}\; \vec B_s ]$ has full rank $p-1$, $\vec B_s$ should not be in the column space of $B_{[-s]}$; hence $c\neq 0$. However, we know $1^{\rm T}_pM = 1_p^{\rm T} - 1_p^{\rm T}B_{[-s]}(B_{[-s]}^{\rm T} B_{[-s]})^{-1} B^{\rm T}_{[-s]}= 1_p^{\rm T}$, as each column of $B_{[-s]}$ adds up to zero --- that is, $1^{\rm T}_pM \vec B_s = 1^{\rm T}_p \vec B_s =0$, which contradicts $\vec 1_p^{\rm T}\vec 1_p c= pc\neq 0$.
\end{proof}

\subsection*{Proof of Lemma 1}
The proof is trivial by checking each element of $B\Psi^{-1}B^{\rm T}$.

\subsection*{Proof of Lemma 2}

\begin{proof}
Introduce augmented matrices $B^* = ( B\; 1_p )$ and $\Lambda^* = \t{diag}\{\Psi^{-1}, \epsilon \}$. As both matrices are square and full rank, we have
\be
|\hat\Omega| & = |B^* \Lambda^* B^{*\rm T}|
 = | B^*| |\Lambda^* | | B^{*\rm T}|
 = | B^* B^{*\rm T} | |\Lambda^* | 
 = | B B\T +J | |\Lambda^* | .
\ee
Using the previous lemma with binary $(A_\phi)_{j,k}\in\{0,1\}$,  $B B\T$ is the Laplacian of an unweighted spanning tree $\tilde T$, with eigenvalues $\tilde \lambda_p \ge \ldots \ge\tilde  \lambda_{2}>\tilde \lambda_1=0$, and
\be
| B B\T +J | = p \prod_{l=2}^p \tilde \lambda_l.
\ee
Using Kirchhoff's matrix tree theorem \citep{buekenhout1998number}, the product of the top $(p-1)$ eigenvalues of the Laplacian for graph $G$ is related to the number of spanning trees $t(G)$ contained in $G$ via
\be
\prod_{l=2}^p  \tilde \lambda_l=  p t(G).
\ee
As $t(\tilde T)=1$, $| B B\T +J |=p^2$. Combining the above, we have $|\hat\Omega| = p^2  \epsilon |\Psi^{-1}|$.
\end{proof}

\subsection*{Proof of Theorem 3}

 \begin{proof}
We first state the theorem from \cite{goh2002duality}.

\begin{theorem}
For a complete graph with edge weights $\{W_{j,k}\}$, a spanning tree
$T$ is a minimum spanning tree if and only if: for every edge $(h,l)\not
\in T$, $W_{j,k} \le W_{h,l}$ for every $(j,k)\in T:(j,k)\in\t{path}(h,l)$.
 \end{theorem}

 We prove by contradiction that we must have  $W_{j,k} \neq W_{h,l}$ for $(h,l)\not \in \cup_{m=1}^M {T^{(m)}_0}$. Suppose $W_{j,k} = W_{h,l}$ for a certain $(h,l)$ not in the tree and a $(j,k) \in \t{path}(h,l)$ in the minimum spanning tree $T^{(m)}_0$. Since $(j,k) \in \t{path}(h,l)$, we can disconnect $(j,k)$ and replace it with $(h,l)$; we have a new path such that $(h,l)\in \t{path}(j,k)$, forming a new minimum spanning tree, which contradicts the condition  $(h,l)\not \in \cup_{m=1}^M {T^{(m)}_0}$.
 \end{proof}

\subsection*{Proof of Theorem 4}

\begin{proof}
For better clarity, in this proof, we use simplified notations $m_{j,k}:=W_{n:j,k}$,  $M_{j,k}:=W_{0:j,k}$, $S:=S_n$ and $\Sigma:=\Sigma_0$.
The posterior mode corresponds to
\be
\hat T=\underset{T}{\arg\min} \sum_{e_s=(j,k)\in T}   (S_{j,j} + S_{k,k}-2 S_{j,k} ),
\ee
where $j\neq k$, and  $m_{j,k}=(S_{j,j} + S_{k,k}-2 S_{j,k} )$. Using $S_{j,k}= 1/n\sum_{i=1}^n y_j^{(i)} y_k^{(i)}$, we 
have
\be
m_{j,k}= n^{-1} \sum_{i=1}^n\{ y^{(i)}_j-y^{(i)}_k \}^2.
\ee
Letting $z^{(i)}_{j,k}=y^{(i)}_j-y^{(i)}_k$, $z_{j,k}^{(i)}$ has mean $0$, and
\be
\mathbb{E} \exp( t z^{(i)}_{j,k}) & \stackrel{(a)}\le \sqrt{\mathbb{E} \exp( 2t y^{(i)}_{j}) \mathbb{E} \exp(- 2t y^{(i)}_{k})}\\
& \stackrel{(b)}\le {\exp( 4t^{2} \lambda^{2}/2)}\\
\ee
where $(a)$ uses Cauchy-Schwarz inequality and $(b)$ uses  $\mathbb{E} \exp(tX) \le \exp(\lambda^{2} t^2/2)\; \forall t\in \mathbb R$ for the $\lambda-$sub-Gaussian random variable. Therefore, $z^{(i)}_{j,k}$ is $2\lambda$-sub-Gaussian.

We have the mean of $\{z^{(i)}_{j,k}\} ^2$ as
\be
M_{j,k}& = \mathbb{E} \{z^{(i)}_{j,k}\} ^2=  \Sigma_{j,j}+\Sigma_{k,k}- 2\Sigma_{j,k},
\ee
and it is not hard to see that $M_{j,k}= \mathbb{E} m_{j,k} =\mathbb{E}   (S_{j,j} + S_{k,k}-2 S_{j,k} )$ as well. 
Since $\Sigma$ is positive definite, letting $x$ be a $p$-element vector with $x_j=1$, $x_k=-1$ and all other elements $0$, we have  $M_{j,k} = x^{\rm T} \Sigma  x >0$. Further, $M_{j,k}\le 4 \max( \Sigma_{j,j},\Sigma_{k,k})$ due to $|\Sigma_{j,k}|\le \sqrt{   \Sigma_{j,j}\Sigma_{k,k}}  \le \max( \Sigma_{j,j},\Sigma_{k,k})$.

{\noindent \bf 1. Show that $(m_{j,k}-M_{j,k})$ is sub-exponential via the Bernstein's condition.} 

Our next goal is to check the Bernstein's moments condition for $w^{(i)}_{j,k}=\{z^{(i)}_{j,k}\}^2-M_{j,k}$, for all $q=2,3,\ldots$:
\bel\label{eq:berstein}
 { |\mathbb{E}\{ w^{(i)}_{j,k}\}^q} |\le  q!2^{-1}{v_{j,k}^2} \beta^{q-2},
\eel
where $v_{j,k}^2$ is the variance of $w^{(i)}_{j,k}$ and we want to find a valid constant $\beta$ that satisfies the inequality.

We now focus on a given index $(i,j,k)$. 
For ease of notation, we omit the subscript $(j,k)$ and superscript $i$. For $q=2$, \eqref{eq:berstein} holds trivially, hence we now focus on $q\ge 3$.
Using Lemma 1.4 from \cite{alma991011107259705251}, for any $q>0$, the moments of a $2\lambda$-sub-Gaussian random variable has
\bel\label{eq:sub-gau-moment-bound}
\mathbb{E} |z|^q \le 2(q/e)^{(q/2)} (2\lambda )^q.
\eel
We have for any $q =3,4,\ldots$:
\bel
\mathbb{E} |z^2|^q /q! & \le 2(2q/e)^{q} (2\lambda )^{2q} \\
& = 2^{3q+1} (q/e)^q \lambda^{2q} /q!\\
& \stackrel{(a)}\le (2/e)(8 \lambda^{2})^q \label{eq:z_q_fac_ineq}
\eel
where $(a)$ uses $q!  \ge e(q/e)^q $.
\be
\big|\frac{\mathbb{E} w^q}{q!} \big |^{1/q} 
& = | \frac{ \mathbb{E}(z^2-M_{j,k})^q}{q!} |^{1/q}\\
& \stackrel{(a)}\le ( \frac{\mathbb{E} |z^2-M_{j,k}|^q}{q!} )^{1/q}\\
& \stackrel{(b)} \le \frac{\mathbb{(E}  z^{2q})^{1/q} +M_{j,k}}{(q!)^{1/q}} \\
&\stackrel{(c)} \le  (2/e)^{1/q}8^{}  \lambda^{2}+ 4\lambda^2/ \{e^{1/q}(q/e)\} \\
& \stackrel{(d)} \le  8^{}  \lambda^{2}+4\lambda^2/ \{e^{1/3}(3/e)\}\\
&<  11\lambda^2.
\ee
where $(a)$ uses $|\mathbb Ex| \le \mathbb E |x|$ ; $(b)$ is due to the Minkowski inequality, $(\mathbb{E}|X+Y|^q)^{1/q}\le (\mathbb{E}|X|^q)^{1/q}+(\mathbb{E}|Y|^q)^{1/q}$ for any $q\ge 1$; $(c)$ uses \eqref{eq:z_q_fac_ineq}, $q!  \ge e(q/e)^q $,
and  $M_{j,k}\le 4 \max( \Sigma_{j,j},\Sigma_{k,k})$ and Lemma 1.2 from \cite{alma991011107259705251}, that for $\lambda$-sub-Gaussian $y^{(i)}_{j}$, its variance $\Sigma_{j,j}\le \lambda^2$; $(d)$ uses $(2/e)^{1/q}<1$, and $e^{1/q}(q/e)$ is increasing in $q\ge 3$. Therefore, 
\be
{ \mathbb{|E}w^q|}  \le q! (11\lambda^2)^q,
\ee
hence the next goal is to find $\beta$ such that $q! (11\lambda^2)^{q}\le q! 2^{-1}{v^2} \beta^{q-2} $.
Slight manipulation yields that, for $q\ge 3$, we need $\beta$ large enough such that
\be
 (11\lambda^2)^{q-2}\le  \frac{{v^2}}{2(11\lambda^2)^2} \beta^{q-2}.
\ee
In addition,
\be
v^2 & = \mathbb{E}w^2= \mathbb{E}(z^2-M_{j,k})^2= \mathbb{E}z^4-M^2 \le\mathbb{E}z^4 \stackrel{(a)}\le 2(4/e)^{2} (2\lambda )^4 \le 70\lambda^4,
\ee
where $(a)$ uses the sub-Gaussian moment bound \eqref{eq:sub-gau-moment-bound}. Therefore, we know $
 {{v^2}} /\{2(11\lambda^2)^2\} <  {{v^2}}/ (70\lambda^4) \le 1$, and a valid  constant is $$\beta \ge  \{ 2(11\lambda^2)^2/({v^2}) \} \times (11\lambda^2)= 2(11\lambda^2)^3/(v^2).$$ Further, note that $\beta\ge 2(11\lambda^2)^3/(70\lambda^4) \ge 38 \lambda^2>v$.

Now including the index $(i,j,k)$, $w^{(i)}_{j,k}$ is sub-exponential with parameters $v_{j,k}$ and $\beta_{j,k}$. This gives the Bernstein inequality for any $\epsilon>0$,
\be
\t{Pr}(  |m_{j,k}-M_{j,k}| \ge \epsilon)=
\t{Pr}(  |\frac{1}{n}\sum_{i=1}^n w^{(i)}_{j,k}| \ge \epsilon) & \le 2\exp\{ - \frac{n\epsilon^2}{2(v_{j,k}^2+ \beta_{j,k}\epsilon)}\}\\
&\stackrel{(a)}\le 2\exp\{ - \frac{n\epsilon^2}{2(\beta_{j,k}^2+ \beta_{j,k}\epsilon)}\},\\
& \stackrel{(b)}\le 2\exp\{ - \frac{n\epsilon^2}{2(\beta_0^2+ \beta_0\epsilon)}\},
\ee
where $(a)$ uses $v\le \beta$, and in $(b)$ we set $\beta_0= \max_{\t{all}(j,k)}\beta_{j,k}$.

{\noindent \bf 2. Analyze Prim's greedy algorithm to find the concentration inequality}

Let $T_0$ be the minimum        spanning tree  based on the oracle covariance:
\be
T_0 = \underset{T}{\arg\min} \sum_{e_s=(j,k)\in T}  M_{j,k},
\ee
where $ M_{j,k} = \Sigma_{j,j} + \Sigma_{k,k}-2 \Sigma_{j,k}$. We can now analyze the Prim's greedy algorithm applied on $m_{j,k}$'s and bound the probability of finding a  spanning tree $\hat T$ different than $T_0$.

For simplicity, let us start with the case when the oracle minimum spanning spanning tree is unique. At the step with two sets of nodes $U$ and $\bar U$, if an edge $(j',k') \in \mathcal E (U,\bar U)$ (the edge set between $U$ and $\bar U$) but $(j',k')\not \in T_0$, then there must be an edge $(j,k)\in \mathcal E (U,\bar U)$, $(j,k)\in T_0$ such that $(j,k)\in \t{path}(j',k')$. By the path optimality of the oracle minimum spanning tree, $M_{(j,k)}<M_{(j',k')}$. The probability of not having $m_{j,k}< m_{j',k'}$ has:
\be
\t{pr}( m_{j,k}\ge m_{j',k'}  )
& =\t{pr} \{ (m_{j,k}-M_{j,k} )- (m_{j',k'}-M_{j',k'}) \ge 
(M_{j',k'}- M_{j,k})\}\\
&\stackrel{(a)}  \le \t{pr} \{ (m_{j,k}-M_{j,k} )- (m_{j',k'}-M_{j',k'}) \ge 
\delta\}\\
& \stackrel{(b)} \le \t{pr} \{ |m_{j,k}-M_{j,k} | + |m_{j',k'}-M_{j',k'}| > 
\delta \}\\
& \stackrel{(c)} \le \t{pr} \{ |m_{j,k}-M_{j,k} |   > 
\delta/2\}
+ \t{pr} \{ |m_{j',k'}-M_{j',k'} |   > 
\delta/2\}
\\
& \le 
 4\exp[ - \frac{n \delta^2}{8(\beta_0^2+ \beta_0\delta} ]
\ee
where $(a)$ uses the condition $M_{(j',k')}-M_{(j,k)}\ge \delta$; $(b)$ is due to the former implies the latter; $(c)$ uses the union bound.

Given a node set $U$ and its complement $\bar U$, denote the event $G(U)$ as picking {\em any} edge $(j',k')$ from  $\mathcal E (U,\bar U)$ but not in $T_0$. Let $\tilde n(U)$ be the number of edges in $\mathcal E (U,\bar U)\cap E_{T_0}$, then using union bound we have:
\bel
\t{pr}  \{ G(U) \}
& \le  \{|\mathcal E (U,\bar U)|- \tilde n(U)\}
  4\exp[ -\frac{n \delta^2 }{8(\beta_0^2+ \beta_0\delta )} ] \\
 & \stackrel{(a)}\le  \{|\mathcal E (U,\bar U)|- 1\}
  4\exp[ -\frac{n \delta^2 }{8(\beta_0^2+ \beta_0\delta )} ], \label{eq:prim_bound}
\eel
where  $(a)$ uses $\tilde n(U)\ge 1$.

Letting $\{ U_1, U_2, \ldots, U_p\}$ be the sequence used to obtain the minimum spanning tree in the Prim's algorithm, with $U_1=\{1\}$ and $U_p=\{1,\ldots,n\}$, we have
\be
\t{pr} & \{ \bigcup_{k=1}^p G(U_k) \} \le 
\sum_{k=1}^p (|\mathcal E (U_k,\bar U_k)|-1)
  4\exp[ -\frac{n \delta^2 }{8(\beta_0^2+ \beta_0\delta )} ]\\
  \stackrel{(a)}\le 
&  \frac{2 p^3}{3}
  \exp[ -\frac{n \delta^2 }{8(\beta_0^2+ \beta_0\delta )} ]
  \\ 
  = 
  &  \frac{2}{3}
  \exp[ -\frac{n \delta^2 }{8(\beta_0^2+ \beta_0\delta )} 
  + 3\log p
  ]
\ee
where $(a)$ uses $\sum_{k=1}^p \{ (p-k)k-1\}= p(p^2-7)/6 \le p^3/6$.

Now consider the case when the oracle minimum spanning tree is not unique.
 Given a node set $U$ and its complement $\bar U$, denote the event $G(U)$ as picking {\em any} edge $(j',k')$ in the edges between $(U,\bar U)$ but not in one of $T_0$'s. Letting $\tilde n(U)$ be the number of edges in $\mathcal E (U,\bar U)\cap (\cup_{k=1}^K E_{T^{(k)}_0})$, clearly,   $\tilde n(U)\ge 1$, hence \eqref{eq:prim_bound} still holds. Therefore, the rest of the result follows.
\end{proof}

\subsection*{Calculation of the normalizing constant in the degree-based prior}

Let $r=\sum_{j=1}^p v_j$. Since $\eta = vv^{\rm T}$, we have $z(\eta)  ={p}^{-1}\prod_{j=2}^p \lambda_{(j)}(L)=|(\t{diag}(rv)-vv^{\rm T})_{2:p,2:p}|$ due  to the proof of Theorem 1, and the equivalence between the $1/p$ of the product of the top $(p-1)$ eigenvalues and the cofactor. Therefore using the matrix determinant lemma
\[
z(\eta)=r^{p-1}(\prod_{j=2}^p v_j) (1 - r^{-1}v_{2:n}^{\rm T} \t{diag}(v^{-1}_{2:p}) v_{2:p}) = r^{p-2}(\prod_{j=2}^p v_j)(r-\sum_{j=2}^p v_j)=r^{p-2}\prod_{j=1}^p v_j.\]

\subsection*{Calculation of the multivariate generalized double Pareto density}

Letting $\vec \beta = \vec y_j-\vec y_k \in \mathbb{R}^p$, we first multiply with $\Pi(\lambda_s^2)$ and integrate out $\lambda^2_s$ 
\be
&\int_{0}^{\infty}\left(\frac{1}{2 \pi \tau^{2} \lambda_s^2}\right)^{n / 2} \exp \left[-\frac{\|{\vec \beta}\|^{2}}{2 \tau^{2} \lambda_s^2}\right] \frac{\left(\frac{\kappa^2_s}{2}\right)^{\frac{n+1}{2}}\left(\lambda_s^2\right)^{\frac{n+1}{2}-1}}{\Gamma\left(\frac{n+1}{2}\right)} \exp \left[-\kappa^2_s \lambda_s^2 / 2\right] \mathrm{d} \lambda_s^2 \\
&= \frac{1}{\tau^{n}} \left(\frac{1}{2 \pi  }\right)^{\frac{n-1}{2}} \frac{1}{\Gamma\left(\frac{n+1}{2}\right)}
\left(\frac{\kappa^2_s}{2}\right)^{\frac{n-1}{2}} \\&
\times \int_{0}^{\infty}\left(\frac{1}{2 \pi\lambda_s^2}\right)^{1 / 2} \exp \left[-\frac{\|{\vec \beta}\|^{2}}{2 \tau^{2} \lambda_s^2}\right] \left(\frac{\kappa^2_s}{2}\right) \exp \left[-\kappa^2_s \lambda_s^2 / 2\right] \mathrm{d} \lambda_s^2 \\
&= \frac{1}{\tau^{n}} \left(\frac{1}{2 \pi  }\right)^{\frac{n-1}{2}} \frac{1}{\Gamma\left(\frac{n+1}{2}\right)}
\left(\frac{\kappa^2_s}{2}\right)^{\frac{n-1}{2}}
 \left(\frac{\kappa_s}{2}\right)
\exp \left[-\kappa_{s}  \frac{ \|{\vec
\beta}\|}{\tau} \right].
\ee

Next, we multiply with $\Pi(\kappa_s)$ and integrate out $\kappa_s$,
\be
& \int_0^{\infty} \frac{1}{\tau^{n}} \left(\frac{1}{2 \pi  }\right)^{\frac{n-1}{2}} \frac{1}{\Gamma\left(\frac{n+1}{2}\right)}
\left(\frac{1}{2}\right)^{\frac{n+1}{2}} \kappa^{n}_s
\exp \left[-\kappa_{s}  \frac{ \|{\vec
\beta}\|}{\tau} \right]
\frac{1}{\Gamma(\alpha)}\kappa_s^{\alpha-1} \exp(-\kappa_s)\mathrm{d} \kappa_s\\
&=\frac{1}{\tau^{n}} \left(\frac{1}{2 \pi  }\right)^{\frac{n-1}{2}} \frac{1}{\Gamma(\alpha)\Gamma\left(\frac{n+1}{2}\right)}
\left(\frac{1}{2}\right)^{\frac{n+1}{2}}
\int_0^{\infty}
\kappa^{n+\alpha-1}_s
\exp \left[-\kappa_{s}  (1+\frac{ \|{\vec
\beta}\|}{\tau}) \right]
\mathrm{d}\kappa_s \\
&= \frac{1}{2^n\Gamma\left(\frac{n+1}{2}\right) { \pi  }^{\frac{n-1}{2}}}   \frac{\Gamma(\alpha+n)}{\Gamma(\alpha)}
\frac{1}{\tau^{n}} \left (1+\frac{ \|{\vec
\beta}\|}{\tau}\right)^{-(\alpha+n)} .
\ee

\subsection*{Efficient Calculation of $(B_{[-s]}^{\rm T}B_{[-s]})^{-1}$ and Computational Complexity}

The matrix inversion can be computationally intensive for large $p$. In order to avoid a direct  matrix inversion at each Gibbs sampling step, we develop a fast  computing method that can: (i) extract $(B_{[-s]}^{\rm T}B_{[-s]})^{-1}$ from $(B^{\rm T}B)^{-1}$, (ii) update $(B^{\rm T}B)^{-1}$ when there is a change in one column of $B$.

For (i), suppose that we have the value of $(B^{\rm T}B)^{-1}$, without loss of generality, let  the matrix $B= [ B_{[-s]} \; \vec B_s]$, using block matrix form:
\be
(B^{\rm T}B)^{-1}=
\begin{bmatrix}
B_{[-s]}^{\rm T}B_{[-s]}	& B_{[-s]}^{\rm T}\vec B_s\\
\vec B_s^{\rm T} B_{[-s]}	& \vec B_s ^{\rm T}\vec B_s\\
\end{bmatrix}^{-1}
=
\begin{bmatrix}
M_{1,1}	& M_{1,2}\\
M^{\rm T}_{1,2} 	& M_{2,2},
\end{bmatrix}
\ee
where $M_{j,k}$ is corresponding block of $(B^{\rm T}B)^{-1}$. Using the block matrix inversion formula, we have: 
\be
(B_{[-s]}^{\rm T}B_{[-s]})^{-1} =M_{1,1}-M_{1,2}M_{2,2}^{-1}M^{\rm T}_{1,2}.
\ee
Since $M_{2,2}$ is a scalar, the above can be evaluated rapidly.

For (ii), supposing that we have updated $\vec B_s$ to $\vec B^*_s$, and denoting $B^*= [ B_{[-s]} \; \vec B^*_s]$, we want to calculate $(B^{*\rm T}B^*)^{-1}$. Note that
\be
(B^{*\rm T}B^*)^{-1}=
\begin{bmatrix}
B_{[-s]}^{\rm T}B_{[-s]}	& B_{[-s]}^{\rm T}\vec B^*_s\\
\vec B^{*\rm T}_s B_{[-s]}	& \vec B^{*\rm T}_s\vec B^*_s\\
\end{bmatrix}^{-1}
=
\begin{bmatrix}
M^*_{1,1}	& M^*_{1,2}\\
M^{*\rm T}_{1,2}	& M^*_{2,2}
\end{bmatrix}.
\ee
We have:
\be
 M^*_{2,2} &=  \{ \vec B^{*\rm T}_s\vec B^*_s- \vec B^{*\rm T}_s B_{[-s]} (B_{[-s]}^{\rm T}B_{[-s]})^{-1} 
B_{[-s]}^{\rm T}\vec B^*_s \}^{-1}\\
& = ( \vec B^{*\rm T}_s   
P_s\vec B^*_s)^{-1},\\
M^*_{1,2}	 & = -(B_{[-s]}^{\rm T}B_{[-s]}	)^{-1} B_{[-s]}^{\rm T}\vec B^*_s M^*_{2,2},\\
M^{*}_{1,1}&=
(B_{[-s]}^{\rm T}B_{[-s]}	)^{-1}
+(B_{[-s]}^{\rm T}B_{[-s]}	)^{-1}
B_{[-s]}^{\rm T}\vec B^{*}_s M^*_{2,2}
\vec B^{*\rm T}_s 
B_{[-s]}
(B_{[-s]}^{\rm T}B_{[-s]}	)^{-1}\\
&= 
(B_{[-s]}^{\rm T}B_{[-s]}	)^{-1}
+M^*_{1,2}	  M^{*-1}_{2,2}
M^{*\rm T }_{1,2}	,
\ee
where $P_s = I-  B_{[-s]} (B_{[-s]}^{\rm T}B_{[-s]})^{-1} B_{[-s]}^{\rm T}$, and we can use step (i) to compute $(B_{[-s]}^{\rm T}B_{[-s]}	)^{-1}$. Since $M_{2,2}$ and $M^*_{2,2}$ are scalars, all matrix inversions are avoided.

Therefore, throughout the posterior estimation, we only need to invert $B^{\rm T}B$ for one time to calculate the initial value. Examining the computational complexity, if using serial computation, the slowest matrix product/addition has a complexity of $O(p^2)$. Since most of the existing linear algebra toolboxes are optimized with some parallelization,  we now check the parallel computing complexity for each term above. Computing $(B_{[-s]}^{\rm T}B_{[-s]})^{-1}$ involves a matrix subtraction and  vector-scalar-vector product, which have complexity $O(1)$. Similarly, computing $ M^*_{1,1}$ has complexity $O(1)$. Computing $ M^*_{2,2} $ and $M^{*}_{1,2}$ involves matrix-vector products with complexity of $O(p)$. Lastly, when computing $\vec \beta_s$, we can bypass the matrix-matrix product by changing the order of multiplication to
 $\vec \beta_s = \{ I- B_{[-s]}(B_{[-s]}^{\rm T} B_{[-s]})^{-1} B_{[-s]}^{\rm T}\}
\vec B_s=  \vec B_s- B_{[-s]}(B_{[-s]}^{\rm T} B_{[-s]})^{-1} (B_{[-s]}^{\rm T}
\vec B_s) $; hence, it involves only matrix-vector product with a complexity of $O(p)$. As the result, overall  the projection has a parallel computing complexity of $O(p)$.

   \subsection*{Rapid Mixing of the Gibbs Sampler}

\begin{figure}[H] \centering
 \begin{subfigure}[t]{.8\textwidth}
 \centering
    \includegraphics[width=1\linewidth]{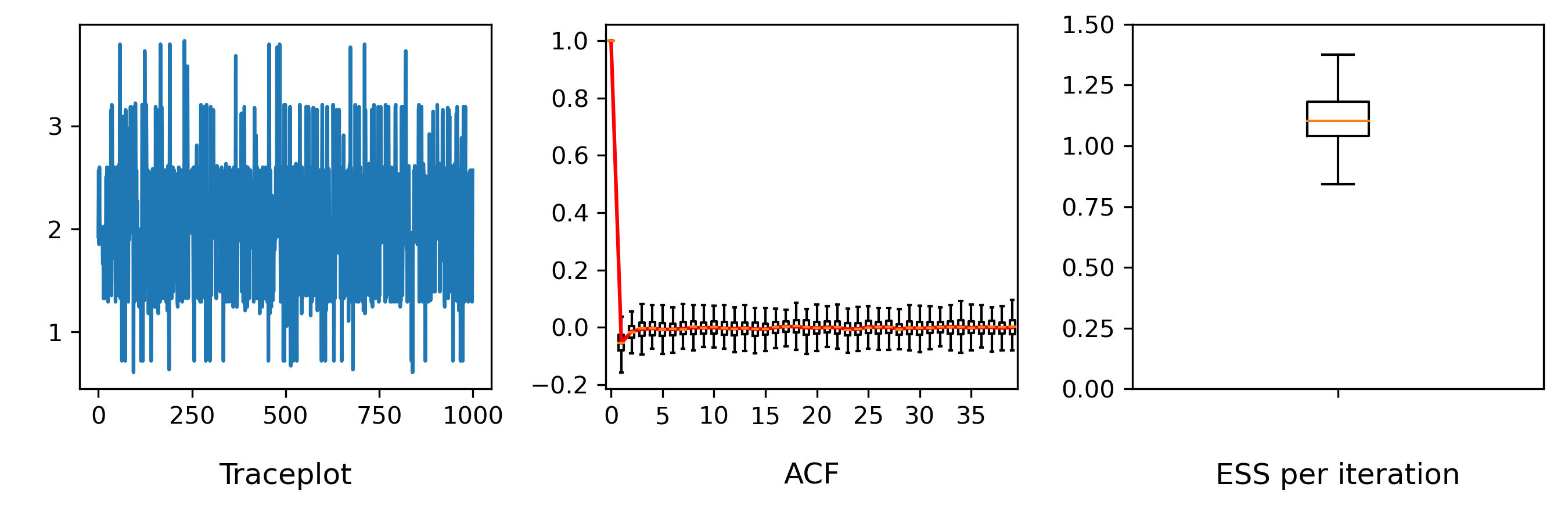}
    \caption{Convergence diagnostics on the degrees of the spanning tree.}
    \end{subfigure}
 \begin{subfigure}[t]{.8\textwidth}
  \centering
        \includegraphics[width=1\linewidth]{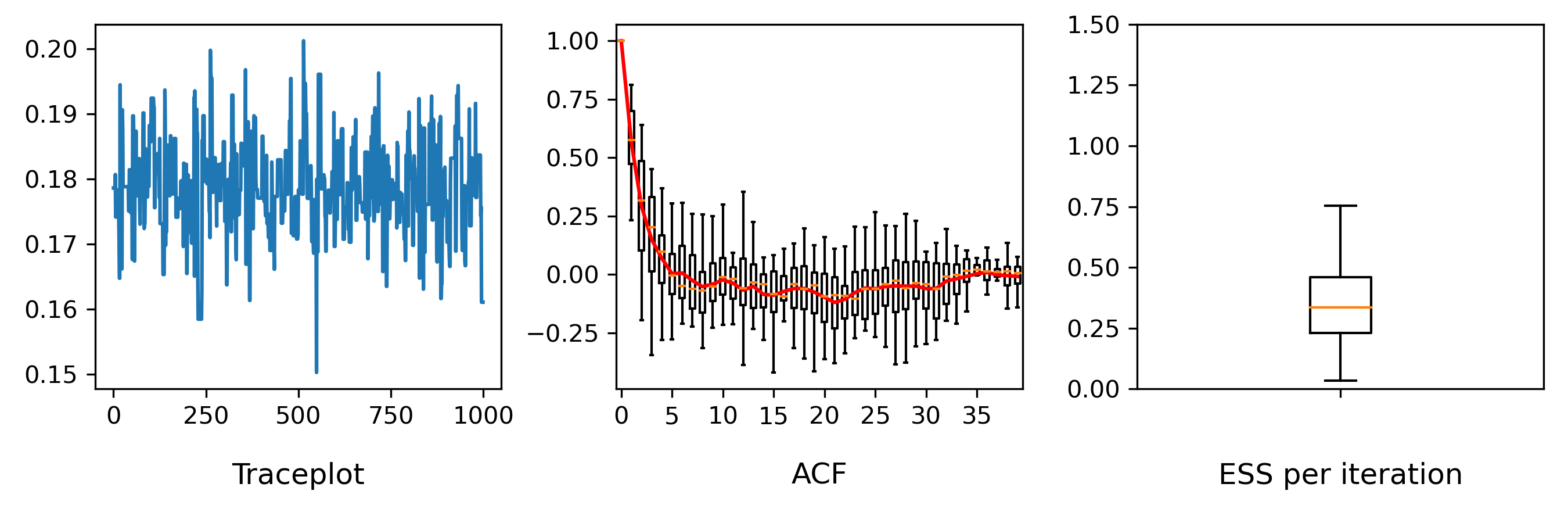}
            \caption{Convergence diagnostics on the scale parameter $\tau$.}
            \end{subfigure}
    \caption{The Gibbs sampler shows a rapid mixing of the Markov chains. Using the MCMC sample collected from the two-moon manifold experiments, we show the traceplot of the degree for one node / the scale parameter in one experiment, and autocorrelation plot computed based on $30$ repeated experiments, and their effective sample sizes per iteration. \label{fig:mcmc_diag}}
 \end{figure}

    The proposed Gibbs sampler exhibits  apparent rapid mixing empirically. As shown in Figure~\ref{fig:mcmc_diag} using the two-moon manifold simulation, the degrees of the tree change quickly from iteration to iteration, with the autocorrelation dropping to near zero almost within 1 lag; the update of the scale parameter $\tau$ using the random-walk Metropolis shows a fast drop to near zero within a lag of 10.  We found similar performance in all of the experiments and data applications demonstrated in the article.

\vspace*{-2cm}

\subsection*{Additional Simulation Details when the Oracle is a Sparse Graph}

\begin{figure}[H]
 \begin{subfigure}[t]{1\textwidth}
    \includegraphics[width=1\linewidth, trim={7cm 0 6cm 0},clip]{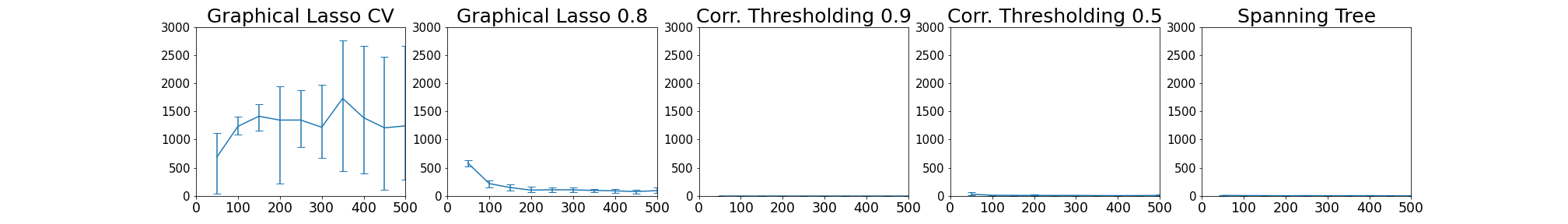}
    \caption{False positive edges comparing with $G_0$.}
\end{subfigure}
 \begin{subfigure}[t]{1\textwidth}
    \includegraphics[width=1\linewidth, trim={7cm 0 6cm 0},clip]{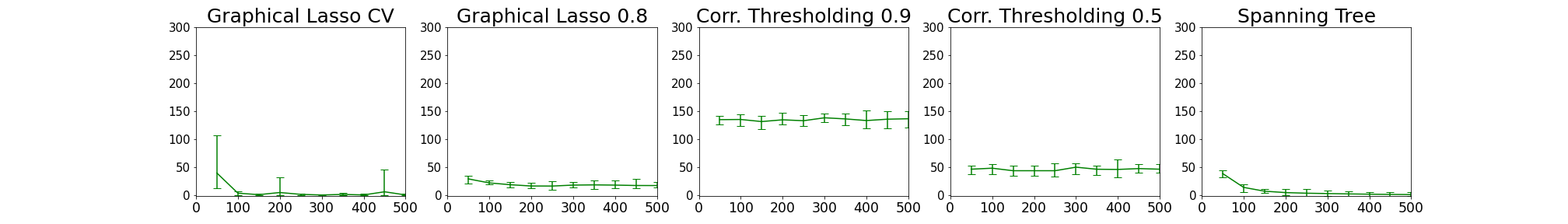}
    \caption{False negative edges comparing with $T_0$.}
\end{subfigure}
    \caption{The finite sample performance of graph estimation with the oracle being a sparse graph with about $600$ edges over $200$ nodes.  \label{fig:sim3b}
    }
 \end{figure}
 
 \subsection*{Additional Simulation Details when the Oracle is a Relatively Dense Graph}

\begin{figure}[H]
 \begin{subfigure}[t]{1\textwidth}
    \includegraphics[width=1\linewidth, trim={7cm 0 6cm 0},clip]{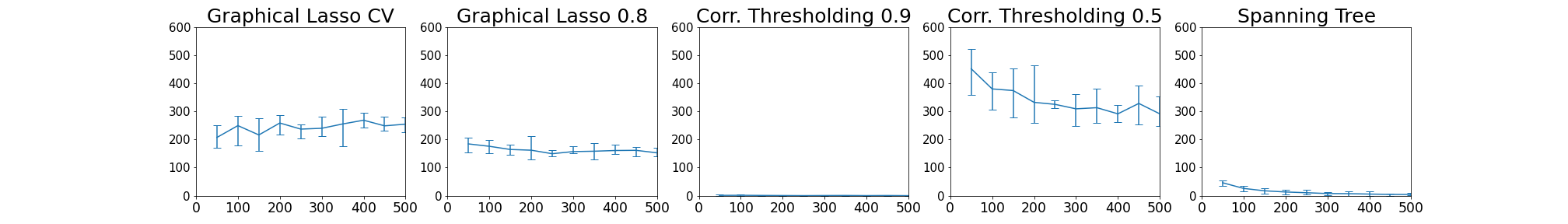}
    \caption{False positive edges comparing with $G_0$.}
\end{subfigure}
 \begin{subfigure}[t]{1\textwidth}
    \includegraphics[width=1\linewidth, trim={7cm 0 6cm 0},clip]{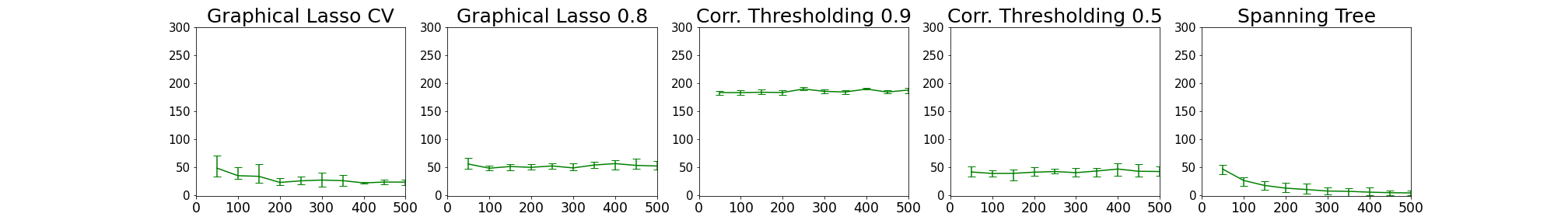}
    \caption{False negative edges comparing with $T_0$.}
\end{subfigure}
    \caption{The finite sample performance of graph estimation with the oracle being a relatively dense graph with about $4,000$ edges over $200$ nodes.  \label{fig:sim4b}
    }
 \end{figure}

 \subsection*{Graph Estimation When the Oracle is a Spanning Tree}

\begin{figure}[H]
\begin{subfigure}[t]{.25\textwidth}
    \includegraphics[width=1\linewidth]{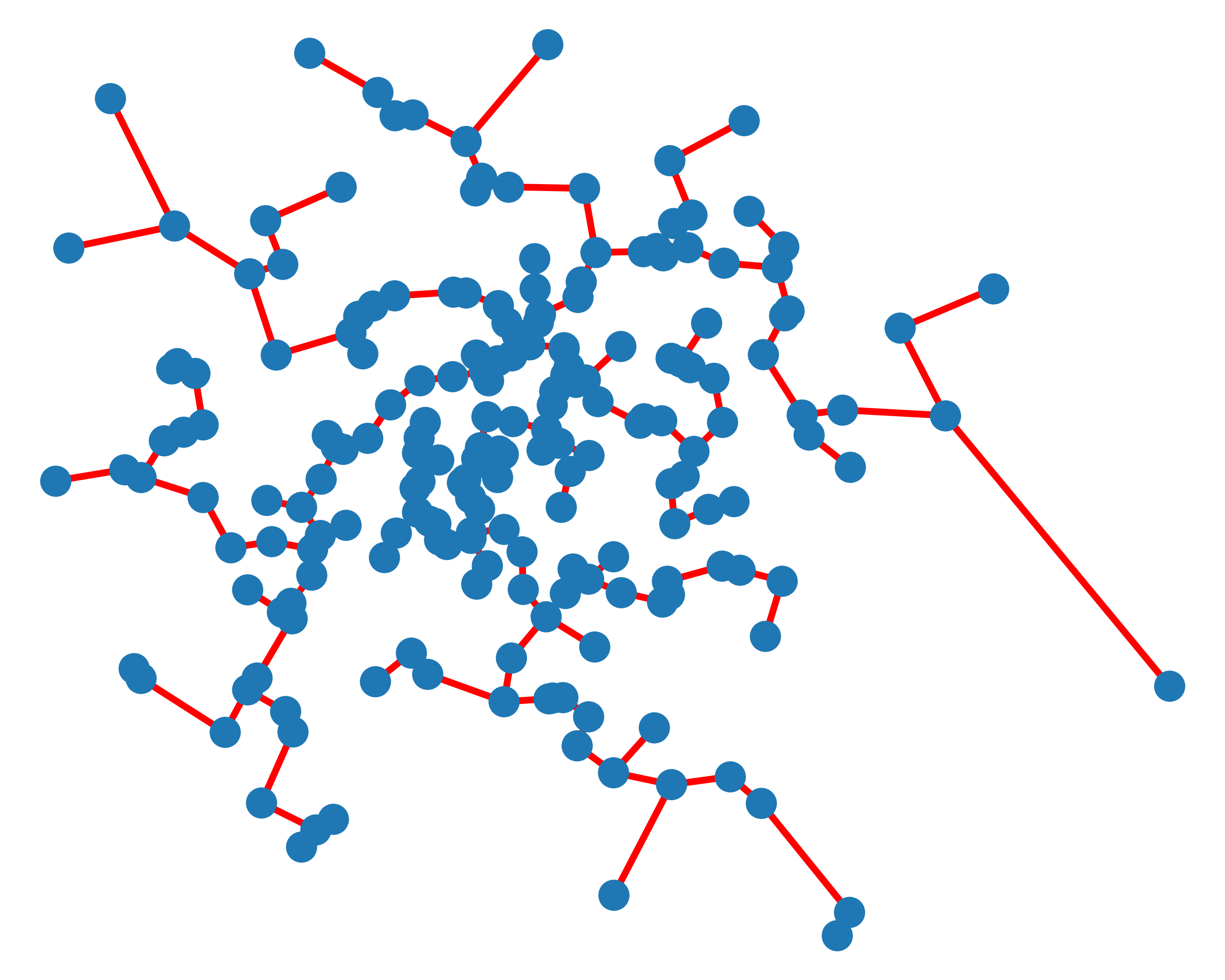}
    \caption{Oracle graph.}
\end{subfigure}
\quad
\begin{subfigure}[t]{.25\textwidth}
    \includegraphics[width=1\linewidth]{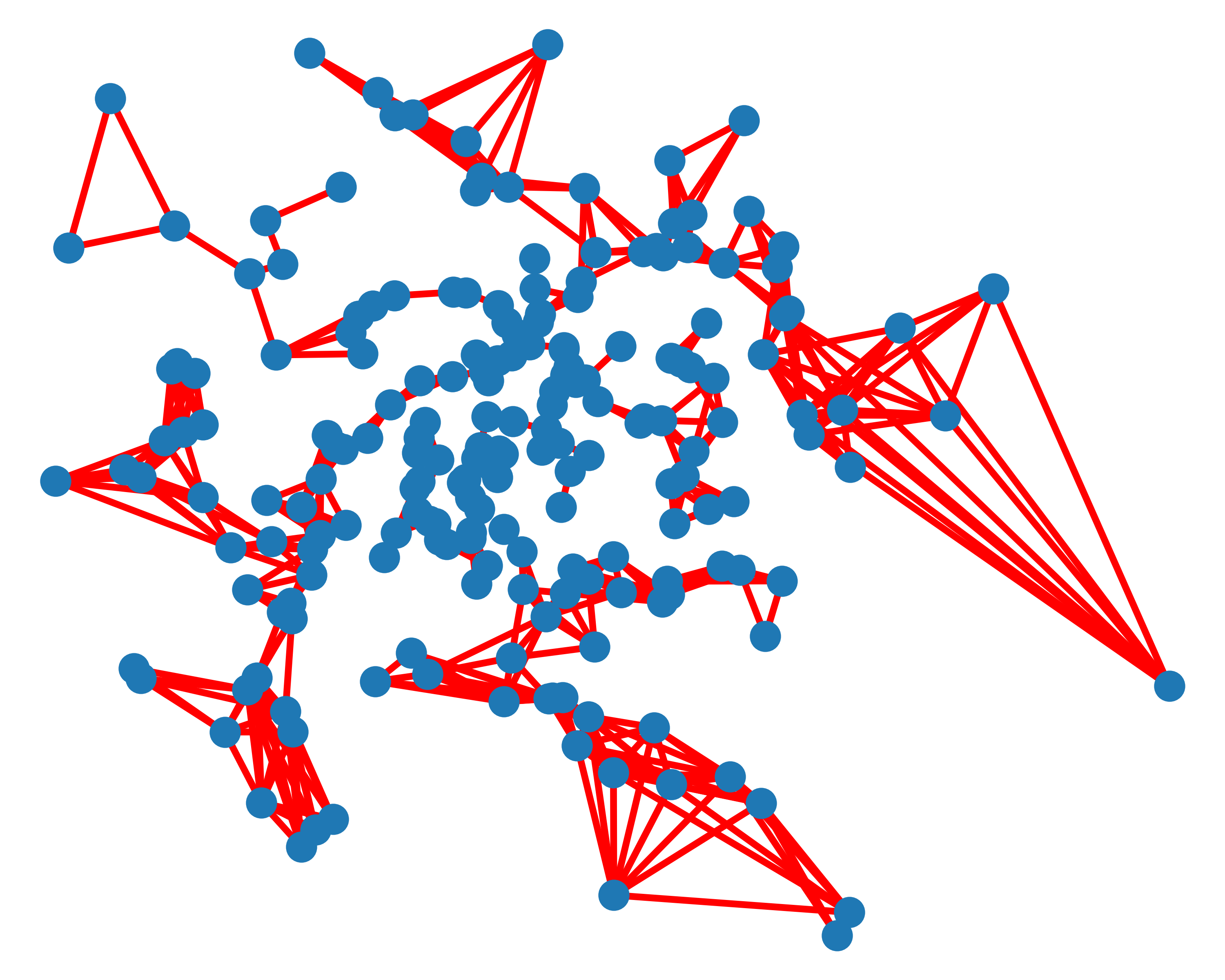}
    \caption{Thresholding absolute correlation at \(0.9\).}
\end{subfigure}
\quad
\begin{subfigure}[t]{.25\textwidth}
    \includegraphics[width=1\linewidth]{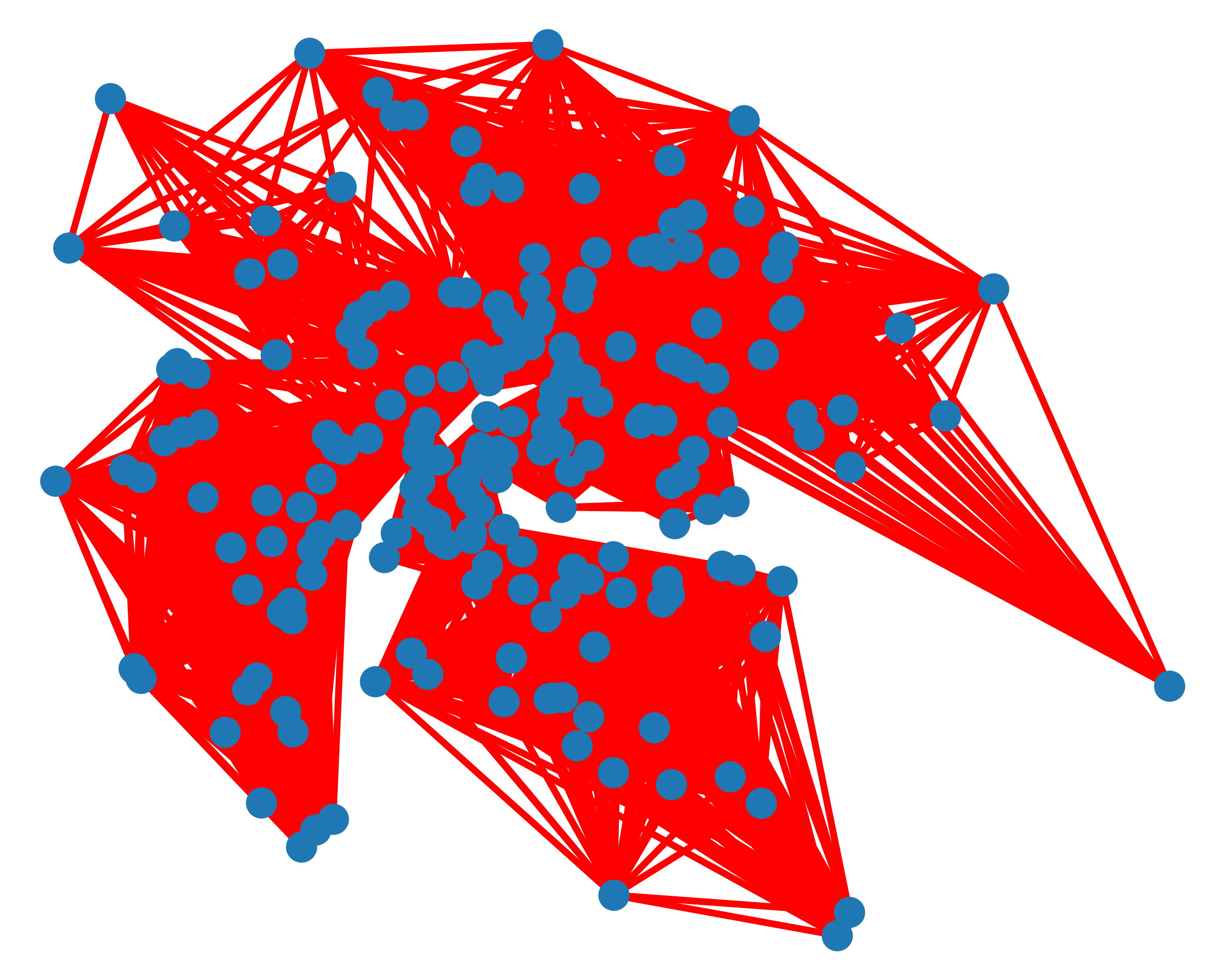}
    \caption{Thresholding absolute correlation at \(0.5\).}
\end{subfigure}\\
\begin{subfigure}[t]{.25\textwidth}
    \includegraphics[width=1\linewidth]{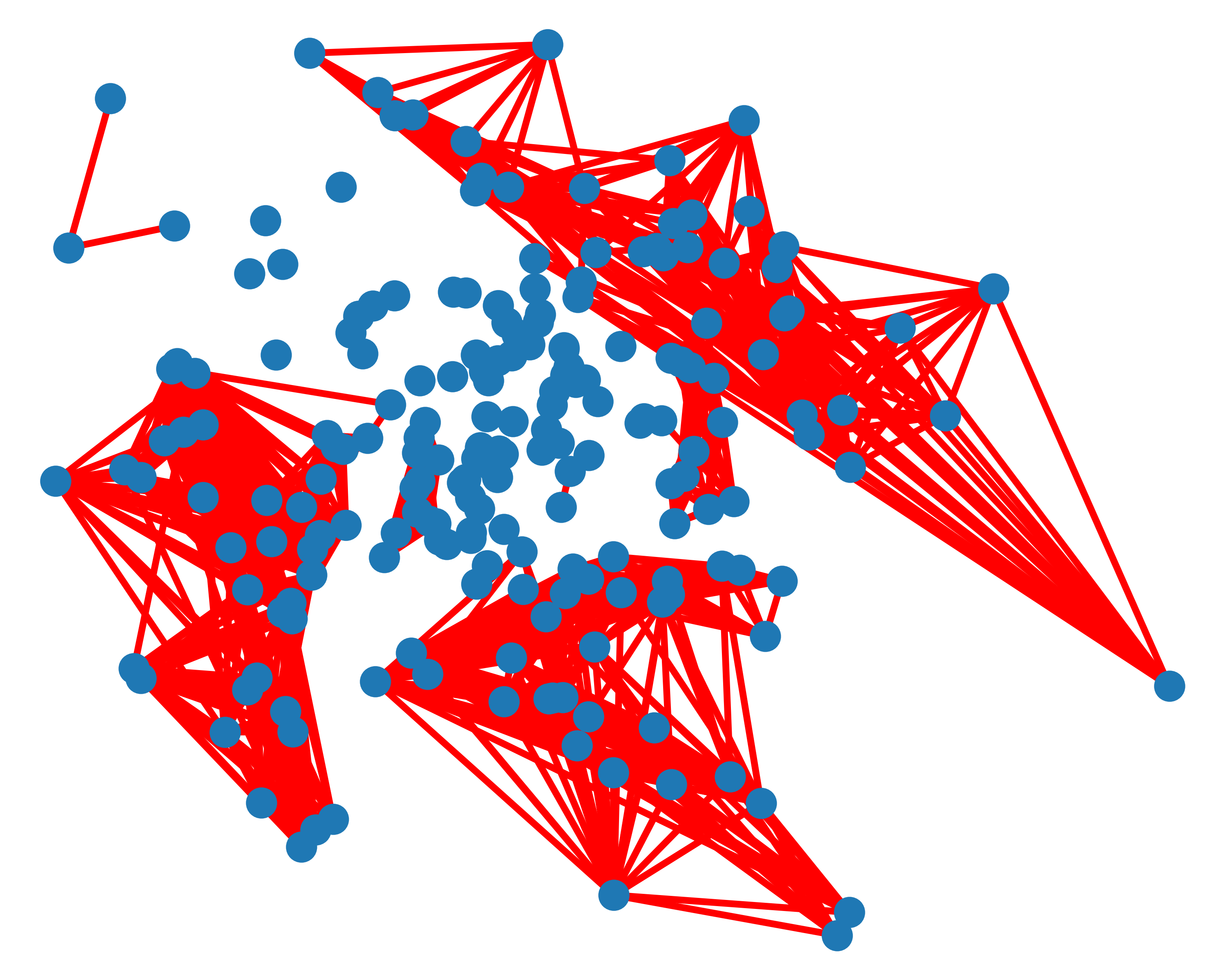}
    \caption{Graphical lasso with \(\alpha\) chosen by cross-validation.}
\end{subfigure}
\quad
\begin{subfigure}[t]{.25\textwidth}
    \includegraphics[width=1\linewidth]{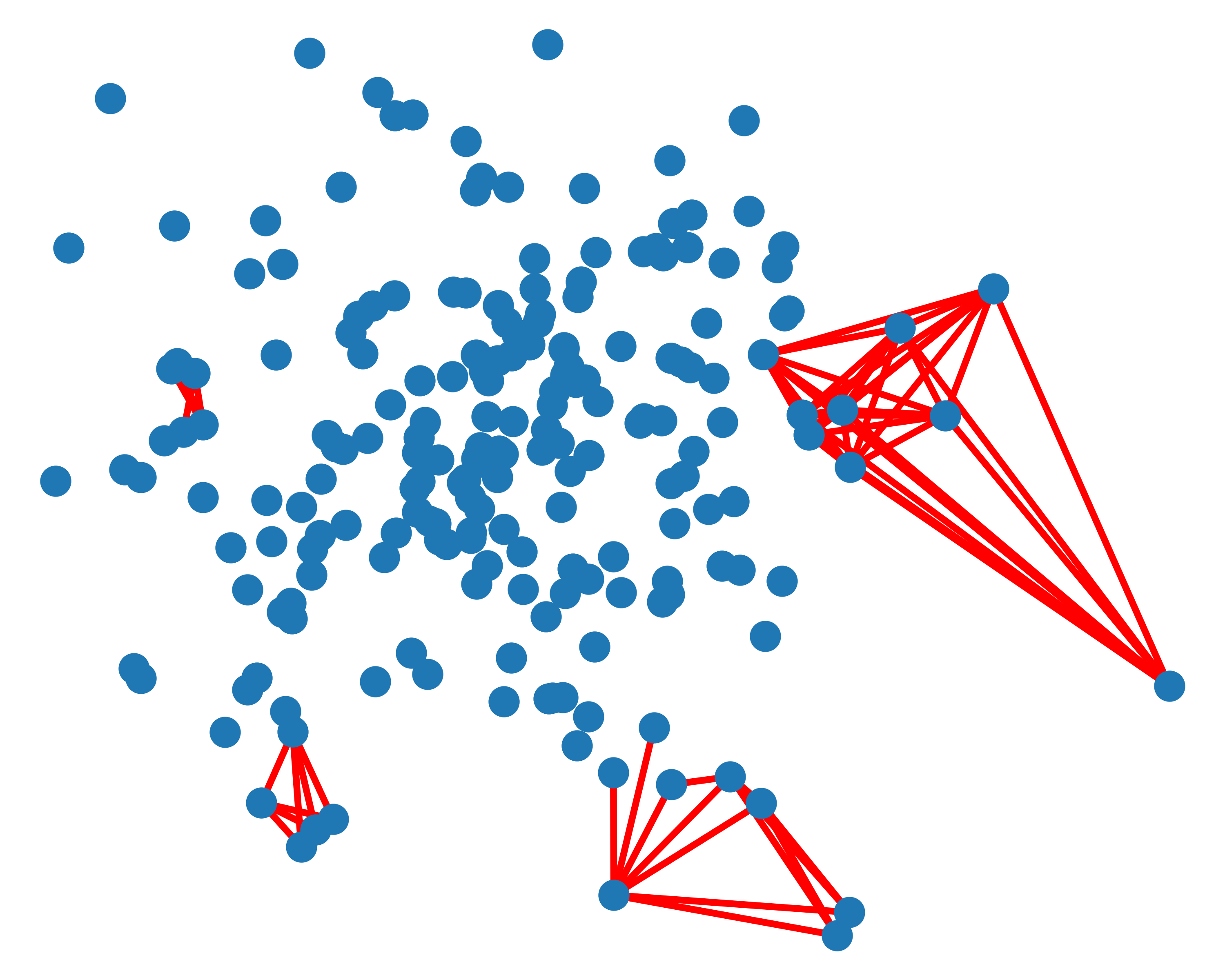}
    \caption{Graphical lasso using \(\alpha=20\).}
\end{subfigure}
\quad
\begin{subfigure}[t]{.25\textwidth}
    \includegraphics[width=1\linewidth]{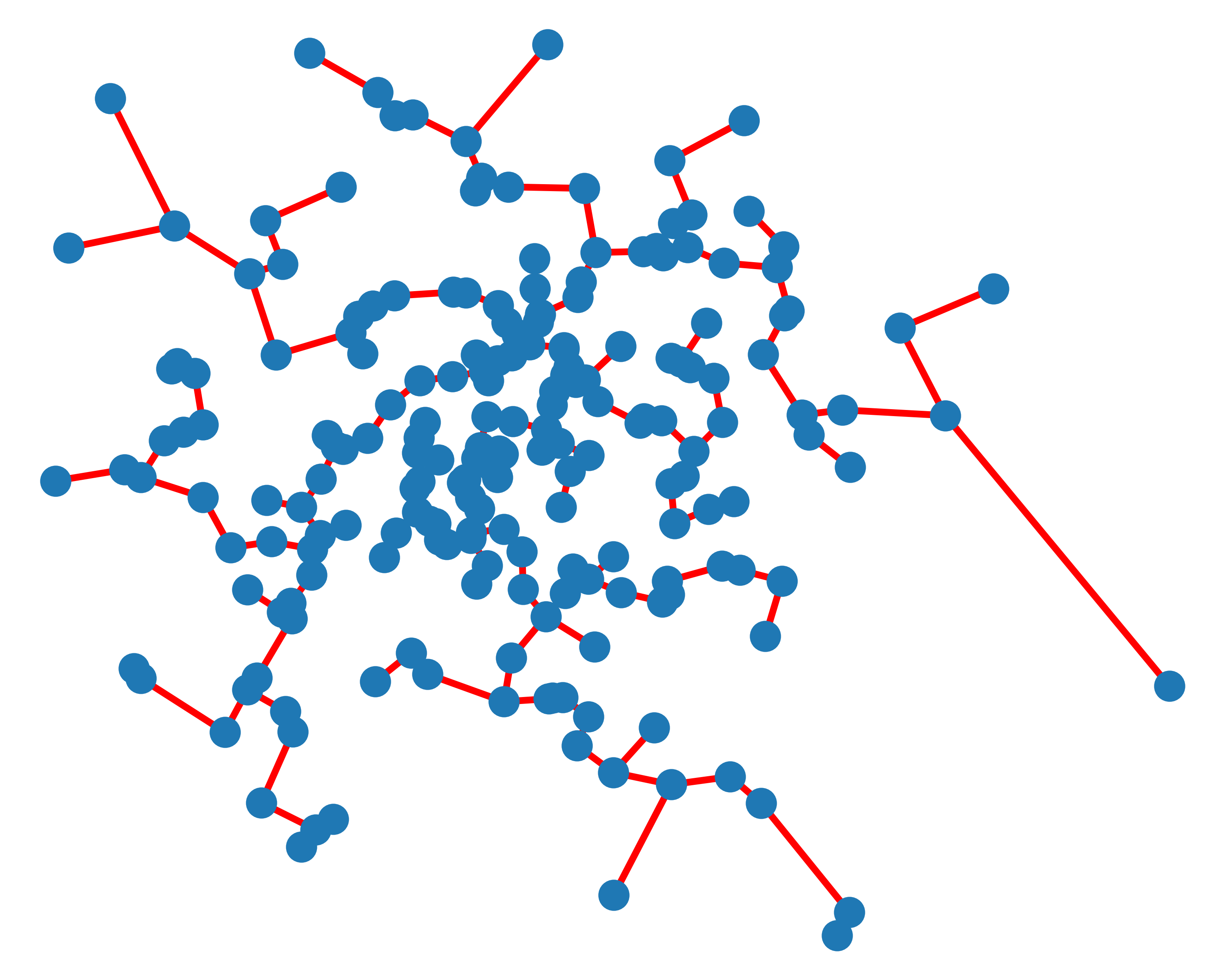}
    \caption{Bayesian spanning tree (posterior mode shown).}
\end{subfigure}
 \caption{Simulated experiments of recovering a graph with \(p=200\) nodes, where the oracle graph is a spanning tree (Panel a). Panels (b-f) are plotted for \(n=50\). Starting around \(n=50\), the Bayesian spanning tree successfully recovers the ground truth with almost no errors, whereas the other approaches show many false positives.
 \label{fig:sim1}}
 \end{figure}

 We consider data generated from a spanning tree [Figure~\eqref{fig:sim1}(a)]. 
The thresholding estimator and graphical lasso show many false positives (Panels b-d). Empirically tuning the graphical lasso to \(\alpha=20\) does somewhat reduce false positives, however, it leads to edge loss and more false negatives (Panel e).
The thresholding estimator has a similar sensitivity issue: thresholding at \(0.5\), as a common ``default'' choice in practice, leads to a severe overestimation of the graph edges, while \(0.9\) reduces this problem to some extent. On the other hand, coherent with the generative model, the Bayesian spanning tree shows good performance (Panel f).

\end{document}